\documentclass[12pt,a4paper]{article}
\usepackage[T1]{fontenc}
\usepackage[utf8]{inputenc}
\usepackage{authblk}
\usepackage{bbding} 

\usepackage{amsmath}
\usepackage{graphicx,float,color}
\usepackage{hyperref}
\hypersetup{colorlinks = true, linkcolor = blue, urlcolor  = blue, citecolor = blue, anchorcolor = blue}
\usepackage{multirow}
\usepackage[numbers,sort&compress]{natbib}
\usepackage{braket}
\usepackage{amsfonts}
\usepackage{amssymb} 
\usepackage{epic}
\usepackage{eepic}
\usepackage{epsfig}
\usepackage{latexsym}
\usepackage{color}
\usepackage{url}  
\usepackage{caption}
\usepackage[caption=false]{subfig}
\usepackage{float}
\usepackage{bm}
\usepackage{geometry}
\usepackage{slashed}
\usepackage[detect-all]{siunitx}

\def\Mp{m_{\mathrm{Pl}}}
\def\lp{\ell_{\mathrm{Pl}}}

\def\Mmin{M_{\mathrm{min}}}
\def\hMmin{\hat{M}_{\mathrm{min}}}
\def\Mini{M_{\mathrm{init}}}
\def\hMini{\hat{M}_{\mathrm{init}}}
\def\Mpeak{M_\textrm{peak}}

\usepackage{lipsum}
\makeatletter
\def\ps@pprintTitle{%
 \let\@oddhead\@empty
 \let\@evenhead\@empty
 \def\@oddfoot{}%
 \let\@evenfoot\@oddfoot}
\makeatother
\usepackage{etoolbox}

\begin{document}
\title{\textbf{Not quite black holes as dark matter}}    

\author[1]{Ufuk Aydemir\thanks{uaydemir@ihep.ac.cn} }
\author[2]{Bob Holdom\thanks{bob.holdom@utoronto.ca} }
\author[1]{Jing Ren\thanks{renjing@ihep.ac.cn}}
\affil[1]{\normalsize Institute of High Energy Physics, Chinese Academy of Sciences, Beijing 100049, P. R. China}
\affil[2]{Department of Physics, University of Toronto, Toronto, Ontario, Canada M5S 1A7}

\maketitle

\begin{abstract}
Primordial black holes that survive until the present have been considered as a dark matter candidate.
In this paper we argue that primordial 2-2-hole remnants provide a more promising and testable option. 2-2-holes arise in quadratic gravity as a new family of classical solutions for ultracompact matter distributions and they possess the black hole exterior without an event horizon. They may serve as the endpoint of gravitational collapse, providing a resolution for the information loss problem. Intriguing thermodynamic behavior is found for these objects when sourced by a thermal gas.
A large 2-2-hole radiates with a Hawking-like temperature and exhibits an entropy-area law. 
At a late stage, the evaporation slows down and essentially stops as the mass asymptotically approaches a minimal value.
This remnant mass is determined by a fundamental scale in quadratic gravity.
We study the cosmological and astrophysical implications of having these remnants as dark matter and derive the corresponding constraints.
A distinctive phenomenon associated with remnant mergers occurs, predicting fluxes of high-energy astrophysical particles due to the spectacular evaporation of the merger product. 
Measurements of high-energy photon and neutrino fluxes could possibly bound the remnant mass to be not far above the Planck mass.
Early-universe physics, on the other hand, requires that 2-2-holes quickly evolve into the remnant state after formation, putting an upper bound on the formation mass.
\end{abstract}

\newpage

\section{Introduction\label{sec:intro}}

With the direct detection of gravitational waves, a new era of testing the strong gravity regime has begun~\cite{LIGOScientific:2019fpa}. The signals, which appear to originate from stellar-mass astrophysical black holes, are so far consistent with General Relativity (GR). However, the implications regarding physics near the black hole horizon are not clear, and it is indeed this regime where deviations from GR might make their first appearance. Such deviations are strongly motivated by the possible resolution of information loss paradox. In particular, it might be a result of the underlying quantum gravity, although at first glance the Planck scale physics is not expected around a macroscopic horizon from naive dimensional arguments. An extraordinary, yet simple, possibility is that quantum gravity effects prevents formation of the horizon, generating horizonless ultracompact objects instead of black holes. These objects appear similar to black holes for current observations, but they may leave distinctive imprints in gravitational wave signals~\cite{Cardoso:2019rvt}. 

Another great puzzle confronting the modern physics for decades is dark matter, the nature of which has so far remained elusive with only evidence coming from gravitational interactions. Among the well studied dark matter candidates in the literature have been Primordial black holes (PBHs)~\cite{Carr:1975qj}. They have recently garnered more attention because of the null results of searches for the dark matter particles as well as the new testing opportunities due to the direct detection of gravitational waves. Yet, the present mass fraction of PHBs in dark matter is heavily constrained. In the standard scenario, where the validity of Hawking radiation is assumed all the way down to complete evaporation, $M_\textrm{PBH}\gtrsim 10^{15}\,$g is required for PBHs to survive until now and account for the dark matter, for which only very few narrow mass windows are still available~\cite{Carr:2009jm,Carr:2016drx}. 

However, it has been conjectured that the evaporation may come to a stop at some stage, and instead of an explosion as the end point, a remnant is left behind and may serve as dark matter~\cite{MacGibbon:1987my,Barrow:1992hq,Carr:1994ar}. For this case, the lower mass range $M_\textrm{PBH}\lesssim 10^{15}\,$g is still allowed. In fact, phenomenological studies show that all the observational constraints can be evaded if PBHs radiate away most of their energy before Big Bang Nucleosynthesis (BBN). When the initial mass satisfies $M_\textrm{PBH}\lesssim 10^{6}\,$g, the leftover Planck mass remnants can account for all of dark matter~\cite{Dalianis:2019asr}. The obvious challenge is then to understand the mechanism responsible for the generation of such remnants. Theoretically, black hole remnants can be realized by either modifying gravity or the matter sector~\cite{Chen:2014jwq}. However, ideas along these lines are often dismissed, mainly because of their apparent ad-hoc nature and their failure to resolve the information loss paradox. Modifications around the macroscopic black hole horizon are already expected before reaching the remnant stage, which constitutes an obstacle in addressing the information loss problem with black hole remnants.

In this paper, a theoretical model for horizonless ultracompact objects as dark matter is investigated. Remarkably,  a remnant naturally arises as a consequence of new physics at a microscopically small distance that in turn determines the mass of the remnant. The underlying theory is quadratic gravity, a candidate for quantum gravity in the framework of quantum field theories. By including all the quadratic curvature terms on top of the Einstein-Hilbert action, i.e. the Weyl term $C^{\mu\nu\rho\sigma}C_{\mu\nu\rho\sigma}$ and the Ricci term $R^2$, quadratic gravity provides a renormalizable and asymptotically free UV completion of GR at dimension four spacetime~\cite{Stelle:1976gc, Voronov:1984kq, Fradkin:1981iu, Avramidi:1985ki}.\footnote{At the classical level the theory suffers from a long-known ghost problem associated with the higher derivative terms. There are proposed solutions to deal with the ghost by taking quantum corrections seriously~\cite{Lee:1969fy, Tomboulis:1977jk, Grinstein:2008bg, Anselmi:2017yux, Donoghue:2018lmc, Bender:2007wu, Salvio:2015gsi, Holdom:2015kbf, Holdom:2016xfn, Salvio:2018crh}. That problem aside, quadratic gravity provides a more tractable framework to study high curvature effects around macroscopic would-be  horizons. }

The horizonless ultracompact object in question,  referred to as a 2-2-hole, emanates as a classical
solution in the theory when sourced by a compact matter distribution~\cite{Holdom:2002xy, Holdom:2016nek}. The 2-2-hole has no analog in GR and is closely related to the Weyl term. Its exterior closely resembles that of a black hole, while in the interior a novel high-curvature solution takes over. A transition region at around the would-be horizon is where significant deviations from a black hole first occur. As the most generic solution in the theory, it may serve as the endpoint of gravitational collapse. In contrast to many other ultracompact objects, a 2-2-hole can be arbitrarily heavy, but it has a minimum allowed mass $\Mmin$, thus indicating the existence of stable remnants. Therefore, not only does the 2-2-hole provide a resolution for information loss paradox due to the absence of horizon, the leftover remnants of primordial 2-2-holes formed in the early universe can very well be considered as a dark matter candidate. 

To investigate 2-2-hole remnants as dark matter, their thermodynamic properties are essential. Recently, solutions sourced by a thermal gas were found in \cite{Holdom:2019ouz} and studied further in \cite{ Ren:2019afg}. This simple form of matter may describe the final state of infalling matter in the high curvature interior. Unlike in GR, in quadratic gravity the thermal gas is able to support an ultracompact configuration without collapsing into a black hole. This model then enables the study of 2-2-hole thermodynamics in terms of properties of a thermal gas on a curved background. 
Thermal 2-2-holes with different masses exhibit qualitatively distinct behaviors~\cite{Holdom:2019ouz, Ren:2019afg}.   A large 2-2-hole with mass away from $\Mmin$ resembles a black hole thermodynamically, notwithstanding its different origin. The temperature is proportional to Hawking temperature up to a constant and the entropy satisfies the area law. A small 2-2-hole with mass quite close to $\Mmin$, on the other hand, behaves more like an ordinary thermodynamic system. With both temperature and entropy approaching zero in the minimal mass limit, it behaves as a stable remnant.

Such a change in thermodynamic behavior in the minimal mass limit is not unprecedented and occurs in various models for black hole remnants, e.g. extremal black holes. What is appealing for the 2-2-hole case is that the absence of horizon and the stabilization mechanism for the small objects both stem from quadratic curvature terms that operate at high energies or curvatures. As a result, a large primordial 2-2-hole starts by radiating like a black hole with increasing temperature and radiation power. After reaching the peak temperature, it enters into the remnant stage with much lower temperature and power. During the course of evaporation, the entropy of 2-2-hole gradually decreases with the information carried out by the thermal radiation, as with any burning object. Therefore, unlike the case of black hole remnants, there is no issue of an arbitrarily large amount of  entropy stored in a small-size object.

The observational constraints for primordial thermal 2-2-hole remnants as dark matter will be explored in this paper. For this purpose, thermodynamic features of 2-2-holes sourced by a thermal gas are elaborated in Sec.~\ref{sec:thermal}, where similarities and differences from PHBs are discussed. 
The present-epoch observations for the 2-2-hole remnants are studied in Sec.~\ref{sec:present}. As a new phenomenon specific to 2-2-holes, the binary merger of two remnants gives rise to a high temperature product, with the excess energy released almost instantly by emitting high-energy particles. We explore this process and its observational consequences in detail in Sec.~\ref{sec:merger}. The early-universe physics for primordial 2-2-holes are investigated in Sec.~\ref{sec:pheno}, including the requirement from the observed relic abundance and observations from BBN and CMB. The later mainly constrains the early-stage evaporation of primordial 2-2-holes. 
The observational constraints and their implications are discussed in Sec.~\ref{sec:discuss}, with the main results summarized in Fig.~\ref{fig:LUcons} and Fig.~\ref{fig:EUcons} for the present-epoch and early-universe constraints, respectively. The paper is concluded in Sec.~\ref{sec:final}.


\section{Thermal 2-2-holes}
\label{sec:thermal}

\begin{figure}[!h]
  \centering%
{ \includegraphics[width=17cm]{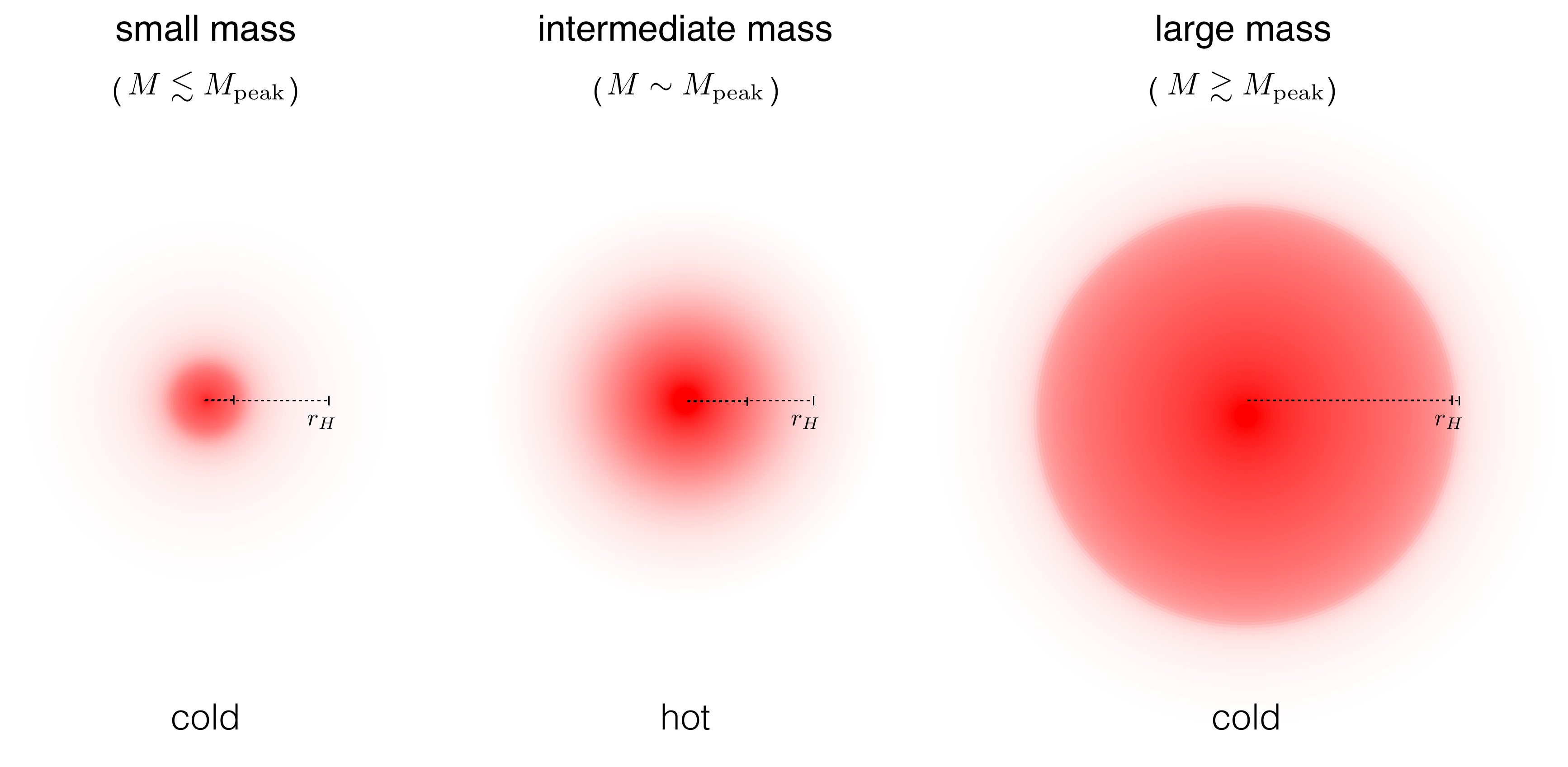}}
\caption{\label{fig:22holes} 
Schematic plots for thermal 2-2-holes with different masses. $M_\textrm{peak}\approx 1.2\Mmin$ denotes the 2-2-hole mass at which the temperature at infinity is maximized. For each plot, the color represents relative magnitude of the temperature (or curvature invariants) as a function of the radial coordinate. The ticks denote the interior size and the would-be horizon size $r_H$.}
\end{figure}

The qualitative features of 2-2-holes are quite simple, although their solutions can only be found numerically due to the nontrivial field equations. Focusing on the spherically symmetric, asymptotically flat and static cases, 
a general 2-2-hole consists of an exterior that resembles the Schwarzschild solution with the same physical mass $M$, an interior characterized by a novel high-curvature region as dominated by the quadratic curvature terms and a transition region around the would-be horizon $r_H=2M\lp^2$ that links the two regimes. 
The existence of 2-2-holes relies on the Weyl term $C^{\mu\nu\rho\sigma}C_{\mu\nu\rho\sigma}$ in the quadratic action, which introduces a new spin-2 mode with mass $m_2$. 
This mass scale determines the minimum mass for the 2-2-hole as
\begin{eqnarray}
\label{eq:Mmin}
\hMmin\equiv\frac{\Mmin}{\Mp}\approx 0.63\frac{\Mp}{ m_2}\approx 0.63 \frac{\lambda_2}{\lp},
\end{eqnarray}
meaning that the size of 2-2-holes is bounded from below by the Compton wavelength $\lambda_2$ of the spin-2 mode.
There are two scenarios for quadratic gravity, as defined by the strength of dimensionless couplings associated with the quadratic curvature terms.  
In the strong coupling scenario, the Planck mass arises dynamically by dimensional transmutation, and there is only one mass scale $m_2\approx \Mp$, i.e. $\hMmin\approx 0.63$. In the weak coupling scenario, the Planck mass can arise either spontaneously through vacuum expectation values of some scalar fields or it can be put in explicitly. For this case, there can be a large mass-hierarchy with $m_2\ll \Mp$, i.e. $\hMmin\gg 1$. 

We find notably different behaviors for thermal 2-2-holes, as shown schematically in Fig.~\ref{fig:22holes}.  
A large hole with $M\gtrsim M_\textrm{peak}$ has an extremely narrow transition region around $r_H$, and it appears very much like a black hole for an outside observer. The high temperature thermal gas filling the interior can be thought as a firewall with a large angular proper length $\sim r_H$ but with a rather small radial proper length $\sim \lambda_2$. This novel interior geometry leads to anomalous thermodynamics e.g. negative heat capacity and the area law for the entropy, as in the case of black holes but with different numerical values. 
A small 2-2-hole with $M\lesssim M_\textrm{peak}$ (``small’’ refers to $M$ being close to $\Mmin$ even when $\Mmin$ is large), on the other hand, has a broader transition region  and a shrinking interior. Thermodynamically it behaves more like the self-gravitating radiation inside a box, with positive heat capacity and the entropy scaling trivially with the interior size. When $M\to \Mmin$, the temperature at infinity, entropy, and the interior size all approach zero. In between these two distinctive behaviors is the intermediate-mass realm, where the 2-2-hole temperature at infinity reaches a maximum.   
In Appendix~\ref{sec:22hole}, we provide more details of the structure of 2-2-holes in these particular cases.

In the following, we first review thermodynamic properties of the thermal 2-2-hole in Sec.~\ref{sec:thermo}, and then derive time evolutions of various quantities during the 2-2-hole evaporation in Sec.~\ref{sec:evapo}. These properties turn out to be quite simple, as mainly determined by the mass of the hole $M$ and its minimum allowed value $\Mmin$.

\subsection{Thermodynamics}
\label{sec:thermo}

The thermal gas that sources the 2-2-hole background may include particles of all kinds. In addition to those from the original infalling matter, any new species will be produced by particle collisions in the high curvature interior. This may include ultra-heavy particles due to the extremely high temperature deep inside. As was found in \cite{Ren:2019afg}, gas particles with large mass can significantly change the interior matter distribution, while the thermodynamic properties of the hole for an outside observer remain quite insensitive to this effect. Therefore, for our
phenomenological study, it is a good approximation to consider the thermal gas model with only massless relativistic particles. 
The energy density and pressure are then given as, 
\begin{equation}
\rho=3p=\frac{\pi^2}{30} \mathcal{N}\, T^{4}\,.
\end{equation}
$T(r)$ is the local measured temperature and 
$\mathcal{N}= g_b+7 g_f/8$, where $g_b$ and  $g_f$ are the number of bosonic and fermionic degrees of freedom. In principle, $\mathcal{N}$ includes particle species of all kinds and could be much larger than its Standard Model value. In the following, we make the $\mathcal{N}$-dependence explicit so that its impact on the results can be clearly seen. 
Given the conservation law of the stress tensor, $T(r)$ satisfies Tolman’s law ($T(r)g_{00}^{1/2}=T_\infty$) and so grows large inside the gravitational potential. The value at spatial infinity $T_\infty$ is roughly the temperature measured by a distant observer. 
When the 2-2-hole is not in thermal equilibrium with its surroundings, $T_\infty$ represents the temperature at which it radiates as a black body. 
The total entropy and energy of the gas are related as, 
\begin{eqnarray}
S=\frac{4}{3}\frac{U}{T_\infty}=\frac{8\pi^3}{45} \mathcal{N} \,T^3_\infty\int dr \sqrt\frac{A(r)}{B(r)^3}r^2\,.
\end{eqnarray}
The numerical solutions for metric functions $A(r), B(r)$ are displayed in Appendix~\ref{sec:22hole}. Being much smaller than unity in the highly curved 2-2-hole interior, they play a significant role in determining the unusual 2-2-hole thermodynamics. 

\begin{figure}[!h]
  \centering%
{ \includegraphics[width=7.8cm]{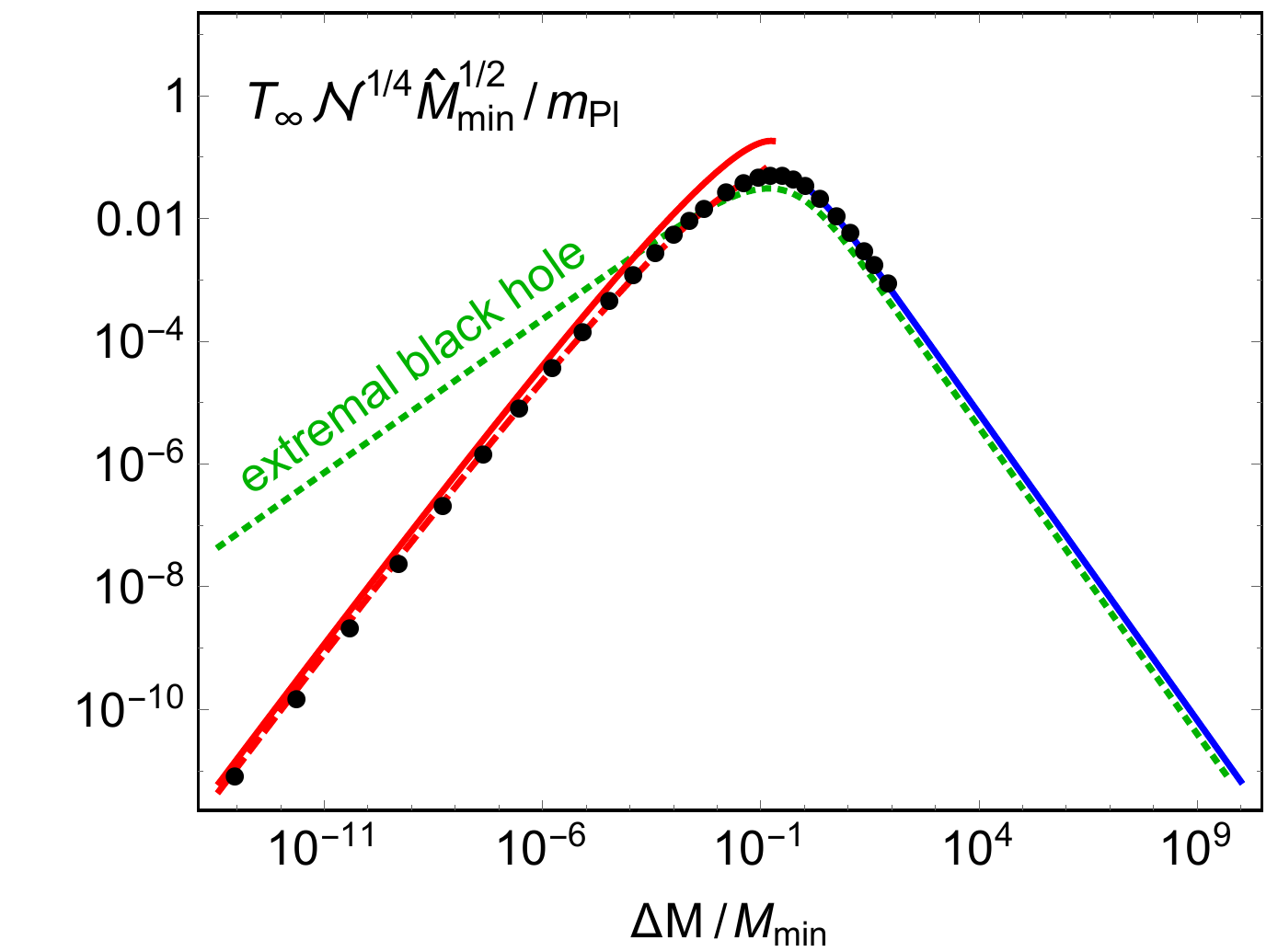}}\quad
{ \includegraphics[width=7.6cm]{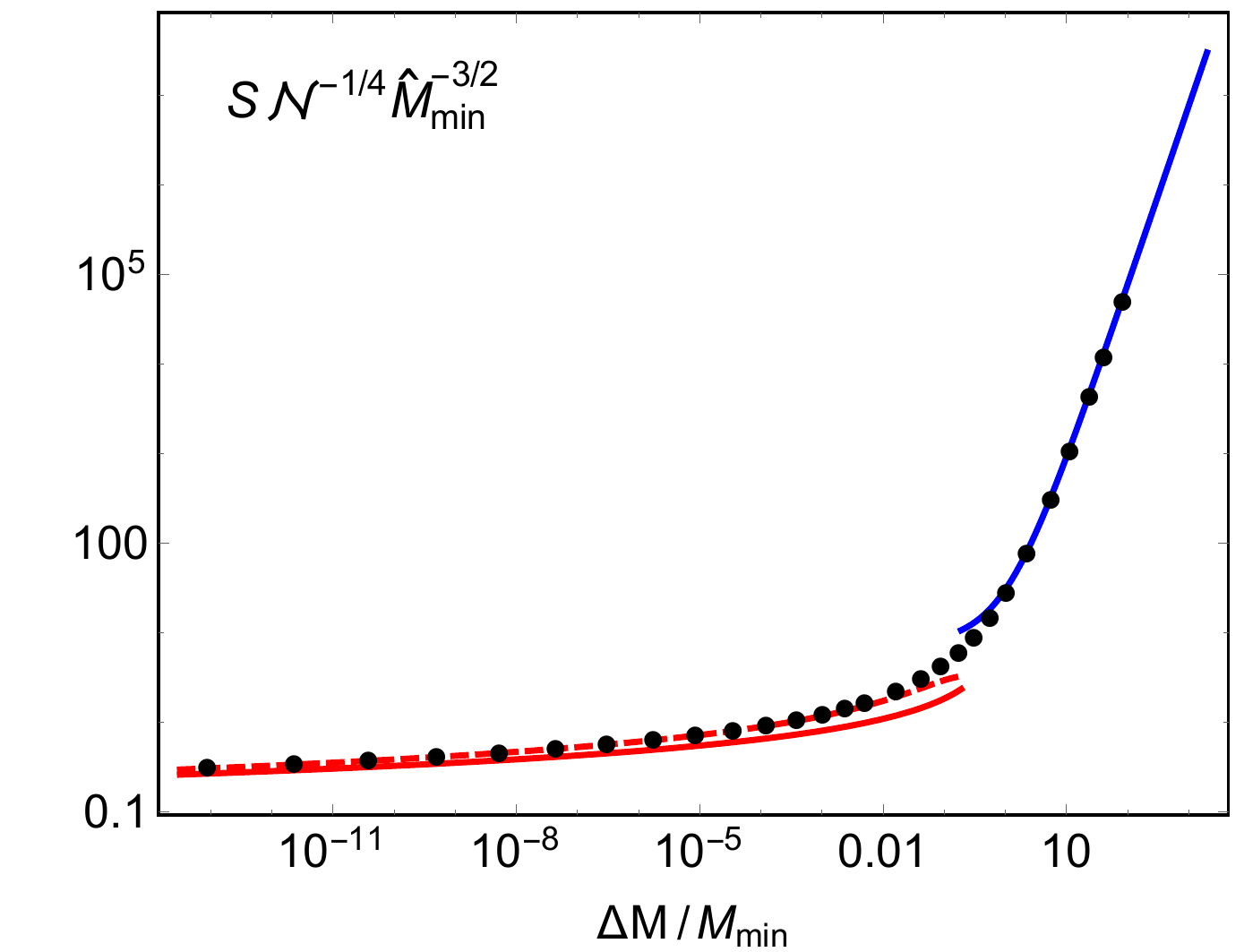}}
\caption{\label{fig:TSdrH} 
The 2-2-hole temperature $T_\infty$ and entropy $S$ as functions of the mass difference $\Delta M=M-\Mmin$. Black dots denote the exact numerical results. The blue and red solid lines are the leading-order analytical approximations for the small- and large-mass ranges, respectively. The red dashed line shows the next-to-leading-order improvement for the small-mass cases. }
\end{figure}

The thermal 2-2-holes exhibit intriguing thermodynamic behavior for the small- and large-mass cases.
Fig.~\ref{fig:TSdrH} shows the temperature $T_\infty$ and the entropy $S$ as functions of the mass, where the plots have been arranged to be independent of the values of $\Mmin$ and $\mathcal{N}$. Given that the mass $M$ can get extremely close to $\Mmin$, we display the dependence on the difference $\Delta M$, instead. 
The exact numerical results, denoted as black dots in Fig.~\ref{fig:TSdrH}, can be well approximated by analytical formulae in the small- and large-mass ranges, as shown by the colored lines. In the following, we discuss this novel behavior in more detail.

In the large-mass range, the temperature and entropy can be well approximated by the following, 
\begin{eqnarray}\label{eq:LMlimit}
T_\infty\approx 1.7\, \mathcal{N}^{-1/4}\hMmin^{1/2}\, T_\textrm{BH},\quad
S\approx 0.60\, \mathcal{N}^{1/4}\hMmin^{-1/2} \,S_\textrm{BH}\,.
\end{eqnarray}
$T_\textrm{BH}=\Mp^2/8\pi M$ is the Hawking temperature and $S_\textrm{BH}=\pi\, r_H^2/\lp^2$ is the Bekenstein-Hawking entropy for a Schwarzschild black hole with the same $M$. 
Anomalous behavior of black hole thermodynamics, i.e. the negative heat capacity and the area law for entropy, now arises from the ordinary thermal gas on a highly curved background spacetime. Therefore, for an outside observer, a large 2-2-hole appears similar to a black hole in terms of its thermodynamic behavior. Yet, the thermodynamic quantities depend on the number of degrees of freedom $\mathcal{N}$ and the minimal mass $\Mmin$. 
In the strong coupling scenario, the difference is mainly from the former. For a reasonable choice of $\mathcal{N}$, e.g. the Standard Model value, we can have $S>S_\textrm{BH}$ with the same mass. This suggests that a 2-2-hole is thermodynamically more stable than a black hole, and would be favored as the endpoint of gravitational collapses. In the weak coupling scenario, where $\Mmin\gg \Mp$,  2-2-holes have much higher temperature and much smaller entropy. Thus, 2-2-holes for this case are no longer entropically favorable and their stability needs to be checked dynamically. 
Note that (\ref{eq:LMlimit}) satisfies $T_\infty S= T_\textrm{BH} S_\textrm{BH}=M/2$ and is consistent with the first law of thermodynamics. This implies $U=3M/8$ for the gas energy, meaning that a sizable fraction of the physical mass for the hole comes from the gas source. 

In the small-mass range, we find the leading-order approximation for the temperature and entropy,\footnote{In the small mass limit, the tiny change of mass $\Delta M$ is hard to identify numerically. Instead, it can be found from the temperature and entropy by using the first law of thermodynamics, $dM=T_\infty dS$~\cite{Ren:2019afg}.}
\begin{eqnarray}\label{eq:SMlimit}
T_\infty\approx 0.39\, \mathcal{N}^{-1/4}\hMmin^{-3/2}\Delta M\left(\ln\frac{M_\textrm{min}}{\Delta M}\right)^{7/4},\,\,
S\approx 3.4\, \mathcal{N}^{1/4}\hMmin^{3/2}\left(\ln\frac{M_\textrm{min}}{\Delta M}\right)^{-3/4}.
\end{eqnarray}
A small 2-2-hole has a positive heat capacity and behaves more like a classical thermodynamic system. In the limit $\Delta M\to 0$, the temperature approaches zero almost linearly in $\Delta M$, while the entropy has a logarithmic dependence and decreases much slower. 
The energy is then dominated by the gravitational field, with negligible contribution from the gas. 
The approximation can be further improved by adding the next-to-leading-order contribution,
\begin{eqnarray}\label{eq:SMlimitp}
T_\infty&\approx& 0.39\, \mathcal{N}^{-1/4}\hMmin^{-3/2}\Delta M\left[\ln\frac{M_\textrm{min}}{\Delta M}-\ln\left(\ln\frac{M_\textrm{min}}{\Delta M}\right)^{3/2}+0.17\right]^{7/4},\,\,\nonumber\\
S&\approx& 3.4\, \mathcal{N}^{1/4}\hMmin^{3/2}\left[\ln\frac{M_\textrm{min}}{\Delta M}-\ln\left(\ln\frac{M_\textrm{min}}{\Delta M}\right)^{3/2}+0.17\right]^{-3/4}.
\end{eqnarray}

As we can see in Fig.~\ref{fig:TSdrH}, with the large-mass approximation (\ref{eq:LMlimit}) applied to $M\gtrsim \Mpeak$ and the small-mass one (\ref{eq:SMlimitp}) applied to $M\lesssim \Mpeak$, the analytical estimations turn out to be quite accurate for the whole mass range, including the estimation in the intermediate region around the temperature peak with 
\begin{eqnarray}
\label{eq:peak}
T_{\infty,\textrm{peak}}\approx 0.050\,\Mp\, \mathcal{N}^{-1/4}\hMmin^{-1/2}\,\,
\textrm{ at }\,\, \Mpeak\approx 1.2 M_\textrm{min}\,. 
\end{eqnarray}
Therefore, although the 2-2-hole solution can only be found numerically, its properties can be expressed in quite simple forms.

It is instructive to compare thermal 2-2-holes to extremal black holes. In the extremal limit, the black hole surface gravity, hence the temperature, approaches zero. So a near-extremal black hole would have suppressed thermal emission rate and may serve as a remnant. In reality, there are complications from non-thermal emission. For example, the Reissner-Nordstrom black hole can emit electrons and positrons due to Schwinger effects. This process tends to discharge the black hole and spoils the possibility for it to be stable or long-lived. To make it relevant, a special arrangement of the matter sector is needed. For example, \cite{Bai:2019zcd} introduces a new dark charge with the lightest charged particle being heavy enough for the non-thermal emission from the charged black hole to be largely suppressed. 
Evidently, in comparison to extremal black holes, the stabilization mechanism for a 2-2-hole is more fundamental and less contrived. 
The temperature for a charged Reissner-Nordstrom black hole is,
\begin{eqnarray}
T_\infty=\frac{\Mp^2}{2\pi}\frac{\sqrt{M^2-\Mmin^2}}{\left(M+\sqrt{M^2-\Mmin^2}\right)^2}\,.
\end{eqnarray}
$\Mmin$ denotes the mass in the extremal limit, i.e. $\Mmin=Q\,\Mp$. 
The green dotted line in Fig.~\ref{fig:TSdrH} shows the combination $\hMmin T_\infty/\Mp $ for this case. For a given mass in the small-mass regime, a 2-2-hole has a lower temperature and a smaller radiation power. 
We also note that the entropy of an extremal black hole has not been fully understood. The semiclassical methods suggest vanishing entropy in the extremal limit, while the string theory calculations find a non-zero value in line with the Bekenstein-Hawking formula. A possible resolution of the discrepancy is discussed in \cite{Carroll:2009maa}. 
Therefore, while entropy remains mysterious for an evaporating black hole, it is clear that entropy and information can simply escape from an evaporating 2-2-hole.

\subsection{Evaporation}
\label{sec:evapo}

A thermal 2-2-hole will radiate if it is hotter than the cosmic microwave background. 
Its mass evolution due to radiation can be described by the Stefan-Boltzmann law, with the power given as
\begin{equation}\label{eq:SBlaw}
-\frac{dM}{dt}
\approx \frac{\pi^2}{120}\, \mathcal{N}_* \, 4\pi r_H^2 \,T_{\infty}^4\;,
\end{equation}
which assumes $4\pi r_H^2$ as the effective emitted area. As in the case of black holes, an effective potential barrier in the exterior region modifies the power spectrum and yields a frequency-dependent absorption cross section. For the large-mass case where $T_\infty r_H\gtrsim 1$, the cross section is roughly a constant and emission can be well described by the black body radiation with an emitted area slightly larger than the would-be horizon one. 
For the small-mass case, $T_\infty r_H$ can be much smaller than unity, and the emission is suppressed by the potential barrier with the effective emitted area being also much smaller. 
As horizonless objects, 2-2-holes exhibit distinct behavior at low frequencies, and the suppression for small $T_\infty r_H$ deserves further study.\footnote{With a reflective boundary condition at the origin, the 2-2-hole features long-lived quasi-normal modes at low frequencies, corresponding to narrow resonance peaks in the spectrum. These resonances have been studied for large 2-2-holes, the features of which determine the signal of gravitational wave echoes generated by the binary merger~\cite{Conklin:2017lwb}. We leave the corresponding discussion of 2-2-hole remnants for future work.} In this paper, we restrict ourselves to the simple form in (\ref{eq:SBlaw}), which provides a conservative order of magnitude estimates for the low energy emission from small 2-2-holes. 
$\mathcal{N}_*$ denotes the number of relativistic degrees of freedom for the radiation~\cite{MacGibbon:1991tj}. It includes particles lighter than $T_{\infty}$ and it could be much smaller than $\mathcal{N}$ for the thermal gas in the interior. For simplicity we treat $\mathcal{N}_*$ as a constant and ignore its temperature dependence. 

Due to the mass dependence of temperature being very different in the large-mass and small-mass ranges, the evaporation of a 2-2-hole can be separated into two different stages. For a large 2-2-hole, the power increases as the object shrinks. As for a black hole, this is due to the negative heat capacity. For a small 2-2-hole, the power drops fast with decreasing mass and approaches zero when $M\to \Mmin$. It then behaves as a slowly decaying cold remnant with mass well approximated by $\Mmin$. When $\Delta M\ll \Mmin$, the remnant radiates so slowly that it appears stable for the age of the universe and can serve as a candidate for dark matter. 
If the initial mass $\Mini$ of a primordial 2-2-hole at formation satisfies $\Mini\gtrsim \Mpeak$, the early and late stages of evaporation are governed by the large-mass and small-mass phases respectively. 

In the large-mass stage, substituting $T_\infty(M)$ in (\ref{eq:LMlimit}) into (\ref{eq:SBlaw}), the time it takes for a 2-2-hole to evolve from $\Mini$ to $M\gtrsim \Mpeak$ is $\Delta t \equiv t-t_\textrm{init}
= 3.8\times 10^3 \,\mathcal{N}\,\mathcal{N}_*^{-1}  \hMmin^{-2}\,\lp^4\, (\Mini^3-M^3)$.
The time spent in the whole range of the large-mass stage, $\tau_L\equiv t_\textrm{peak}-t_\textrm{init}$, is then
\begin{eqnarray}\label{eq:tauL}
\tau_L
\,\approx \,\mathcal{N}\,\mathcal{N_*}^{-1}  \hMmin^{-2} \left(\frac{\Mini}{3.7\times 10^8\,\textrm{g}}\right)^3\textrm{s}\,
\approx\, \mathcal{N}\, \mathcal{N_*}^{-1}  \hMmin^{-1/2} \left(\frac{4.8\times 10^4\,\textrm{GeV}}{\mathcal{N}^{1/4}\,T_{\infty,\textrm{init}}}\right)^3\textrm{s}\,.
\end{eqnarray}
The time dependences of the temperature and mass take the same form as a black hole, 
\begin{eqnarray}\label{eq:LMlimitTime0}
T_\infty(t)\approx T_{\infty,\textrm{init}}\left(1-\frac{\Delta t}{\tau_L}\right)^{-1/3},\quad
M(t)\approx \Mini\left(1-\frac{\Delta t}{\tau_L}\right)^{1/3}.
\end{eqnarray}
In comparison to a primordial black hole with the lifetime $\tau_\textrm{BH}=\tau_L$, the time dependence differs only by an overall constant. Substituting $\Mini$ and $T_{\infty,\textrm{init}}$ as functions of $\tau_L$, (\ref{eq:LMlimitTime0}) becomes
\begin{eqnarray}\label{eq:LMlimitTime}
T_\infty(t)&\approx& 1.1\,\Mp\, \mathcal{N}^{1/12}\mathcal{N_*}^{-1/3} \hMmin^{-1/6}\left(\frac{\tau_L-\Delta t}{\lp}\right)^{-1/3},\nonumber\\
M(t)&\approx& 0.064\,\Mp \,\mathcal{N}^{-1/3}\mathcal{N_*}^{1/3}\hMmin^{2/3}\left(\frac{\tau_L-\Delta t}{\lp}\right)^{1/3},
\end{eqnarray}
where we can see the explicit $\Mmin$ dependence. 

In the small-mass stage, we use the leading order approximation in (\ref{eq:SMlimit}) to find the time dependence of the temperature.
Rewriting (\ref{eq:SBlaw}) as the following,   
\begin{eqnarray}
\frac{dT_\infty}{dt}=-\frac{\pi^2}{120}\, \mathcal{N_*}\, 4\pi r_H^2\, T_{\infty}^4\frac{dT_\infty}{dM}
\approx-\frac{2\pi^3}{15}\,0.4 \,\mathcal{N}^{-1/4} \mathcal{N_*} \,\hMmin^{1/2}\; T_\infty^4\,\lp^2\left(\ln \frac{\lp}{T_\infty}\right)^{7/4},
\end{eqnarray}
we obtain, at the leading order, 
\begin{eqnarray}\label{eq:SMlimitTime}
T_\infty(t)\approx 1.1\,\Mp\,  \mathcal{N}^{1/12} \mathcal{N_*}^{-1/3} \hMmin^{-1/6} \left(\frac{\Delta t-\tau_L}{\lp}\right)^{-1/3}\left(\ln\frac{\Delta t-\tau_L}{\lp\,\hMmin}\right)^{-7/12}.
\end{eqnarray}
The behavior of the tiny mass difference with the minimum value $\Delta M(t)$ can be found from (\ref{eq:SMlimit}) and (\ref{eq:SMlimitTime}) as
\begin{eqnarray}
\Delta M(t)\approx 19\, \Mp\, \mathcal{N}^{1/3} \mathcal{N_*}^{-1/3} \hat{M}_{\mathrm{min}}^{4/3}\left(\frac{\Delta t-\tau_L}{\lp}\right)^{-1/3}\left(\ln \frac{\Delta t-\tau_L}{\lp\,\hat{M}_{\mathrm{min}}} \right)^{-7/3}\;.
\end{eqnarray}
Compared to a near-extremal black hole with temperature $T_\infty\propto (\Mmin/t)^{1/2}$~\cite{Bai:2019zcd}, the 2-2-hole remnant has $T_\infty\propto\Mmin^{-1/6}t^{-1/3}$, which decreases more slowly with time and decreases for increasing remnant mass.

\begin{figure}[!h]
  \centering%
{ \includegraphics[width=11cm]{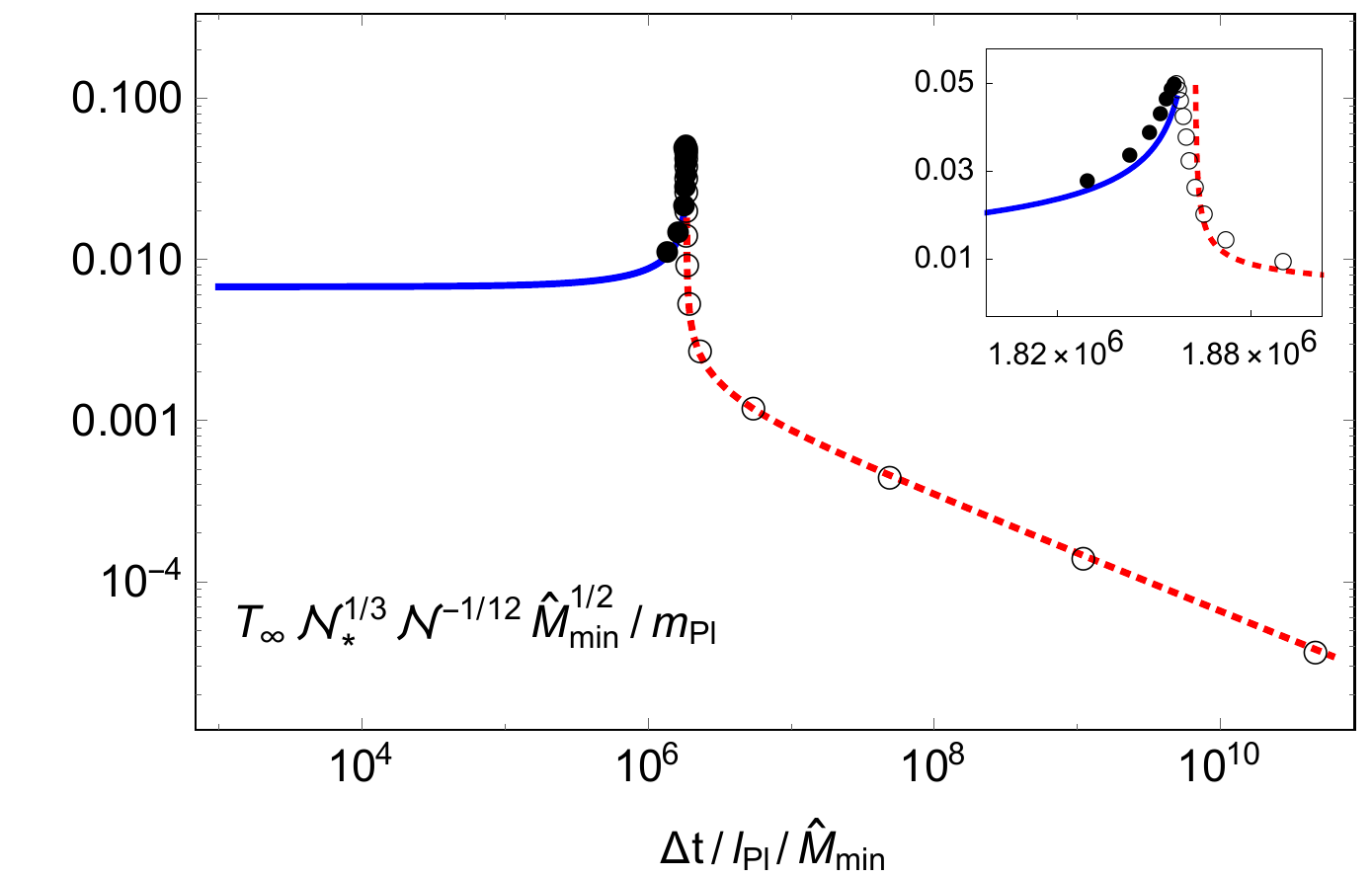}}
\caption{\label{fig:TinfDt} 
The 2-2-hole temperature $T_\infty$ as function of $\Delta t$ for $\Mini/\Mmin=10$. The dots denote the numerical results and the blue and red lines are analytical approximations (\ref{eq:LMlimitTime}) and (\ref{eq:SMlimitTime}). The inset shows the peak shape around $\Delta t\approx\tau_L$ with the linear scale.}
\end{figure}

Fig.~\ref{fig:TinfDt} compares the numerical and analytical results for the temperature of a 2-2-hole as a function of time. At early times, the temperature changes very slowly with time for a long period. For $t\lesssim t_\textrm{peak}$, it is well approximated by the large-mass analytical results (\ref{eq:LMlimitTime}) as given by the blue line. 
The temperature drops right after $t_\textrm{peak}$ and approaches the late-time behavior that is insensitive to $\Mini$ and determined solely by $\Mmin$. This part is well described by the small-mass approximation (\ref{eq:SMlimitTime}), as given by the red dotted line. 
From the inset plot, we can see that the analytical approximations become worse only in a quite narrow range around the peak temperature.
Since temperature of the remnant falls rapidly below the initial value, radiation from the 2-2-hole remnant quickly becomes negligible in comparison to its early-time radiation.

We also need to compare the 2-2-hole temperature $T_\infty$ with the cosmic background temperature $T_\textrm{bkg}$. The cosmic temperature drops faster with time than the remnant temperature, i.e. $T_\textrm{bkg}\propto t^{-1/2}\,( t^{-2/3})$ in the radiation (matter) era in comparison to $T_\infty\propto t^{-1/3}$ in (\ref{eq:SMlimitTime}). Thus, accretion of cosmic radiation onto primordial 2-2-holes need only be considered at the earliest times after formation when $T_\textrm{bkg}>T_\infty$. Assuming formation in the radiation era as in Sec.~\ref{sec:pheno}, we find that the growth in the mass is at most of order one and the influence on $\tau_L$ is negligible.

 
\section{Present epoch observations for 2-2-hole remnants}
\label{sec:present}

Our interest in this paper is primordial 2-2-holes that have already become remnants today, i.e. $\tau_L\lesssim t_0\approx 4.3 \times 10^{17}\,$s (the age of the universe). 
Being insensitive to details of the formation, observations for the remnants at the present epoch can be used to probe the remnant mass, which in turn is related to the fundamental mass scale $\Mmin$ in quadratic gravity. This mass has a theoretical lower bound, $\Mmin\gtrsim 0.63\,\Mp$, corresponding to  the strong coupling scenario. Considering the precise solar system test of GR, there is a rough upper bound $\Mmin\lesssim 10^{33}\,\textrm{g}\sim M_\odot$, by requiring that the Compton wavelength $\lambda_2$ be no larger than $\mathcal{O}(\textrm{km})$. An isolated remnant could be detected through its gravitational interaction in the same way as a PHB.\footnote{Most of these studies assume a Newtonian force for the object, so a 2-2-hole remnant that  deviates at $r\sim \mathcal{O}(r_H)$ still appears indistinguishable from a black hole.}
The parameter space starts to be constrained  for $\Mmin\gtrsim 10^{17}$\,g, with some examples summarized in Fig.~\ref{fig:LUcons}. Thus only smaller remnants with feeble gravitational interactions are able to constitute the entirety of dark matter.

Possible thermal radiation from isolated 2-2-hole remnants is expected to be weak.
A conservative estimation in Appendix~\ref{sec:remrad} shows that the remnant radiation with the dark matter abundance can safely evade BBN, CMB constraints as well as measurements for the diffuse photon flux at present. On the other hand, if two remnants form a binary and merge, then the merger product is hot and this can produce spectacular radiation. In this section we study the high-energy particle emission  from this process, which is better understood for 2-2-holes than for black hole remnants of ad-hoc nature~\cite{Bai:2019zcd, Barrau:2019cuo}. The corresponding experimental constraints turn out to be significant. 
This opens a new window onto small size dark matter that only interacts gravitationally with normal matter. The latter is usually considered to be the untestable nightmare scenario.

\subsection{High-energy particles from binary mergers}
\label{sec:merger}

As we can see in Fig.~\ref{fig:TSdrH}, a 2-2-hole remnant can be pushed away from the remnant stage if it is able to absorb sufficient mass. The merger of two 2-2-hole remnants or the accretion of ordinary matter onto a remnant can both contribute. Larger remnants that accrete matter more easily have already been strongly constrained by their gravitational interactions. Smaller remnants, on the other hand, may have a cross section with normal matter that is too small. The more likely mechanism is instead the merger of two remnants after forming a gravitationally bound state.
After reaching equilibrium, the merger product would acquire a high temperature and radiate away the absorbed mass within a short time. This process provides a significant source of high-energy astrophysical particles, as we will show below by calculating the flux.

The binary merger of 2-2-hole remnants generates a 2-2-hole with $M_\textrm{merger}\approx 2\Mmin>\Mpeak$.
The structure of a 2-2-hole with this mass is explicitly displayed in Appendix~\ref{sec:22hole}.
From the analytical approximation for the large-mass cases (\ref{eq:LMlimit}), the merger product has 
\begin{eqnarray}
\label{Temp-merger}
T_{\infty,\textrm{merger}}=3.4\times 10^{-2}\,\Mp\,\mathcal{N}^{-1/4}\hMmin^{-1/2}
=1.9\times10^{15} \mathcal{N}^{-1/4}\left(\frac{\Mmin}{\textrm{g}}\right)^{-1/2}\,\textrm{GeV}\,.
\end{eqnarray}
As approximated by the temperature, the average energy of emitted particles drops as $\Mmin$ increases and it spans a wide range of values. For a Planck mass remnant, the particles could have roughly the Planck energy, whereas for a large remnant with $\Mmin\sim 10^{23}\,$g, the energy is around TeV scale. With the lifetime being much smaller than a second for this mass range, it is assured that such mergers release their significant amount of excess energy almost instantly to get back to the remnant phase.

Observations in high-energy cosmic rays, gamma rays and neutrinos turn out to probe energies that are quite appropriate for the high-energy particle flux from mergers and can be used to constrain the fundamental scale $\Mmin$ in the theory. Ultra-high energy cosmic rays with energy beyond the GZK cut-off have long been observed. But a clear high-energy suppression around $10^{11}\,$GeV is now seen in the latest observations, and the need for new physics explanations is no longer as strongly motivated. Recently the photon flux around the same energy has also been measured with improved precision. High-energy neutrino experiments, on the other hand, probe a much wider energy range from $10^{3}\,$GeV to $10^{16}\,$GeV.

Notice that we are dealing with high energy emissions up to the Planck scale, where the strengths of both electroweak interaction and strong interaction could be sizable. Therefore, rather than a small number of very energetic particles as characterized by the temperature, the  flux generally receives the dominant contribution from high multiplicity final states with a broad energy spectrum. These are generalized parton showers of highly off-shell initial particles and they have been studied in detail for the super-heavy particle decay~\cite{Barbot:2002gt, Aloisio:2003xj}. 
The fragmentation of initial quarks or gluons generates nucleons, and then photons and neutrinos from decay of hadrons. Initial particles with only the electroweak charges can also initiate showers, depending on the strength of relevant couplings. 
A thorough study of the high-energy particle spectrum for the 2-2-hole evaporation is beyond the scope of this paper. 
Instead, we first estimate the neutrino flux by assuming only direct on-shell production, given that the shower spectrum of neutrinos peaks around the maximum energies~\cite{Barbot:2002gt}. This already strongly bounds $\Mmin$ to be much smaller than the range accessible by the conventional PBH searches. Then we study the contribution from showers of initial quarks to cosmic ray, gamma ray and neutrino fluxes, and this limits the size of $\Mmin$ even further. As we will see, depending on fragmentation functions at small energies,  the current experiments are not far from detecting the small flux from even a Planck mass remnant.

The neutrino flux from binary mergers of 2-2-hole remnants in the Milky way can be estimated as 
\begin{eqnarray}
\Phi_{\nu}=\frac{\mathcal{D}}{2 M_{\textrm{min}}}\,\frac{d N_{\nu}}{d E_\nu dt}\,.
\end{eqnarray}
The solid-angle-averaged $\mathcal{D}$-factor, used as in the case for dark matter decay, is given as~\cite{Evans:2016xwx} 
\begin{equation}
\label{D-factor}
\mathcal{D}=\frac{f}{4\pi}\int \rho_{\textrm{DM}}\left(r(s),\Omega\right)ds\; d\Omega\;,
\end{equation}
where $f\equiv \rho(t_0)/\rho_\textrm{DM}(t_0)$ denotes the mass fraction of 2-2-hole remnants in dark matter at present.
The integral is taken along the line of sight, with $d \Omega=\cos b\,db\,dl$ and $r^2= s^2 +R^2_{\odot}-2s R_{\odot} \cos b \cos l$ in spherical heliocentric coordinates. Here, $-\pi/2\leqslant b \leqslant \pi/2$ and $0\leqslant l \leqslant 2\pi$
are the galactic latitudinal and longitudinal angles respectively, $R_{\odot}$ is the distance of the Sun to the center of the galaxy, $s$ is the line of sight distance, and $0\lesssim r \lesssim 100$\,kpc denotes the distance to the galactic center.  
Using the Einasto density profile for the Milky way, $\rho_{\textrm{DM}}(r)= \rho_{s}  \exp(-\frac{2}{\alpha}[(r/r_s)^{\alpha}-1])$, with $\rho_{s}=0.077\mbox{ GeV}/\mbox{cm}^3=0.002 M_{\odot}/\mbox{pc}^3$, $r_s= 20$\,kpc, $R_{\odot}=8.0$\,kpc, and $\alpha=0.17$~\cite{Pato:2015dua}, we find that $\mathcal{D}\approx (0.04\,f)\, \textrm{g}\,\mbox{cm}^{-2}\,\mbox{sr}^{-1}$. 
The uncertainty from choosing different density profiles~\cite{Navarro:1996gj, Burkert:1995yz}, sizes of the halo~\cite{1989ApJ...345..759Z} or smaller ranges of $b$ as for some experiments is less than $10\%$, and this remains insignificant in our analysis.

The neutrino emission rate from the discharge of the excess energy, in the amount of one remnant mass, is roughly 
\begin{eqnarray}
\frac{d N_{\nu}}{d E_\nu dt}\approx \eta_{\nu} \frac{M_{\textrm{min}}}{\langle{E}_{\nu}\rangle^2}\;\Gamma\,,
\end{eqnarray}
where we approximate the spectrum by on-shell emission at the average energy $\langle{E}_{\nu}\rangle\approx 4.2$ $\,T_{\infty,\textrm{merger}}$. Here, $\eta_{\nu}\approx 0.058$ denotes the fraction of the total energy as neutrinos for the Standard Model~\cite{MacGibbon:1991tj}, which can be smaller if there are new active particles. $\eta_\nu \Mmin/\langle{E}_{\nu}\rangle$ then gives roughly the number of neutrinos emitted from one merger event, which increases with the minimal mass as $\Mmin^{3/2}$. $\Gamma$ denotes the merger rate for 2-2-hole remnants, and is a function of $f$ and  $\Mmin$. The quantity used to compare with experimental data is the following,
\begin{eqnarray}
E_\nu^2\, \Phi_{\nu}\approx \frac{1}{2}\,\eta_\nu\,\mathcal{D}\,\Gamma
\approx 6.5\times 10^{20} f\,\frac{\Gamma}{\textrm{s}^{-1}}\;\textrm{GeV}\;\textrm{cm}^{-2}\textrm{ s}^{-1}\textrm{ sr}^{-1}\,,
\end{eqnarray}
which shows no explicit dependence on $\langle{E}_{\nu}\rangle$ and $\Mmin$ other than in the merger rate. 

For the estimate of the merger rate $\Gamma$, we make use of the fact that 2-2-holes are like PBHs in forming gravitationally bound binaries~\cite{Sasaki:2018dmp}.\footnote{We assume that the 2-2-hole has already become a remnant at the time of binary formation, and so further evaporation has no influence on the merger rate estimation for PBHs.} In the early universe, this can happen after the 2-2-holes have decoupled from the cosmic expansion. Binaries with high eccentricities then contribute to the merger rate today. At the present epoch, binaries can form due to accidental encounters in a halo, with the rate enhanced by a small relative velocity. 
It has been found that formation in the early universe is the dominant scenario. 
As suggested recently in \cite{Raidal:2018bbj,Vaskonen:2019jpv}, earlier studies might overestimate this merger rate by ignoring disruptions of the binaries from nearby holes, especially when they constitute a significant fraction of dark matter. 
In the case of disruption, the total rate includes contributions from  the non-perturbed binaries and the perturbed ones $\Gamma=\Gamma_{\textrm{np}}\textrm{P}_{\textrm{np}} +\Gamma_{\textrm{p}}$, where  $\textrm{P}_{\textrm{np}}$ is the fraction of binaries remaining unperturbed.
From the total rate per volume given in \cite{Vaskonen:2019jpv}, we find
\begin{eqnarray}
\Gamma_{\textrm{np}} &=&4.7\times 10^{-26}\left(1+5.8\times 10^{-5}f^{-2}\right)^{-21/74}\,f^{16/37}\left(\frac{\Mmin}{\textrm{g}}\right)^{5/37} \left(\frac{t}{t_0}\right)^{-34/37} \textrm{s}^{-1}\;,\nonumber\\
\Gamma_{\textrm{p}}&=&
\begin{cases}
&\Gamma_{\textrm{p}}^{(1)}= 4.7\times10^{-32}\;f^{214/259}\left(\frac{\Mmin}{\textrm{g}}\right)^{10/37} \left(\frac{t}{t_0}\right)^{-6/7} \textrm{s}^{-1}\;,\\
&\Gamma_{\textrm{p}}^{(2)}= 4.0\times10^{-37}\;f^{358/259}\left(\frac{\Mmin}{\textrm{g}}\right)^{15/37} \left(\frac{t}{t_0}\right)^{-5/7} \textrm{s}^{-1}\;.
\end{cases}
\label{mergerrates}
\end{eqnarray}
The disruption effects come into play for $f\gtrsim 4\times 10^{-3}$, in which case the fitted relation $\textrm{P}_{\textrm{np}}\approx8.2\times10^{-3} f^{-4/5}$ can be obtained; otherwise the no-disruption case is recovered with  $\textrm{P}_{\textrm{np}}\approx 1$. The rate for perturbed binaries is bounded above and below  by $\Gamma_{\textrm{p}}^{(1)}$ and $\Gamma_{\textrm{p}}^{(2)}$ in (\ref{mergerrates}).

\begin{figure}[h!]
\captionsetup[subfigure]{labelformat=empty}
\centering
\hspace{-1.0cm}
\begin{tabular}{rrr}
{\includegraphics[width=12cm]{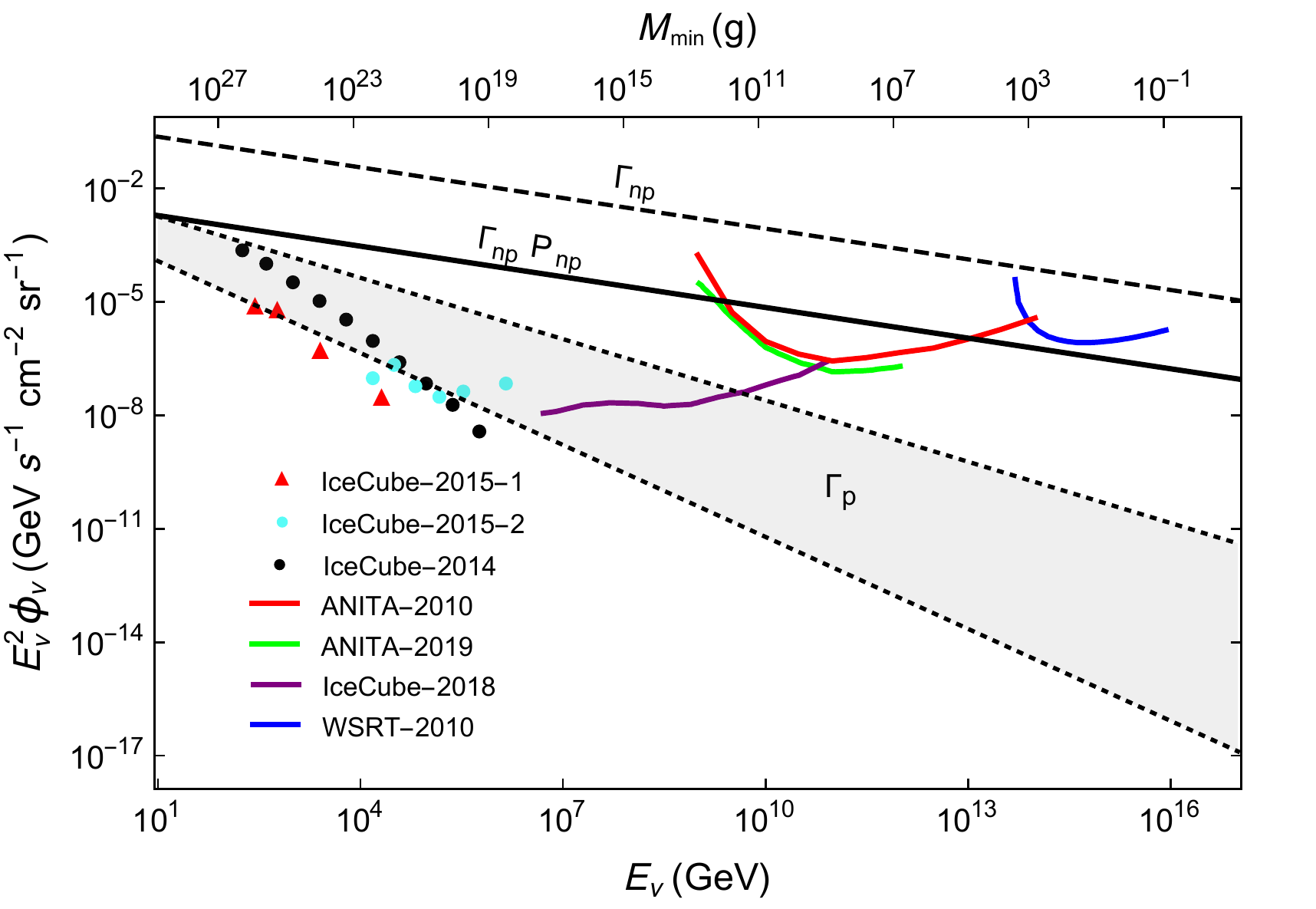}\label{fig:largeD}}&
\end{tabular}
\caption[]{The neutrino flux from direct on-shell production for binary mergers of 2-2-hole remnants  with $f=1$ and different estimations of the merger rate. In the case with disruption, the dominant contribution comes from non-perturbed binaries (solid line), with a suppressed rate in comparison to the earlier no-disruption estimation (dashed line).  For a given $\Mmin$, the energy is approximated by the average value $\langle{E}_{\nu}\rangle\approx 4.2 \,T_{\infty,\textrm{merger}}$.
Colored solid lines represent the experimental upper bounds on the neutrino flux from \cite{Buitink:2010qn} (NuMoon-WSRT-2010), \cite{Gorham:2019guw} (ANITA-2019), \cite{Gorham:2010kv} (ANITA-2010), \cite{Aartsen:2018vtx} (IceCube-2018), whereas the close symbols denote the observed signals from \cite{Aartsen:2015xup,Aartsen:2015knd} (IceCube-2015), and \cite{Aartsen:2014qna} (IceCube-2014). 
}\label{fig:neutrinoflux22}
\end{figure}

The comparison between the experimental data of high-energy neutrino flux and the on-shell production of neutrinos from the 2-2-hole binary mergers is displayed in Fig.~\ref{fig:neutrinoflux22}, assuming $f=1$ and $\mathcal{N}\approx 107$. 
The experimental upper bounds for energies $10^{7}\,$GeV--$10^{16}\,$GeV were obtained from lack of signals~\cite{Buitink:2010qn,Gorham:2019guw,Gorham:2010kv,Aartsen:2018vtx} , whereas the IceCube data for energies $10^{3}\,$GeV--$10^{7}\,$GeV represent the detected signals~\cite{Aartsen:2015xup,Aartsen:2015knd,Aartsen:2014qna}.\footnote{Translating the detected signals from IceCube into bounds on the flux have large uncertainties~\cite{Aartsen:2015knd} that we do not show in Fig.~\ref{fig:LUcons}.}  In order to see uncertainties associated with the merger rate, we show in Fig.~\ref{fig:neutrinoflux22} the theoretical predictions for $\Gamma_{\textrm{np}}$, $\Gamma_{\textrm{np}}\textrm{P}_{\textrm{np}}$ and the band of $\Gamma_{\textrm{p}}$. 
The 2-2-hole contribution increases with $\Mmin$ due to the $\Mmin$ dependence in the merger rate. With no disruption (dashed line), the 2-2-hole prediction with the dark matter abundance exceeds experimental upper bounds at all available energies, so only $\Mmin\lesssim 0.1\,$g is viable given the bound from the conventional PBH searches. After taking into account the suppression from disruption, with the merger rate still dominated by the non-perturbed binaries (solid line), the constraint is considerably relaxed and the 2-2-hole remnants with $\Mmin\gtrsim 10^{5}$\,g are excluded.

We next consider showers of the highly off-shell initial quarks (initial gluons are ten times less abundant). 
The flux of proton, photon, neutrino can then be well approximated by the flux of initial quarks multiplied with the fragmentation function $D_q^j(x)$, $j=p,\,\gamma,\,\nu$. For the quantities of interest, we have
\begin{eqnarray}
E_j^2\,\Phi_{j}=\frac{\mathcal{D}}{2 M_{\textrm{min}}}\left[E_q^2\frac{d N_{q}}{d E_q dt}\right]\Big[x^2D_q^{j}(x)\Big]_{x=E_j/E_q}
\approx \frac{1}{2}\,\eta_q\,\mathcal{D}\,\Gamma \,\Big[x^2D_q^{j}(x)\Big]_{x=E_j/\langle E_q \rangle} 
\end{eqnarray}
for the galactic contribution, where $\langle{E_q}\rangle\approx 4.2\,T_{\infty,\textrm{merger}}$ and $\eta_q\approx 0.67$.
For a given $\Mmin$, the shower generates a broad spectrum extending below $\langle{E_q}\rangle$. The explicit shape of the distribution depends on the coupling strength at high energy, and it becomes skewed more towards small $x$ for a larger coupling.

An extragalactic flux from merger products generated at an earlier time also needs to be included in the case of neutrinos, due to their negligible interaction with background photons as they propagate through the universe. 
The present flux can be defined as~\cite{Carr:2009jm}
\begin{eqnarray}
\label{flux-EG}
\Phi_\nu^\textrm{EG}=\frac{c}{4\pi}\frac{n_{\nu }}{E_{\nu  }}
= \frac{c}{4\pi}\frac{n(t_0)}{E_{\nu}} \int^{t_0}_{t_\textrm{min}} E_{\nu}(t)  \frac{d N_\nu}{dE_\nu dt} e^{-S_\nu(E_\nu(t),z)}\;dt\,,
\end{eqnarray}
where $E_{\nu }=E_\nu(t)/(1+z(t))$ is the redshifted energy. $n_{\nu }$ is the number density of neutrinos at present, where emissions extending back to $t_\textrm{min}$ (such that $E_\nu(t_\textrm{min})=\langle E_q \rangle$) are summed up for a given $E_{\nu }$. $S_\nu(E, z)$ denotes the neutrino opacity of the Universe that increases with $E$ and $z$~\cite{Gondolo:1991rn}. 
The quantity of interest is then given as
\begin{eqnarray}
E_\nu^2\,\Phi_\nu^\textrm{EG}
\approx \frac{c}{4\pi}f \,\rho_\textrm{DM}(t_0)\,\eta_q\,\frac{E_\nu}{\langle E_q \rangle}  \int_{t_\textrm{min}}^{t_0} \Gamma\, \Big[x\, D_q^\nu(x)\Big]_{x=E_\nu(t)/\langle E_q \rangle}e^{-S_\nu(E_\nu(t),z(t))} \,,
\end{eqnarray}
where the merger rate $\Gamma$ is given in (\ref{mergerrates}) and increases back in time. We find that the extragalactic flux at $E_\nu\ll \langle E_q\rangle$ could be a few times larger than the galactic one. 

\begin{figure}[h!]
\captionsetup[subfigure]{labelformat=empty}
\centering
\hspace{-1.0cm}
\begin{tabular}{rrr}
{\includegraphics[width=12cm]{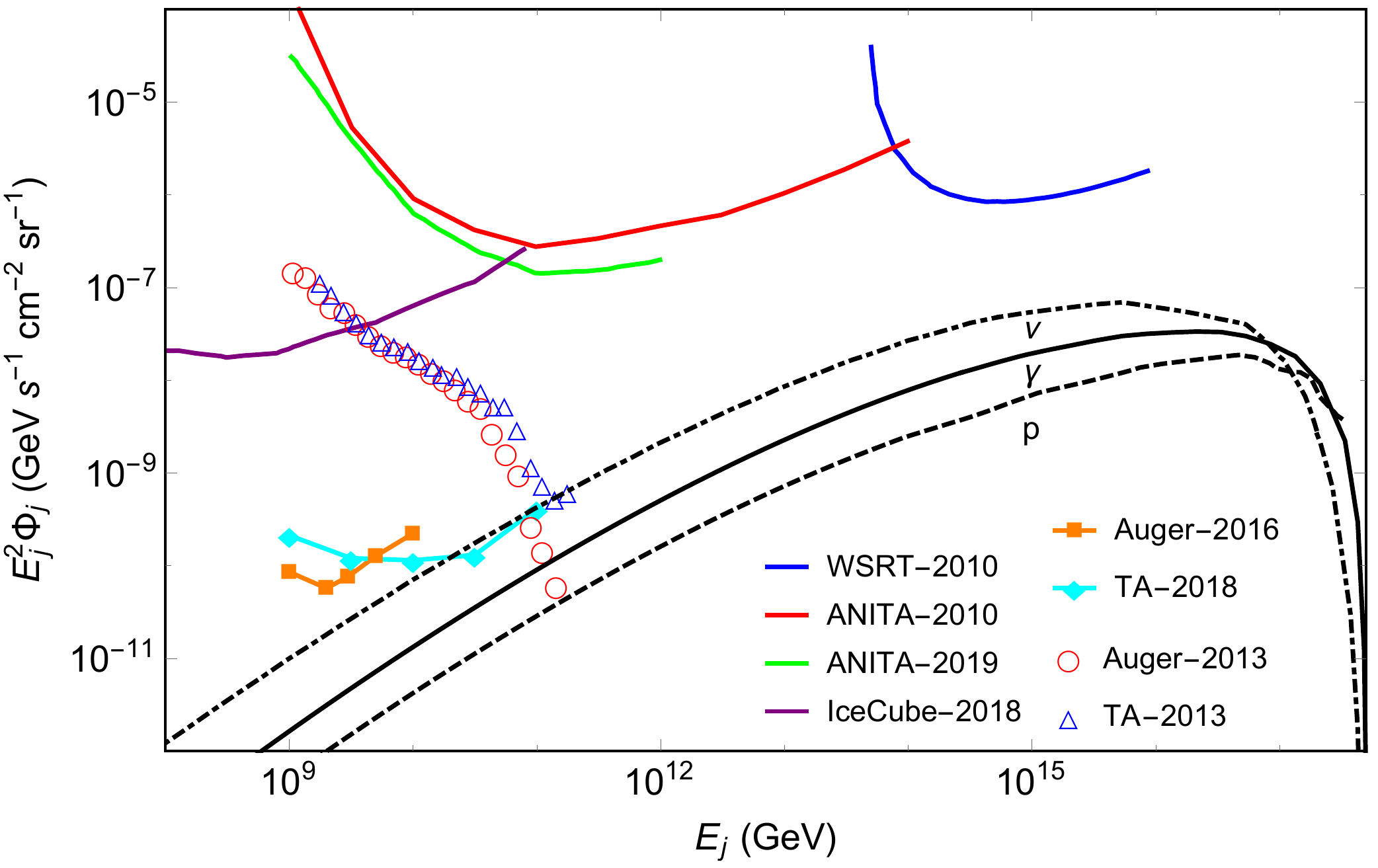}\label{fig:shower}}&
\end{tabular}
\caption[]{The proton flux (black dashed line), photon flux (black solid line), neutrino flux (black dotdashed line) from parton shower of initial quarks emission for binary mergers of 2-2-hole remnants, with $\Mmin=\Mp$, $f=1$ and $\Gamma=\Gamma_{\textrm{np}}\textrm{P}_{\textrm{np}}$. Extragalactic contribution is included for the neutrino flux.  Colored solid lines (without symbols) are upper bounds from neutrino experiments as in Fig.~\ref{fig:neutrinoflux22}. Solid lines with symbols denote constraints on the diffuse photon flux from \cite{Aab:2016agp} (Auger-2016) and \cite{Abbasi:2018ywn} (TA-2018). Open symbols show the observed signals of high-energy comic rays from \cite{ThePierreAuger:2013eja} (Auger-2013) and \cite{Abu-Zayyad:2013qwa} (TA-2013).}
\label{fig:showerflux22}
\end{figure}

Fig.~\ref{fig:showerflux22} compares the experimental data for high-energy particle flux with the parton-shower prediction from a Planck mass remnant. It is important to note that the quantity $E_j^2\,\Phi_j$ increases with $\Mmin$ for a given $x$ due to the $\Mmin$ dependence in the merger rate. For illustration, we use the fragmentation function in ordinary QCD, where the numerical results~\cite{Aloisio:2003xj} and an analytical approximation~\cite{Barbot:2002gt} are used for the large and small $x$ respectively.\footnote{The numerical results for $10^{-5}\lesssim x<1$ are found by two methods (a Monte Carlo simulation and the evolution based on the DGLAP equations) that agree well. At very small $x$, a modified leading log approximation is used to account for the color coherence effects. Its normalization is determined by matching with the numerical results at $x_0\sim 10^{-3}\text{--}10^{-4}$, with both the function and its first derivative being continuous.} Given the dominance of pions in the parton shower, there are many more photons and neutrinos than protons due to pion decay. The latest observations for the diffuse photon flux around $10^{11}\,$GeV impose the strongest constraint. 2-2-hole remnants with $\Mmin\gtrsim 10\,\Mp$ are excluded from being the entirety of dark matter. This bound relies on the fragmentation function within a narrow range of quite small $x$, i.e. $10^{-9}\lesssim x\lesssim 10^{-7}$, and this brings in theoretical uncertainties. 
The observations for neutrinos from showers have poorer sensitivities, and can only exclude $\Mmin$ larger than $1$\,g. But this bound comes from a wider range of energy, and is more robust against the variation of the distributions. The cosmic ray data could also provide a potentially interesting constraint. But this data has quite large uncertainties, in particular for the mass composition, and we do not consider it.

There is another effect involving the fragmentation function that is special to 2-2-holes. The shower may be occurring in the deep gravitational potential well within the would-be horizon, which thus implies blue-shifted energies. This reduces the couplings in the shower and gives additional suppression for the flux at lower energy. Further studies are required for a more conclusive analysis of the viable parameter space. Overall, we can see the complementarity between different observations in probing the small-mass range for the 2-2-hole remnant.

 
\section{Early-universe constraints for primordial 2-2-holes}
\label{sec:pheno}

Primordial thermal 2-2-holes, just like PBHs, can form in the early universe when parts of the universe stop expanding and re-collapse, either due to density inhomogeneities seeded by inflation or due to a first order phase transition.
In this paper, we focus on formation due to density inhomogeneities in the radiation era. 
The initial mass for the 2-2-hole $\Mini$ can be no larger than the horizon mass $1/2H(t_\textrm{init})\approx 4\times 10^{38}\left(t_\textrm{init}/\textrm{s}\right)\,\textrm{g}$ at the time of formation $t_\textrm{init}$. 
The horizon mass ranges from $\sim 1\,$g at the end of inflation if the reheating temperature is no larger than $10^{16}\,$GeV~\cite{Dalianis:2019asr}, to $\sim 10^{50}$\,g at matter-radiation equality.

It should be typical for 2-2-holes to be formed with the initial mass $\Mini$ much larger than $\Mmin$.
 The phenomenology then strongly depends on the duration of the early stage of evaporation $\tau_L$,  as given in (\ref{eq:tauL}). For later discussion, it is convenient to define the following critical masses for $\Mini$,
\begin{eqnarray}\label{eq:Mi}
\left(M_\textrm{uni},\, M_\textrm{rec},\, M_\textrm{BBN}\right)
=\left(2.8\times 10^{14},\, 8.8\times 10^{12},\,3.7\times 10^8\right)\hMmin^{2/3}\;\mathcal{N}^{-1/3}\,\mathcal{N_*}^{1/3} \;\textrm{g} \,,
\end{eqnarray} 
corresponding to $\tau_L\approx t_0$, $10^{13}\,$s (recombination), $1\,$s (BBN). 
Note that in the strong coupling scenario the mass values above are comparable to those for PBHs, while they can be much larger in the weak coupling scenario given the $\hMmin^{2/3}$ dependence. 
The mass range of interest in this paper is then $\Mini\lesssim M_\textrm{uni}$,
where the primordial 2-2-hole has already become a remnant at present. 
The hole with larger $\Mini$ stays more or less the same as it was at its formation over the history of the universe, and such holes, like PBHs, have already been constrained by their gravitational interactions.

In this section, assuming a monochromatic mass spectrum, we discuss the early-universe constraints in terms of the formation mass $\Mini$ for a given remnant mass $\Mmin$.
We study the requirement of the dark matter relic abundance in Sec.~\ref{sec:relic}. The mass fraction of 2-2-holes at the present epoch turns out to have a maximum as a function of $\Mini$. This in turn imposes an upper bound $\Mini\lesssim M_\textrm{DM}$ (see (\ref{eq:Mrelic}) below) if the 2-2-hole remnants constitute all of dark matter. 
Constraints from the BBN and the cosmic microwave background (CMB) observations are explored in Sec.~\ref{sec:BBN} and Sec.~\ref{sec:CMB}. When $M_\textrm{BBN}\lesssim\Mini\lesssim M_\textrm{rec}$, the object is in the early stage of evaporation at the BBN epoch or afterwards. The radiation is then strongly constrained by the relic abundance of light elements. For $\Mini\lesssim M_\textrm{BBN}$ or $\Mini\gtrsim M_\textrm{rec}$, there are constraints from the baryon-to-photon ratio or the CMB anisotropies. 
A large portion of the mass range relevant to BBN and CMB turns out to be larger than $M_\textrm{DM}$. The observations then constrain the parameter space when an explanation of dark matter is absent.

\subsection{Dark matter relic abundance}
\label{sec:relic}

We start with the cosmic evolution of primordial ultracompact objects. The following discussion applies to both PBHs and 2-2-holes unless otherwise specified. 

The mass fraction of primordial objects at formation in the radiation era is
\begin{eqnarray}\label{eq:beta}
\beta \equiv \frac{\rho(t_\textrm{init})}{\rho_\textrm{tot}(t_\textrm{init})}
=\frac{4 \,M(t_\textrm{init})\,n(t_\textrm{init})}{3 \,T(t_\textrm{init})\,s(t_\textrm{init})}
=2.5\, g_*^{1/4}\,\gamma^{-1/2}\,\hMini^{3/2}\,\frac{n(t_\textrm{init})}{s(t_\textrm{init})}\,,
\end{eqnarray}
where $\hMini\equiv \Mini/\Mp$. $\rho(t), \,n(t)$ denote the energy density and number density for primordial objects, and $\rho_\textrm{tot}(t_\textrm{init})\approx \rho_\textrm{rad}(t_\textrm{init})$ for $\beta$ of interest. 
For the last expression, we use $T(t)=0.55 \,\Mp\,g_*^{-1/4} (t/\lp)^{-1/2}$ and $\Mini\approx 4\times 10^{38}\gamma\left(t_\textrm{init}/\textrm{s}\right)$\,g. 
$\gamma$ denotes the fraction of the horizon mass that enters into the 2-2-hole. A typical value is $\gamma\approx 0.2$, but this is quite uncertain~\cite{Carr:2009jm}. 
As we will see below, observations are determined by the number density to entropy density ratio $n(t_\textrm{init})/s(t_\textrm{init})$, namely, the combination $\beta\,\gamma^{1/2}g_*^{-1/4}$, which is insensitive to $\gamma$. 

The mass fraction of primordial objects in dark matter at present is,
\begin{eqnarray}\label{eq:betaf}
f=\frac{M(t_0)\,n(t_0)}{\rho_\textrm{DM}(t_0)}
=\frac{M(t_0)\,s(t_0)}{\rho_\textrm{DM}(t_0)}\frac{n(t_0)}{s(t_0)}\,,
\end{eqnarray} 
where $s(t_0)=2.9\times 10^3\,\textrm{cm}^{-3}$, $\rho_{\textrm{DM}}(t_0)\approx 0.26\rho_c$, $\rho_c=9.5\times10^{-30}\,\textrm{g}\,\textrm{cm}^{-3}$. $M(t_0)$ is $\Mini$ for the large-mass case with negligible evaporating rate, and $M(t_0)$ is $\Mmin$ for the small-mass case where a remnant is left behind.

\begin{figure}[!h]
  \centering%
{ \includegraphics[width=7.7cm]{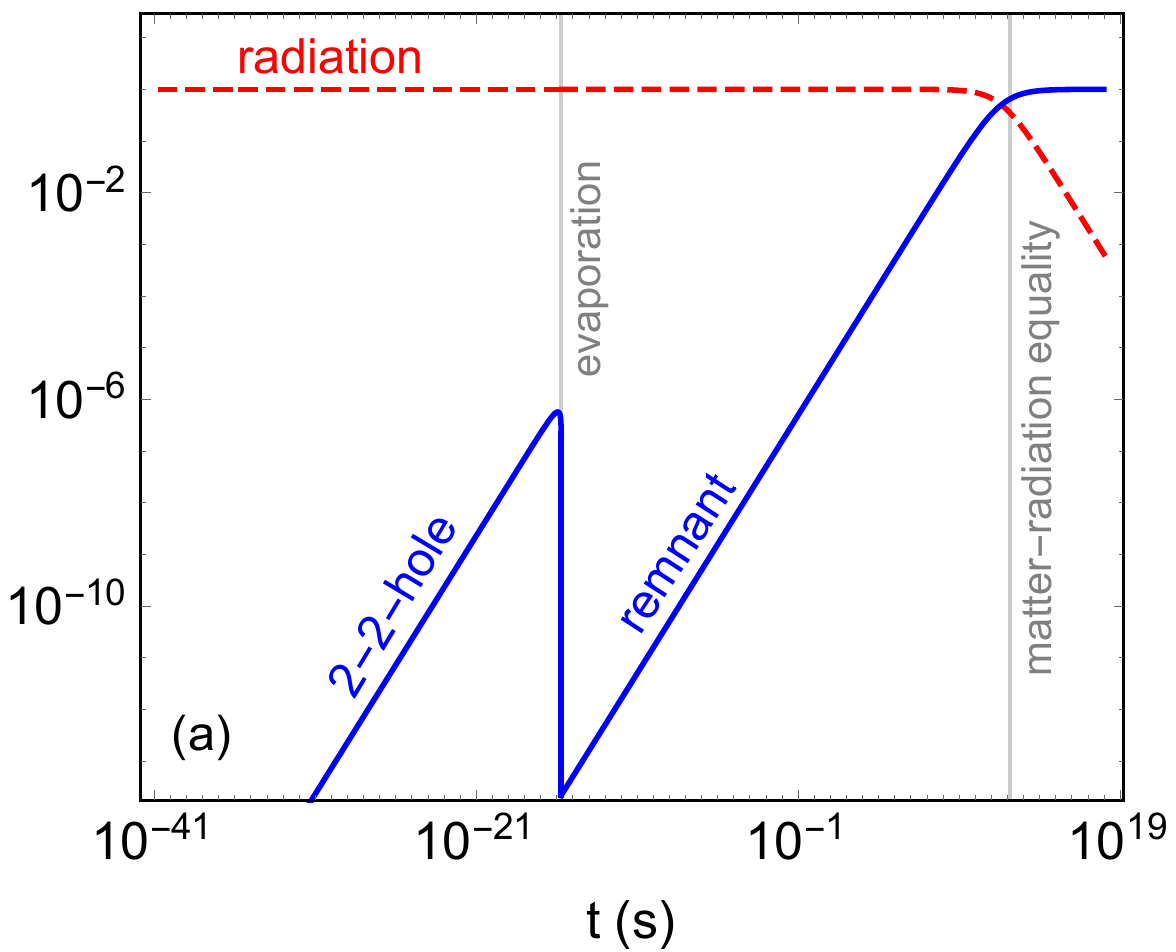}}\quad
{ \includegraphics[width=7.5cm]{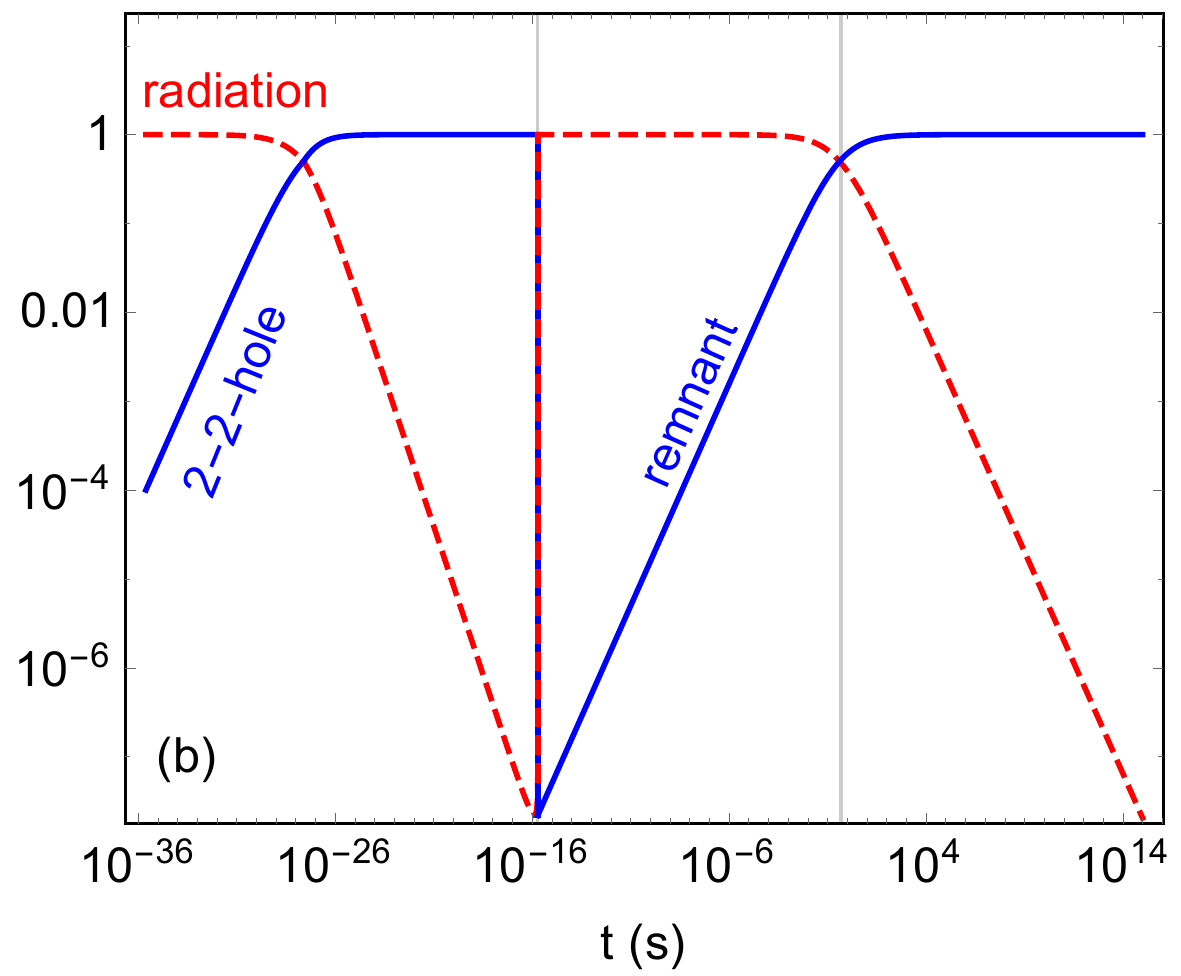}}
\caption{\label{fig:evolution} 
Time evolution for the fractions of energy densities for radiation (red) and primordial 2-2-hole with $\Mmin=\Mp$, $\Mini=10^3$\,g (blue).  (a) $n(t_\textrm{init})<n_c(t_\textrm{init})$, the 2-2-hole density never dominates before $t_\textrm{eva}$ and the remnants constitute all of dark matter at present. (b) $n(t_\textrm{init})>n_c(t_\textrm{init})$, the 2-2-hole density becomes dominant before $t_\textrm{eva}$ and too many remnants are left behind with $\Mini< M_\textrm{DM}$ in this example.}
\end{figure}

After formation the mass fraction of primordial objects increases with time. As $M(t)$ varies slowly with time in comparison to other quantities at both early and late time, we approximate the evaporation as 
an instantaneous radiation of energy at $t_\textrm{eva}\equiv t_\textrm{init}+\tau_L\approx \tau_L$ (or $\tau_\textrm{BH}$) for the cosmic evolution studies in this subsection. Thus, we use $M(t)\approx\Mini$ at $t\leq t_\textrm{eva}$ and $M(t)\approx\Mmin$ at $t>t_\textrm{eva}$. 
It is then convenient to define a critical number density at formation, 
\begin{eqnarray}
n_c(t_\textrm{init})=\frac{\rho_\textrm{tot}(t_\textrm{init})}{\Mini}\sqrt{\frac{t_\textrm{init}}{t_\textrm{eva}}}\,,
\end{eqnarray}
which corresponds to the equality of energy densities $\rho=\rho_\textrm{rad}$ at $t_\textrm{eva}$ for a given $\Mini$. 
When $n(t_\textrm{init})\lesssim n_c(t_\textrm{init})$, 
the primordial objects are always subdominant in the energy budget before the instantaneous evaporation. The entropy injection from evaporation is then negligible, and $n(t)/s(t)$ remains constant till the present. In this case, the mass fraction for the remnant at present $f$ is related to the number density at formation with $n(t_0)/s(t_0)=n(t_\textrm{init})/s(t_\textrm{init})$,
\begin{eqnarray}\label{eq:betaf}
f=2.6\times 10^{28}\hMmin \frac{n(t_\textrm{init})}{s(t_\textrm{init})}\,.
\end{eqnarray} 
In comparison to large PBHs, it receives a suppression factor of $\Mmin/\Mini$. 
Fig.~\ref{fig:evolution}\,(a) shows the time variation of energy densities in this case and for $f=1$. 

For the other case with $n(t_\textrm{init})\gtrsim n_c(t_\textrm{init})$, the primordial objects become dominant at some earlier time $t_\textrm{dom}\approx t_\textrm{init}\,\rho_\textrm{tot}^2(t_\textrm{init})/(\Mini \,n(t_\textrm{init}))^2$, and there is a new era of matter domination for $t_\textrm{dom}<t<t_\textrm{eva}$. This introduces an extra redshift of the number density,
\begin{eqnarray}\label{eq:njeva}
n(t_\textrm{eva})\approx n(t_\textrm{init})\left(\frac{t_\textrm{eva}}{t_\textrm{dom}}\right)^{-2}\left(\frac{t_\textrm{dom}}{t_\textrm{init}}\right)^{-3/2}
=\frac{\rho_\textrm{tot}(t_\textrm{init})}{\Mini}\left(\frac{t_\textrm{eva}}{t_\textrm{init}}\right)^{-2}
=n_c(t_\textrm{init})\left(\frac{t_\textrm{eva}}{t_\textrm{init}}\right)^{-3/2},
\end{eqnarray}
which cancels with the enhancement from $n(t_\textrm{init})/n_c(t_\textrm{init})$ so that $n(t_\textrm{eva})$ remains the same as the one with $n_c(t_\textrm{init})$.
To find the mass fraction at present $f\propto n(t_0)/s(t_0)=n(t_\textrm{eva})/s(t_\textrm{eva})$, 
we also need the entropy density after evaporation, as given dominantly by radiation from primordial objects,
\begin{eqnarray}\label{eq:entropy}
s(t_\textrm{eva})\approx \mathcal{N}_*^{1/4}\rho^{3/4}(t_\textrm{eva})
= \mathcal{N}_*^{1/4} \left(\rho_\textrm{tot}(t_\textrm{init})t_\textrm{init}^2t_\textrm{eva}^{-2}\right)^{3/4}
\approx 0.07\, \mathcal{N}_*^{1/4}\, \left(t_\textrm{eva}\lp\right)^{-3/2}\,.
\end{eqnarray}
This can be compared with the entropy density without the dominance of primordial objects,   
\begin{eqnarray}
s(t_\textrm{eva})\approx g_*^{1/4}\rho_\textrm{rad}^{3/4}(t_\textrm{eva})
= g_*^{1/4} \left(\rho_\textrm{rad}(t_\textrm{init})t_\textrm{init}^2t_\textrm{eva}^{-2}\right)^{3/4}
\approx 0.07\, g_*^{1/4}\,\left(t_\textrm{eva}\lp\right)^{-3/2}\,.
\end{eqnarray}
The two differ only by a factor of the number of degrees of freedom, where $\mathcal{N}_*\geq g_*$ due to the possibly new particle species from primordial objects radiation. 
Therefore, the mass fraction at present has a maximum $f_{\textrm{max}}$ as given by  (\ref{eq:betaf}) with $n(t_\textrm{init})=n_c(t_\textrm{init})$, and $f\approx f_{\textrm{max}}\,(g_*/\mathcal{N}_*)^{1/4}$ when the primordial objects actually dominate.

For the 2-2-hole, with $\tau_L$ in (\ref{eq:tauL}), we obtain the maximal mass fraction as
\begin{eqnarray}
f_{\textrm{max}}=2.6\times 10^{28}\hMmin \frac{n_c(t_\textrm{init})}{s(t_\textrm{init})}
\approx 1.7\times 10^{26}\,\mathcal{N}^{-1/2}\mathcal{N}_*^{1/2}g_*^{-1/4}\hMmin^2\hMini^{-5/2}\,.
\label{eq:betacfmax}
\end{eqnarray}
For a given $\Mmin$, having the 2-2-hole remnant to account for all of dark matter then requires $f_{\textrm{max}}$ to be greater than unity. This imposes an upper bound on the initial mass as
\begin{eqnarray}\label{eq:Mrelic}
\Mini\lesssim M_\textrm{DM}
\equiv 
6.8\times 10^5\,
\hMmin^{4/5}\;
\mathcal{N}^{-1/5}\,\mathcal{N}_*^{1/5}\,g_*^{-1/10}\, \,\textrm{g}\,.
\end{eqnarray}
In comparison to PBHs, the bound is relaxed if $\hMmin$ is large as for the weak coupling scenario.
If 2-2-holes dominate before $t_\textrm{eva}$, then the bound on $\Mini$ needs to be saturated, up to the  $g_*/\mathcal{N}_*$ factor. Fig.~\ref{fig:evolution}\,(b) shows the time variation of energy densities for an example with 2-2-hole dominance and where $\Mini$ is too small.

\subsection{BBN constraints}
\label{sec:BBN}

The investigation of the effects of 2-2-hole evaporation on BBN can be tied to the analyses of PBHs evaporation that has been a subject of heavy interest in the literature~\cite{1978SvA....22..138V,10.1143/PTP.59.1012,1977SvAL....3..110Z,1978SvAL....4..185V,Kohri:1999ex,Carr:2009jm}. We first briefly review the analysis for PBHs. The emitted particles can affect BBN in several ways~\cite{Carr:2009jm}. First, high energy mesons with long enough lifetime scatter off the ambient nucleons inducing additional interconversion between protons and neutrons, changing the freeze-out value of $\textrm{n}/\textrm{p}$ for $t\sim 10^{-2}\text{--}10^{2}$\,s. Second, high energy hadrons disassociate background nuclei, predominantly ${^4}$He, thus reducing its abundance and increasing the abundance of D, T, ${^3}$He, ${^6}$Li, and ${^7}$Li for $t\sim 10^2\text{--}10^{4}$\,s. Finally, high energy photons generated indirectly through scattering involving the initial high energy quarks and gluons
cause further disassociation of ${^4}$He and enhance the abundance of the lighter elements for $t\sim 10^{4}\text{--}10^{12}$\,s.
The time evolution of number densities are governed by the Boltzmann equation~\cite{Kohri:1999ex}
\begin{eqnarray}
\label{Boltzmann}
\frac{dn_i}{dt}+3H(t)n_i=\left[\frac{d n_i}{dt}\right]_{{\mbox{\tiny SBBN}}}-\quad \sum_{\mbox{\scriptsize proc.}}\Gamma_h(t) K_i\,,
\end{eqnarray}
where  $i$ denotes the particle species ($n, p$, D, T, $^3$He, $^4$He, $^6$Li, $^7$Li). The first term on the right-hand side denotes the contribution from the standard BBN scenario, and the second one sums over the available processes relevant to PBHs evaporation. $K_{i}$ is the average number of interactions with background per emission for the relevant process, and $\Gamma_h(t)$ denotes the  emission rate for hadronic particles,
\begin{eqnarray}\label{eq:emissionrate}
\Gamma_h (t)= \textrm{B}_h \,n(t) \,\frac{1}{\langle E_h(t)\rangle} \frac{dM}{dt}\,,
\end{eqnarray}
where $\textrm{B}_h$ is the hadronic branching ratio and $\langle E_h (t)\rangle$ is roughly $T_\infty(t)$ up to an $\mathcal{O}(1)$ factor.

Since the effects of radiating holes on the BBN processes are directly proportional to the emission rate $\Gamma_h(t)$, constraints on the mass fraction of primordial 2-2-holes can be inferred from the analysis performed for PBHs~\cite{Carr:2009jm} by computing the ratio $\Gamma_{h,\textrm{BH}}/\Gamma_{h,22}$. During the relevant time-period, \textit{i.e.} $10^{-2}$ s $\lesssim t \lesssim 10^{12}$\,s, 2-2-holes with $M_\textrm{BBN} \lesssim \Mini \lesssim M_\textrm{rec}$ remain at the early stage  and produce radiation much like PBHs. As from (\ref{eq:LMlimitTime0}), if we compare the 2-2-hole in this stage to black hole with $\tau_L=\tau_{\textrm{BH}}$, $\Gamma_i$ only differs by a overall constant and $\Gamma_{\textrm{BH}}/\Gamma_{22}$ is time independent. 
Radiation from the remnant stage is much weaker and can be safely ignored.

Therefore, BBN constraints for 2-2-holes can be found by a simple scaling of the corresponding constraints for PBHs with an appropriate choice of masses. Given $ \tau_L= \tau_{\textrm{BH}}$, objects used for rescaling have different initial masses $\Mini^{22}=A^{4/3} \Mini^{\mathrm{BH}}$,  where $A\equiv1.7\; \mathcal{N}^{-1/4} \,\hMmin^{1/2}$ is the factor that appears in $T_\infty$ in (\ref{eq:LMlimit}). 
Using (\ref{eq:emissionrate}), we obtain the ratio of the emission rates as
\begin{eqnarray}
\label{ratio}
\frac{\Gamma_{h,22}}{\Gamma_{h,\textrm{BH}}}= \frac{\left(\textrm{B}_h\,\mathcal{N}_*\right)_{22}}{\left(\textrm{B}_h\,\mathcal{N}_*\right)_\textrm{BH}}\frac{n_{22}(t)}{n_{\textrm{BH}}(t)} A^{5/3}\;.
\end{eqnarray}
The factor $\textrm{B}_h\,\mathcal{N}_*$ denotes the number of active hadronic degrees of freedom, which depends on the temperature. As we can see from (\ref{eq:tauL}), 2-2-holes with $\tau_L=\tau_\textrm{BH}$ have lower temperature for larger $\Mmin$. Thus, $\left(\textrm{B}_h\,\mathcal{N}_*\right)_{22}$ will in general be smaller than its counterpart for the black-hole case.    
Assuming $\Gamma_{h,\textrm{BH}}=\Gamma_{h,22}$, we obtain the following relation,  
\begin{eqnarray}
\label{betas}
\frac{n_{22}(t_\textrm{init})}{s(t_\textrm{init})}=\frac{n_\textrm{BH}(t_\textrm{init})}{s(t_\textrm{init})}\,  \frac{\left(\textrm{B}_h\,\mathcal{N}_*\right)_\textrm{BH}}{\left(\textrm{B}_h\,\mathcal{N}_*\right)_{22}}  A^{-5/3}
\approx 0.4\Big(\hMini^{22}\Big)^{-3/2}\beta_\textrm{BH}\,\gamma^{1/2}g_*^{-1/4}\,  A^{1/3}\,,
\end{eqnarray}
given (\ref{eq:beta}) and $n(t)/s(t)=n(t_\textrm{init})/s(t_\textrm{init})$.
In the last expression, we convert the ratio of the number density to entropy density for black holes to the quantity $\beta_\textrm{BH}\,\gamma^{1/2}g_*^{-1/4}$, for which the latest constraints are given in \cite{Carr:2009jm}. We also assume $\left(\textrm{B}_h\,\mathcal{N}_*\right)_{22}\approx \left(\textrm{B}_h\,\mathcal{N}_*\right)_\textrm{BH}$ for simplicity. This leads to a conservative upper bound for 2-2-holes. 
In comparison to PBHs case,  this conservative constraint already enjoys some relaxation for larger $\Mmin$ with $A\gg 1$. It could be further relaxed if $\left(\textrm{B}_h\,\mathcal{N}_*\right)_{22}$ becomes considerably smaller at lower temperature. 

\begin{figure}[h!]
\captionsetup[subfigure]{labelformat=empty}
\centering
\hspace{-0.8cm}
\begin{tabular}{rrr}
\subfloat[\quad(a) $\Mmin=m_{\mathrm{Pl}}$]{\includegraphics[width=8.0cm]{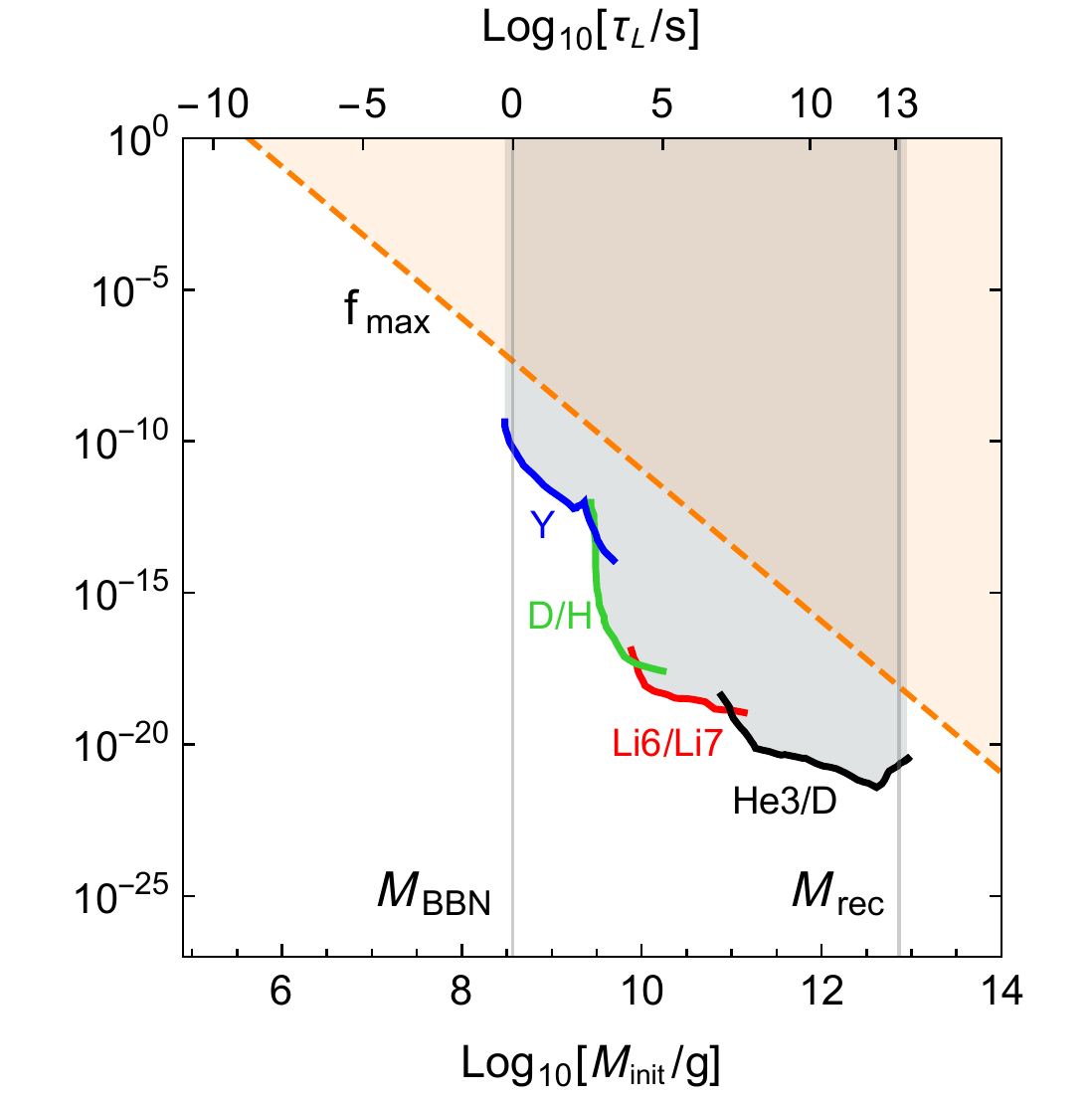}\label{fig:f1}}&
\subfloat[\quad(b) $\Mmin=10^{20}$ g]{\includegraphics[width=8.0cm]{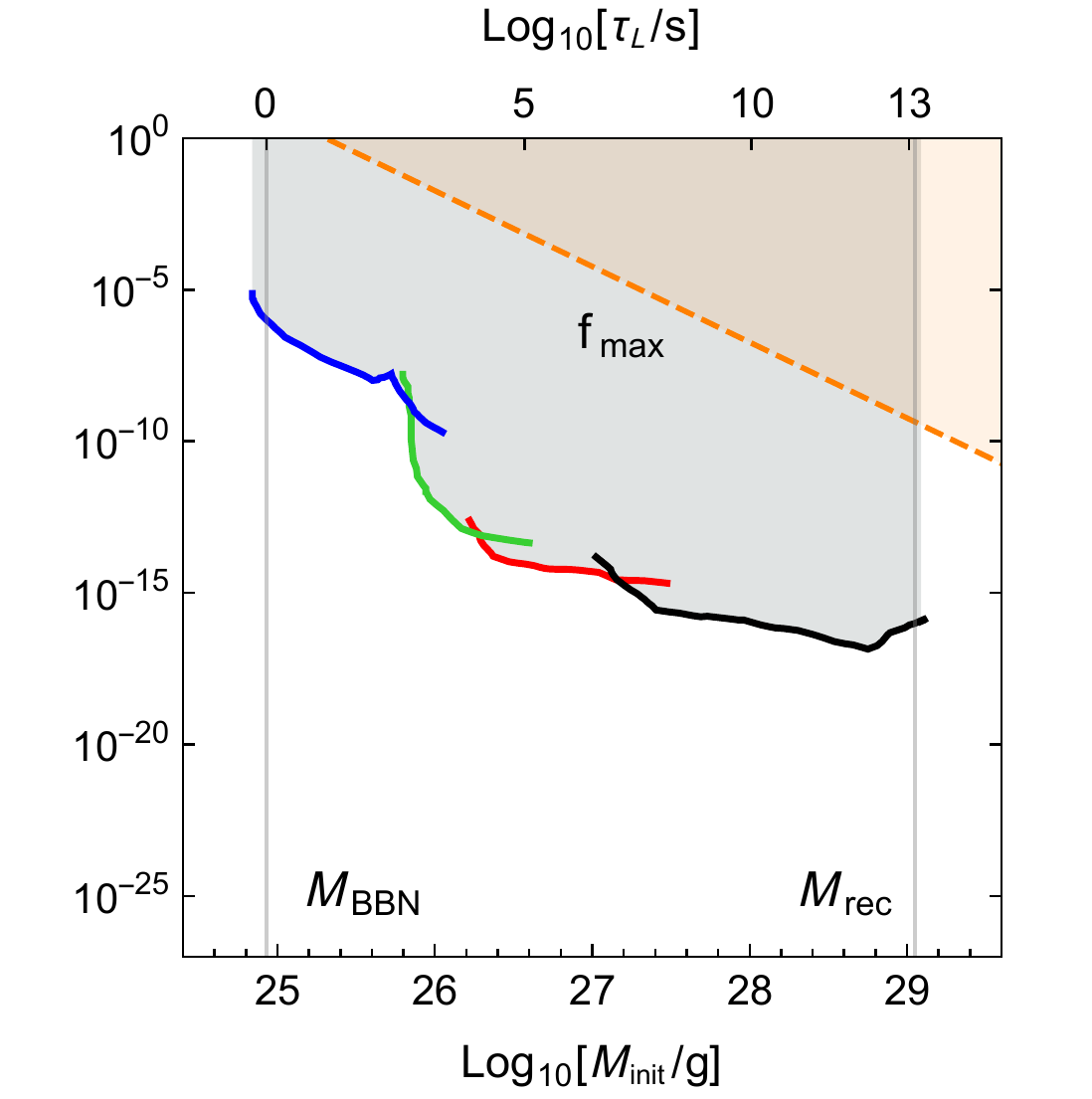}\label{fig:f2}}
\end{tabular}
\caption{BBN constraints (solid lines) on the present mass fraction of 2-2-hole remnants in dark matter $f=\rho(t_0)/\rho_\textrm{DM}(t_0)$, for two benchmark values of $\Mmin$. The 2-2-holes in this case are in the early-radiation stage during BBN epoch ($M_\textrm{BBN} \lesssim \Mini \lesssim M_\textrm{rec}$). The dashed line represents the maximum value of $f$ as given in (\ref{eq:betacfmax}).  }
\label{fig:BBN-early}
\end{figure}

The constraints turn out to be so strong that 2-2-holes cannot dominate the energy density before $t_\textrm{eva}$, i.e. $n(t_\textrm{init})<n_c(t_\textrm{init})$ as defined in Sec.~\ref{sec:relic}. Thus the upper bound on the mass fraction at present can be found by using  (\ref{eq:betaf}), 
 \begin{eqnarray}
 \label{fs}
 f_{22}\approx A^{-5/3} f_{\textrm{BH}}\;,
 \end{eqnarray}
where BBN puts stronger constraints for  2-2 hole remnants in comparison to a PBH remnant with the mass $\Mmin$. The constraints are displayed in Fig.~\ref{fig:BBN-early}, assuming $\mathcal{N}\approx 107$.
The dashed line on each panel shows the maximal mass fraction $f_\textrm{max}$ in (\ref{eq:betacfmax}) as derived from the cosmic evolution. For smaller $\Mmin$, say in Fig.~\ref{fig:f1}, $f_\textrm{max}<1$ for $\Mini$ relevant for BBN constraints, and 2-2-hole cannot constitute all of dark matter independent of the BBN observations. It is for larger $\Mmin$, as in Fig.~\ref{fig:f2}, that there is a small region with $f_\textrm{max}>1$ being excluded directly by the BBN observations. 
 

\subsection{CMB constraints}
\label{sec:CMB}

Depending on the mass, the evaporation of 2-2-holes may influence the CMB observations in various different ways, as in the case for PBHs. 

For $\Mini\lesssim M_\textrm{BBN}$, photons emitted before BBN are completely thermalized and only contribute to the density of background radiation. If the baryon asymmetry is generated purely in the early universe, the observed baryon-to-photon ratio could be used to impose an upper bound on the entropy injection from the 2-2-hole dominant phase~\cite{1976Zel}. The baryon number density is bounded from above by the radiation density right before the final evaporation, i.e. $n_B(t_\textrm{eva})\lesssim g^{1/4}_* \rho_\textrm{rad}^{3/4}(t_\textrm{eva})$ with $\rho_\textrm{rad}(t_\textrm{eva})\approx0.03\,(t_\textrm{eva}\lp)^{-2}(t_\textrm{eva}/t_\textrm{dom})^{-2/3}$, while the entropy density is bounded from below by the 2-2-hole entropy injection $s(t_\textrm{eva})$ given in (\ref{eq:entropy}). The ratio $g^{1/4}_* \rho_\textrm{rad}^{3/4}(t_\textrm{eva})/s(t_\textrm{eva})\approx 6.4\times 10^{-3}\,\mathcal{N}^{-1/2}\mathcal{N}_*^{1/4} \hMmin \,\hMini^{-5/2}s(t_\textrm{init})/n(t_\textrm{init})$ then provides an upper bound on the  baryon-to-photon ratio $n_B/n_\gamma$, which should be larger than the observed value $6\times 10^{-10}$. This imposes an upper bound on the 2-2-hole number density to entropy density ratio as
\begin{eqnarray}\label{eq:entropybound}
\frac{n(t_\textrm{init})}{s(t_\textrm{init})}\lesssim 1.2\times 10^7\,\mathcal{N}^{-1/2}\mathcal{N}_*^{1/4} \hMmin \, \hMini^{-5/2}\,.
\end{eqnarray}
This entropy constraint is weaker than the requirement of generating the observed relic abundance, and is relevant only for $M_\textrm{DM}\lesssim \Mini \lesssim M_\textrm{BBN}$.

It is possible that the baryon asymmetry can be generated by the evaporation of primordial objects, as been discussed for PBHs in \cite{Carr:1976zz, Zeldovich:1976vw, Toussaint:1978br,Turner:1979bt, Grillo:1980rt}. This requires that the initial temperature is above the electroweak scale, namely, $\Mini\lesssim 10^{12}\left(\Mmin/\textrm{g}\right)^{1/2}\,$g. For $\Mmin$ of interest, this upper bound is comparable to $\Mini\lesssim M_\textrm{BBN}$. Thus, 2-2-holes that complete the early-time evaporation before the BBN era may account for the observed baryon asymmetry. In this case the entropy bound derived in the previous paragraph does not apply.

The emission after BBN, but before the time of recombination, can produce distortions in the CMB spectrum. Yet, given that the parameter space has already been strongly constrained by the BBN observations discussed before, these constraints are of less interest.   

The emission after recombination causes the damping of small scale CMB anisotropies, providing a new constraint on the number density of 2-2-holes for $\Mini \gtrsim M_\textrm{rec}$. With the dominant contribution from  the early stage of evaporation as before, the bounds would be similar to the  PBH case, which were found in \cite{Zhang:2007zzh} by modifying the calculation for decaying dark matter particles. The bound depends on the PBH  lifetime, the mass fraction, and the branching ratio for electrons and positrons $\textrm{B}_e$, which dominates the energy that goes into heating the matter~\cite{Carr:2009jm}. So, in the case of 2-2-holes we can make a simple replacement with
\begin{eqnarray}\label{eq:CMB}
\log_{10}\left(\textrm{B}_e \,f\right) < -10.8-0.50 \,x+0.085 \,x^2+0.0045 \,x^3,\quad
x=\log_{10}\left(\frac{10^{13}\,\textrm{s}}{\tau_L}\right).
\end{eqnarray} 
Here, $f$ denotes the mass fraction in dark matter at the time of recombination, which can be related to $n(t_\textrm{init})/s(t_\textrm{init})$ by (\ref{eq:betaf}), with $\Mmin$ replaced by $\Mini$, since the 2-2-hole mass remains close to the initial value. 
It turns out that the constraint on $n(t_\textrm{init})/s(t_\textrm{init})$ can be well approximated by a simple form,
\begin{eqnarray}
\frac{n(t_\textrm{init})}{s(t_\textrm{init})}\lesssim 3\times 10^{-80}\,\textrm{B}_e^{-1}\mathcal{N}^{\,0.8}\,\mathcal{N}_*^{-0.8}\hMmin^{-1.5}\,\hMini^{1.3}\,.
\end{eqnarray}
As in the case for PBHs, the bound is quite strong and it becomes weaker as $\Mini$ increases. 
For an order of magnitude estimation, we use $\textrm{B}_e\approx 0.1$ for 2-2-holes with $\Mmin\lesssim 10^{20\,}$g and $T_{\infty,\textrm{init}}\gtrsim 0.1\,$MeV, as for PBHs with $M_\textrm{BH}^\textrm{init} \sim 10^{13}\text{--}10^{14}\,$g. 
For much larger $\Mmin$ with $T_{\infty,\textrm{init}}\ll 0.1\,$MeV, there is only heating from photons, thus the constraint would be much more relaxed.


\section{Discussions}
\label{sec:discuss}

\begin{figure}[!h]
  \centering%
{ \includegraphics[width=11cm]{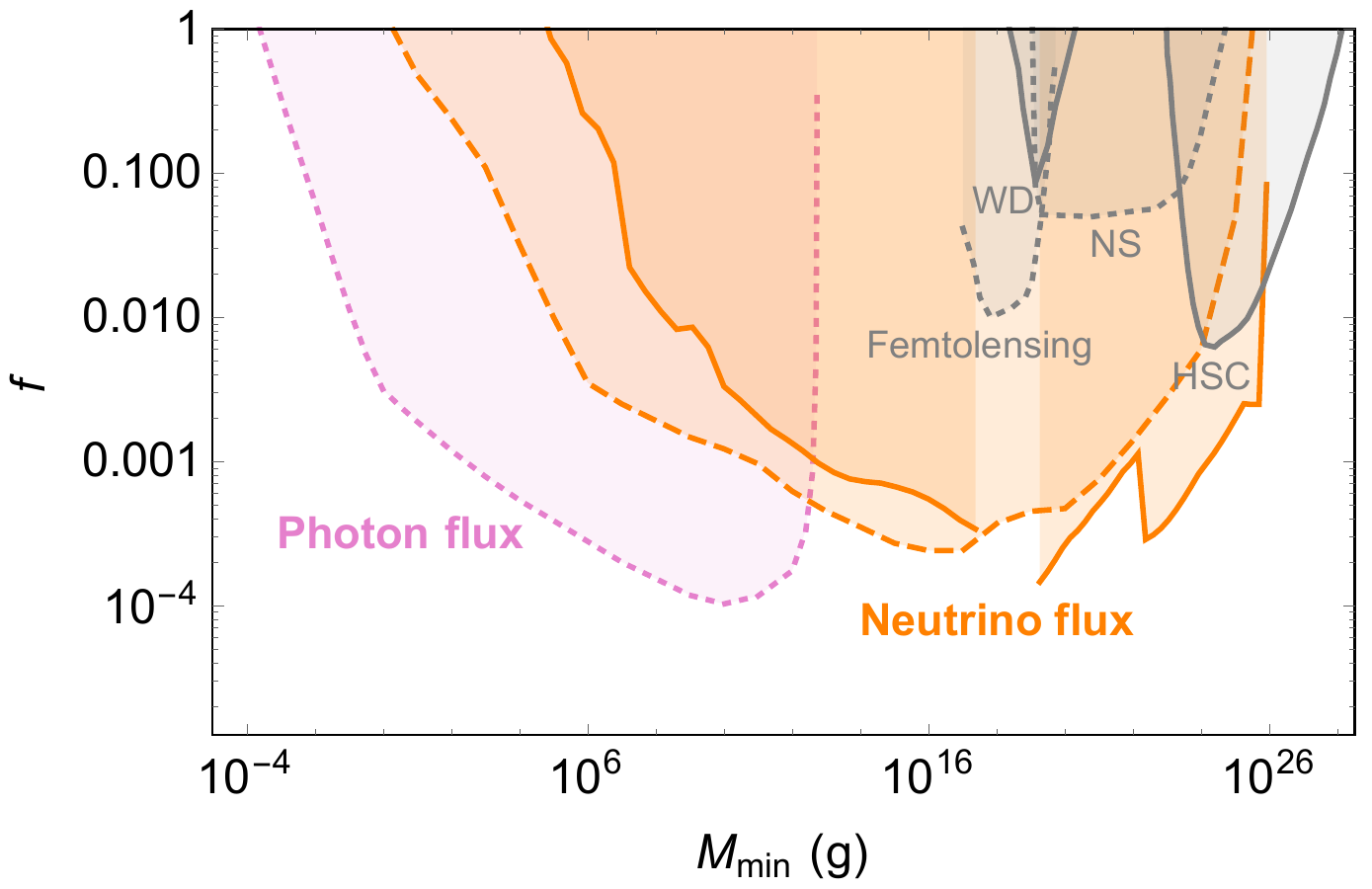}}
\caption{\label{fig:LUcons} 
Summary of constraints on the mass fraction of 2-2-hole remnants in dark matter at present $f\equiv \rho(t_0)/\rho_\textrm{DM}(t_0)$ as a function of $\Mmin$. The colored lines show the constraints on the high-energy particle flux produced by binary mergers of 2-2-hole remnants. The orange solid lines denote the upper bound from neutrino observations by considering only the on-shell production, whereas the orange dashed line includes contribution from the parton shower of initial quarks. The pink dotted line shows the strongest constraints from the diffuse photon flux at energies much smaller than the average value. The gray lines present a few upper bounds from purely gravitational interactions as directly adapted from PBHs with the same mass. Femtolensing of gamma-ray bursts may constrain $\Mmin$ down to $10^{17}\,$g~\cite{Barnacka:2012bm}. The dynamical constraints from disruptions of white dwarfs (WD) and neutron stars (NS) become relevant for $\Mmin$ around $10^{20}\,$g~\cite{Graham:2015apa, Capela:2013yf}. The microlensing observations, e.g. HSC~\cite{Niikura:2017zjd, Smyth:2019whb}, cover a wide mass range from $10^{23}$\,g all the way up to $M_\odot$. (The validity of the gray dotted lines has been questioned recently~\cite{Katz:2018zrn, Conroy:2010bs, Ibata:2012eq}.)}
\end{figure}

\begin{figure}[h!]
\captionsetup[subfigure]{labelformat=empty}
\centering
\hspace{-0.8cm}
\begin{tabular}{rrr}
\subfloat[\quad(a) $\Mmin=m_{\mathrm{Pl}}$]{\includegraphics[width=8.3cm]{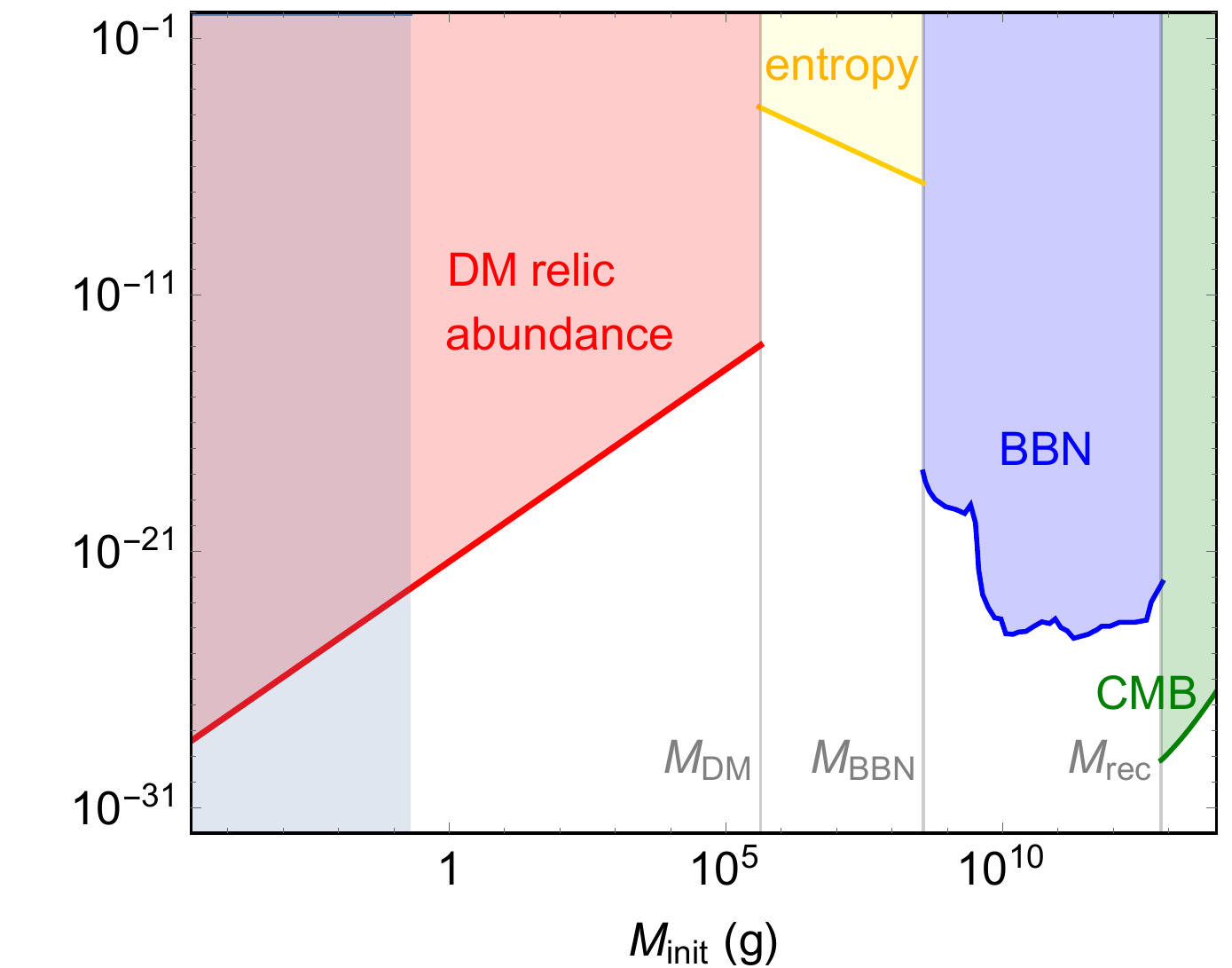}\label{fig:EU1}}&
\subfloat[\quad(b) $\Mmin=10^{20}$\,g]{\includegraphics[width=8.3cm]{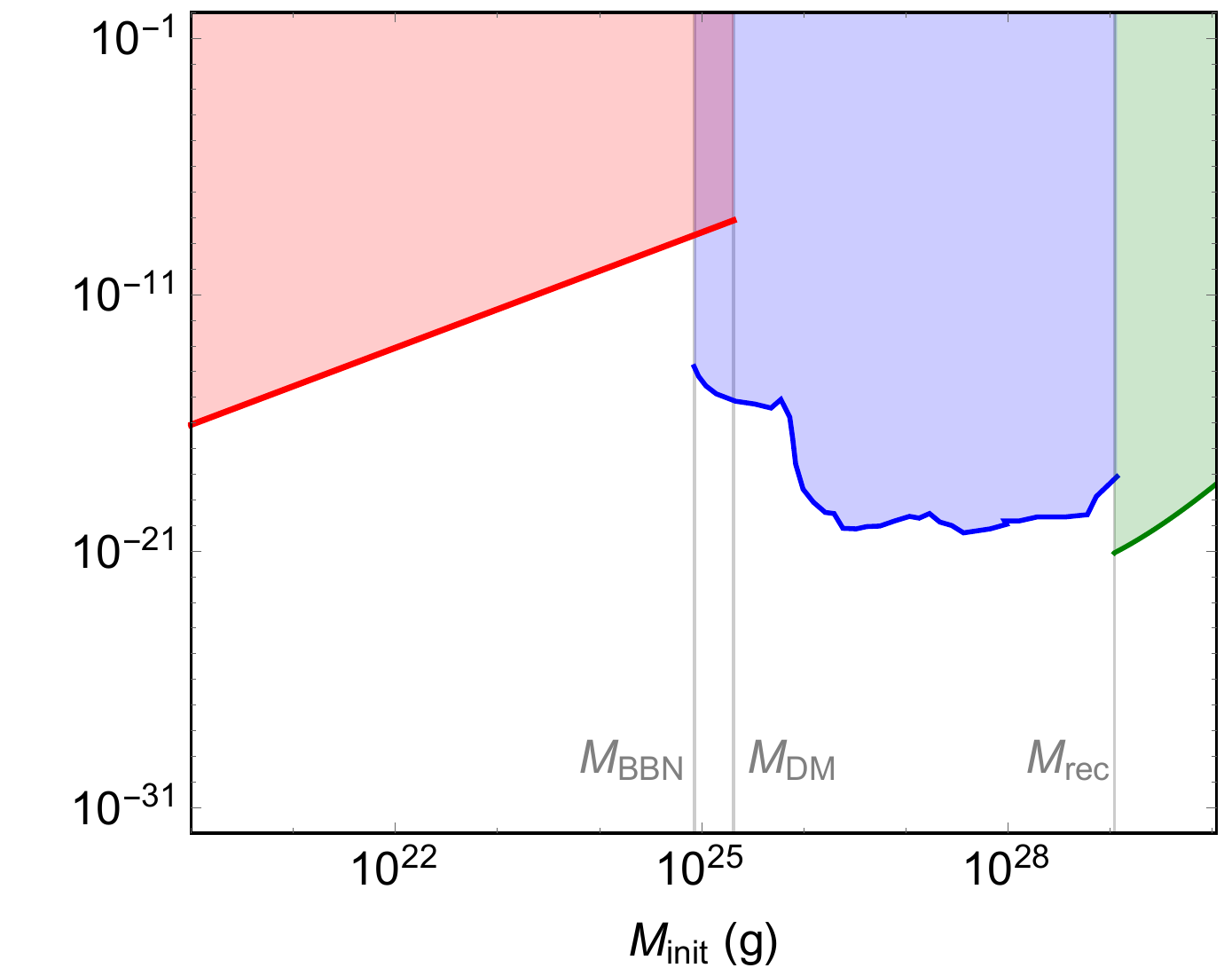}\label{fig:EU3}}
\end{tabular}
\caption{\label{fig:EUcons} 
Summary of the early-universe constraints on the number density to entropy density ratio at formation $(\Mini/\Mp)^{3/2} \,n(t_\textrm{init})/s(t_\textrm{init})$ as a function of $\Mini$ for two benchmark values of $\Mmin$. This quantity is related to the mass fraction at formation, $\beta$, as defined in (\ref{eq:beta}). 
We restrict to the small-mass range, i.e. $\Mmin \lesssim \Mini\lesssim M_\textrm{uni}$, where the primordial 2-2-holes have already become remnants today. 
The red curve denotes the parameter space that produces the observed relic abundance of dark matter. Exclusions from the photon-to-baryon ratio through entropy injection in (\ref{eq:entropybound}) (invalid if the baryon asymmetry is generated by the 2-2-hole evaporation), light element abundance formed in BBN in (\ref{betas}) and CMB anisotropy in (\ref{eq:CMB}) are shown in yellow, blue and green respectively.  
The critical masses $M_\textrm{BBN},\, M_\textrm{rec}$ in (\ref{eq:Mi}) and $M_\textrm{DM}$ in (\ref{eq:Mrelic}), as shown by the gray vertical lines, specify the mass ranges relevant to various observations. 
The gray vertical band excludes low values of $\Mini$ due to the minimal horizon mass in the radiation era, assuming the upper bound on the reheating temperature say around $10^{16}\,$GeV.}
\end{figure}

As we have seen, the behavior of primordial thermal 2-2-holes is mainly determined by the initial mass at formation $\Mini$ and the minimal mass $\Mmin$, with minor dependence on the number of degrees of freedom for both the gas in the interior ($\mathcal{N}$) and the radiation ($\mathcal{N}_*$). As discussed in detail in Sec.~\ref{sec:present} and Sec.~\ref{sec:pheno}, various observations  can be used to probe $\Mini$ and $\Mmin$. The results are summarized in Fig.~\ref{fig:LUcons} and Fig.~\ref{fig:EUcons}, assuming the Standard Model values for $\mathcal{N}$ and $\mathcal{N}_*$.

The present epoch observations for the remnants can be used to probe the fundamental mass scale $\Mmin$ in the theory. Constraints on the present mass fraction in dark matter $f$ as a function of $\Mmin$ are displayed in Fig.~\ref{fig:LUcons}. 
The most relevant constraints come from high-energy particles emitted by the merger product of the remnant binaries.\footnote{Radiation from the single remnants is rather weak and does not yield signals sensitive to either the early universe or present epoch observations, as detailed in Appendix~\ref{sec:remrad}.} 
When using a recent estimation for the binary merger rate we find that the current observations have a good coverage for the neutrino and photon fluxes for a wide range of $\Mmin$. 
By considering only on-shell production of neutrinos from 2-2-hole evaporation, remnants with $\Mmin\gtrsim 10^5\,$g are excluded from making up all of dark matter. 
We then consider parton showers of highly off-shell initial particles, where a significant amount of lower energy neutrinos and photons are generated from hadron decay. This enables us to probe smaller $\Mmin$.
By assuming the fragmentation functions in ordinary QCD, the neutrino observations push the bounds down to $\sim 1$\,g, and the measured photon flux further strengthens the constraint to be $\Mmin\lesssim  10\,\Mp$.\footnote{For illustration,  we use the fragmentation function for temperature around $10^{16}\,$GeV. The larger $\Mmin$ cases with much lower temperature would have stronger couplings and a different shape of the distribution functions. The exact constraint on $f$ in Fig.~\ref{fig:LUcons} will then change for large $\Mmin$. But it remains true that remnants cannot be all of dark matter for large $\Mmin$.}

It is instructive to compare these novel constraints with the existing bounds from gravitational interaction, which become relevant only for relatively large remnants, i.e. $\Mmin$ no smaller than $10^{17}\,$g. 
Apparently, we have opened a new window onto the range of dark matter parameter space that was previously considered untestable. 
The strongest photon bound nonetheless depends on the fragmentation function at quite small energy fraction and so may suffer more from the theoretical uncertainties. A better understanding of parton showers for 2-2-hole evaporation, in particular at lower energy,  and a more reliable estimation of the merger rate will help clarify the observational consequences of small 2-2-hole remnants comprising the dark matter.

The fact that only small $\Mmin$ is allowed excludes a large portion of parameter space for the weak coupling scenario of quadratic gravity. One motivation for this scenario is to resolve the Higgs naturalness problem with feeble gravitational interactions. The relevant parameter range, i.e. $m_2\lesssim 10^{-8}\,\Mp$~\cite{Salvio:2014soa}, has been excluded if the constraints from the parton shower are taken seriously.  Recently, new experimental setups have been proposed for the gravitational direct detection of dark matter in the small-mass range. Ref.~\cite{Carney:2019pza} considered an array of quantum-limited impulse sensors. A meter-scale apparatus may be capable of detecting Planck mass remnants, whereas a heavier mass range can be reached with a sparse, larger array of detectors.

Next, we review the constraints obtained from early-universe physics, assuming formation of primordial 2-2-holes in the radiation era. Fig.~\ref{fig:EU1} shows the constraints for Planck mass remnants in the strong coupling scenario of quadratic gravity, which resemble closely the constraints for PBH relics of the Planck mass~\cite{Carr:2009jm, Dalianis:2019asr}. The red curve shows the requirement of generating the observed dark matter relic abundance for $\Mini$ extending up to $M_\textrm{DM}\approx 4\times 10^5\,\textrm{g}\lesssim  M_\textrm{BBN}$, meaning that the early stage of evaporation ends way before the BBN epoch. Larger 2-2-holes have too small remnant abundance, due to suppression from an extra 2-2-hole dominant phase as discussed in Sec.~\ref{sec:relic}, but their number density can still be constrained by early universe observations, as shown for the photon-to-baryon ratio, BBN and CMB observations. In the weak coupling scenario, with the remnants being heavier, the constraints in general differ from the black hole remnants. For increasing $\Mmin$, the parameter region constrained by entropy injection shrinks. When $\Mmin\gtrsim 10^{17}\,$g, we have $M_\textrm{DM}\gtrsim M_\textrm{BBN}$ and the red and blue regions overlap, as shown in Fig.~\ref{fig:EU3}. This implies that the parameter space relevant to dark matter starts to be excluded by BBN observations.

With the remnant mass bounded to be small as in Fig.~\ref{fig:LUcons}, the upper bound on the formation mass as required for remnant dark matter is $\Mini\lesssim 10^{13}$\,g for $\Mmin\lesssim10^5$\,g and $\Mini\lesssim 10^{6}\,$g for $\Mmin\lesssim 10\,\Mp$, respectively. 
This  upper bound has implications for the inflation model that gives rise to  density inhomogeneities responsible for the production of primordial 2-2-holes. For instance, our bounds are smaller than the lower bound on $\Mini$ derived from some conjectures for the UV physics~\cite{Cai:2019igo}. The density inhomogeneities can generate a stochastic gravitational-wave background, but the peak frequency for $\Mini\lesssim 10^{13}\,$g is above kHz~\cite{Wang:2019kaf} and is beyond the current reach of ground based detectors. As in the PBH case, the demanded mass fraction for 2-2-hole dark matter implies quite large density perturbations at small scales in comparison to CMB observations at large scales. This makes the model building of inflation more contrived, and other production mechanisms deserve to be further explored in this new context.


\section{Summary}
\label{sec:final}

As horizonless ultracompact objects, remnants from primordial thermal 2-2-holes constitute a well-motivated candidate for dark matter. They arise in quadratic gravity, a candidate theory for quantum gravity~\cite{Holdom:2002xy, Holdom:2016nek, Holdom:2019ouz, Ren:2019afg}. The fact that remnants appear naturally in the theory puts them in a favorable position over PBH remnants. Moreover, the 2-2-hole is a probable endpoint of gravitational collapse instead of the black hole, offering a resolution to the information loss conundrum due to the absence of a horizon. 

The remnant mass $\Mmin$, the minimum allowed mass for a 2-2-hole, is linked to the mass of the spin-2 mode in the theory. Therefore, any information on $\Mmin$ directly connects to the underlying theory of quantum gravity. In understanding the observational implications of 2-2-hole remnants as dark matter and exploring the available parameter space for $\Mmin$, the main determinant is the thermodynamic behavior of 2-2-holes. The case of a relativistic thermal gas as the matter source was investigated in \cite{Holdom:2019ouz,Ren:2019afg}, and this provides a realistic scheme to work in.

Thermodynamic properties of a thermal 2-2-hole in the large-mass range have the same form as a black hole, thus the evaporation in the early stage shares most of the features of the black hole evaporation. Once the temperature reaches the peak value, the 2-2-hole enters into the remnant stage with close to the minimal mass, where drastic changes occur in the thermodynamic behavior; heat capacity becomes positive, the evaporation significantly slows down and asymptotically halts. It is this small 2-2-hole that behaves as a cold, stable remnant, and serves as dark matter.

As we have shown in this paper, 2-2-hole remnants can account for all of dark matter and satisfy all observational constraints if both the remnant and formation masses are relatively small. 
The formation mass is bounded to be small mainly by the requirement of generating the observed dark matter abundance. The early stage of evaporation in turn  ends way before BBN begins, with little influence on other early-universe observations. The remnant mass, on the other hand, can be probed by the observations of high-energy astrophysical particles, in additional to the conventional PBH searches through gravitational interactions. When two remnants form a binary and merge, the merger product is no longer a remnant state, but a very hot 2-2-hole that produces a strong flux of high energy particles before settling back down to a cold remnant. With the latest estimation for the binary merger rate, the predicted signals turn out to be strong enough to be confronted by the data, especially from the photon and neutrino observations. This enables us to constrain the remnant mass to be far below the range accessible by the conventional PBH searches. The neutrino bounds are more robust against theoretical uncertainties in the distribution function from parton showers, with $\Mmin\lesssim 10^5\,$g being a conservative estimate. The photon bounds are stronger, and they narrow down the viable parameter space to be $\Mmin\lesssim 10\,\Mp$. In this way, our constraints are tending to push the theory of quantum gravity towards the strong coupling regime, and thus towards a theory with only one fundamental scale.


\vspace{0.1cm}
\section*{Acknowledgements} 
\vspace{-0.1cm}
We thank Hardi Veermae, Sai Wang, Shun Zhou for valuable discussions. Work of U.A. is supported in part by the Chinese Academy of Sciences President's International Fellowship Initiative (PIFI) under Grant No. 2020PM0019, and the Institute of High Energy Physics, Chinese Academy of Sciences, under Contract No.~Y9291120K2. B.H. is supported in part by the Natural Sciences and Engineering Research Council of Canada. J.R. is supported in part by the Institute of High Energy Physics under Contract No. Y9291120K2.

\vspace{0.5cm}
\noindent
\emph{Note added.}— Recently the authors of Ref.~\cite{Salvio:2019llz} have also discussed the possibility that horizonless ultracompact objects in quadratic gravity serve as dark matter. However, they consider a different object, which is a new limit of the regular solution that can reach large compactness only for small masses, i.e. $M\lesssim \Mmin$. 
Such solutions are rather disconnected from solutions that can serve as the endpoint of gravitational collapse of large masses, which are black holes, or as we argue here, 2-2-holes.


\appendix

\section{Structure of thermal 2-2-holes}
\label{sec:22hole}

With no horizon, the whole spacetime for a 2-2-hole can be described by one coordinate system. For a static and spherically symmetric case, we can choose the form of the line element as
\begin{eqnarray}\label{eq:metric}
ds^2=-B(r)dt^2+A(r)dr^2+r^2d\Omega^2\,.
\end{eqnarray}
A 2-2-hole is then defined by the following characteristic leading-order behavior at small $r$,
\begin{eqnarray}
A(r)=a_2 r^2+...,\quad 
B(r)=b_2 r^2+... \,,
\end{eqnarray}
which correspond to a vanishing metric at the origin. As a new family of solutions in quadratic gravity, it is also the most generic one with a much larger parameter space than for either the regular solutions or the Schwarzschild-like solutions~\cite{Holdom:2002xy, Holdom:2016nek}.

As the Ricci term $R^2$ is inessential, we find numerical solutions for the thermal 2-2-holes in Einstein-Weyl gravity as described by  two mass scales, the Planck mass and  the mass of the spin-2 mode. The latter is related to the minimum allowed mass $\Mmin$ for the thermal 2-2-holes as in (\ref{eq:Mmin}). The relativistic thermal gas model is governed by the metric functions $A(r), B(r)$ and the locally measured gas temperature $T(r)$. With two field equations and the conservation law $T(r)B(r)^{1/2}=T_\infty$, a one-parameter family of solutions exists and is characterized by $M/\Mmin$~\cite{ Ren:2019afg}.

\begin{figure}[!h]
  \centering%
{ \includegraphics[width=8.0cm]{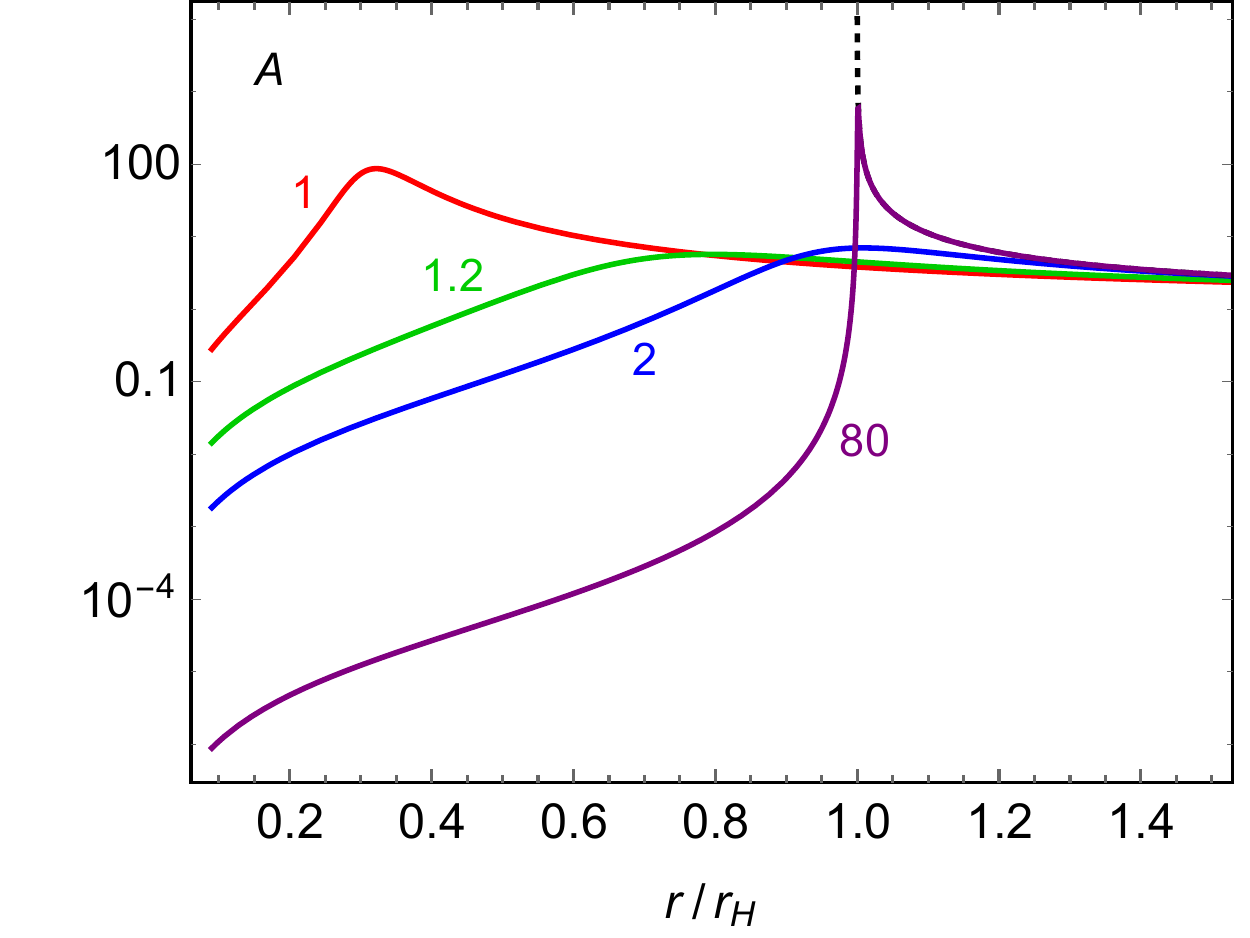}}\quad
{ \includegraphics[width=8.0cm]{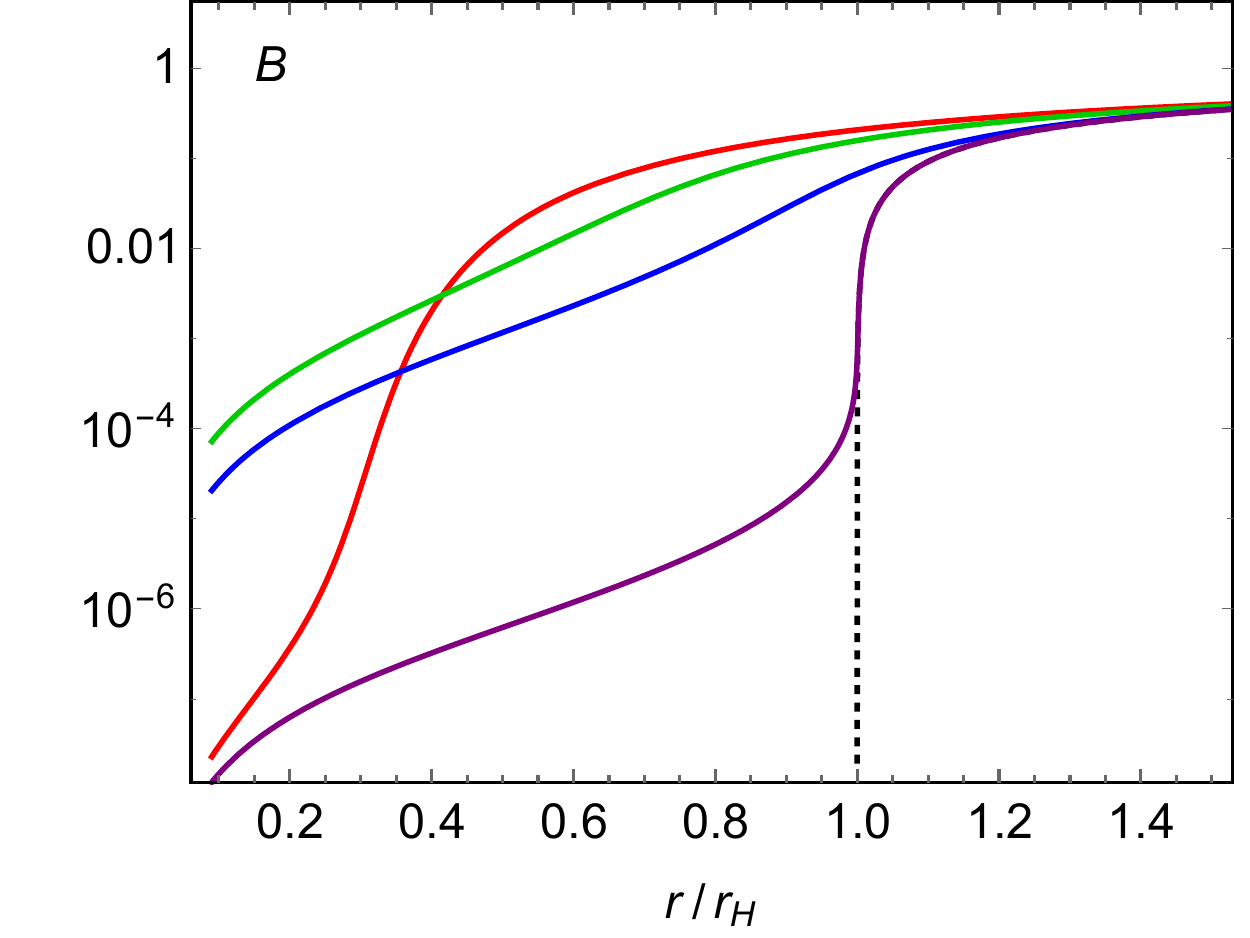}}\\
{ \includegraphics[width=8cm]{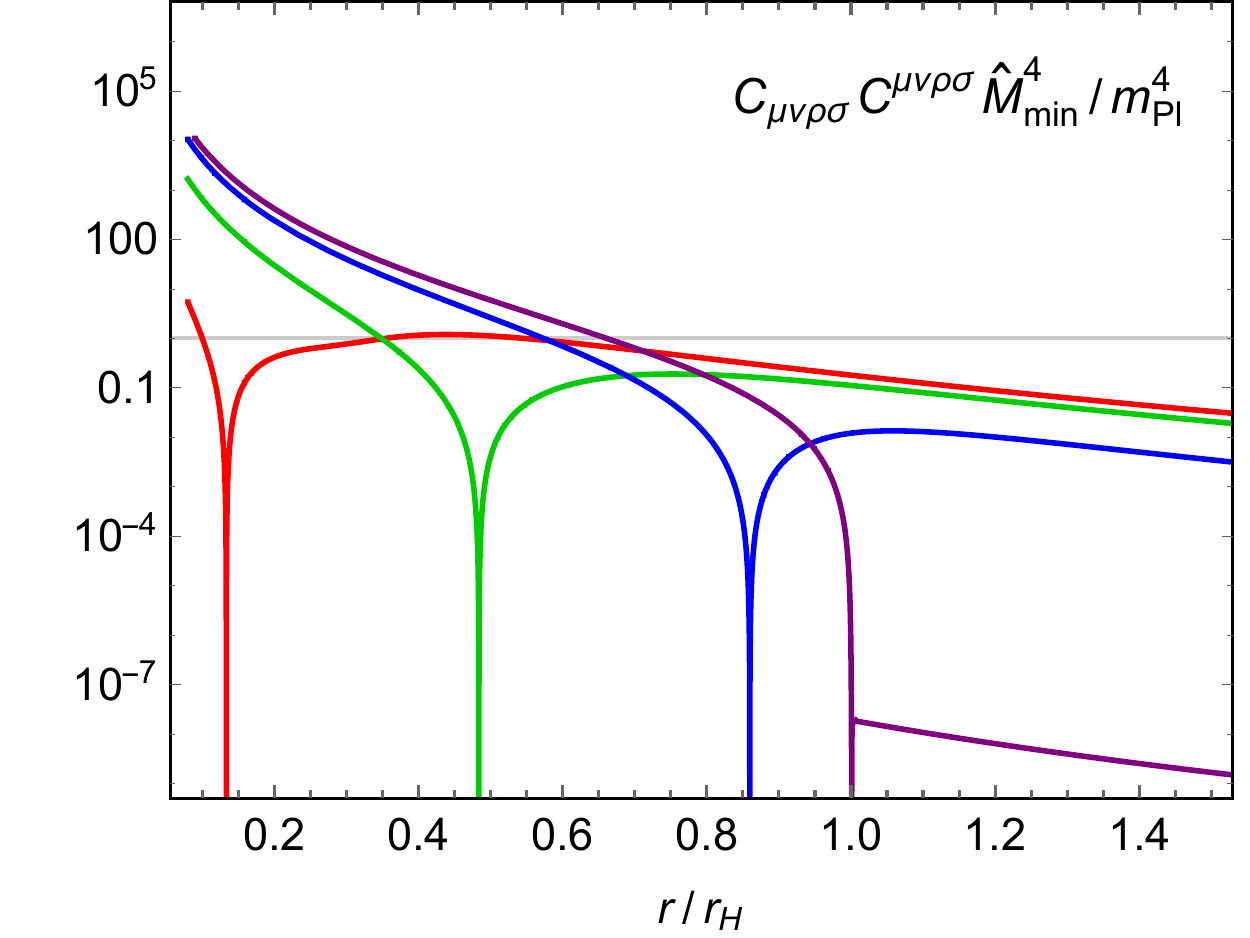}}\quad
{ \includegraphics[width=8cm]{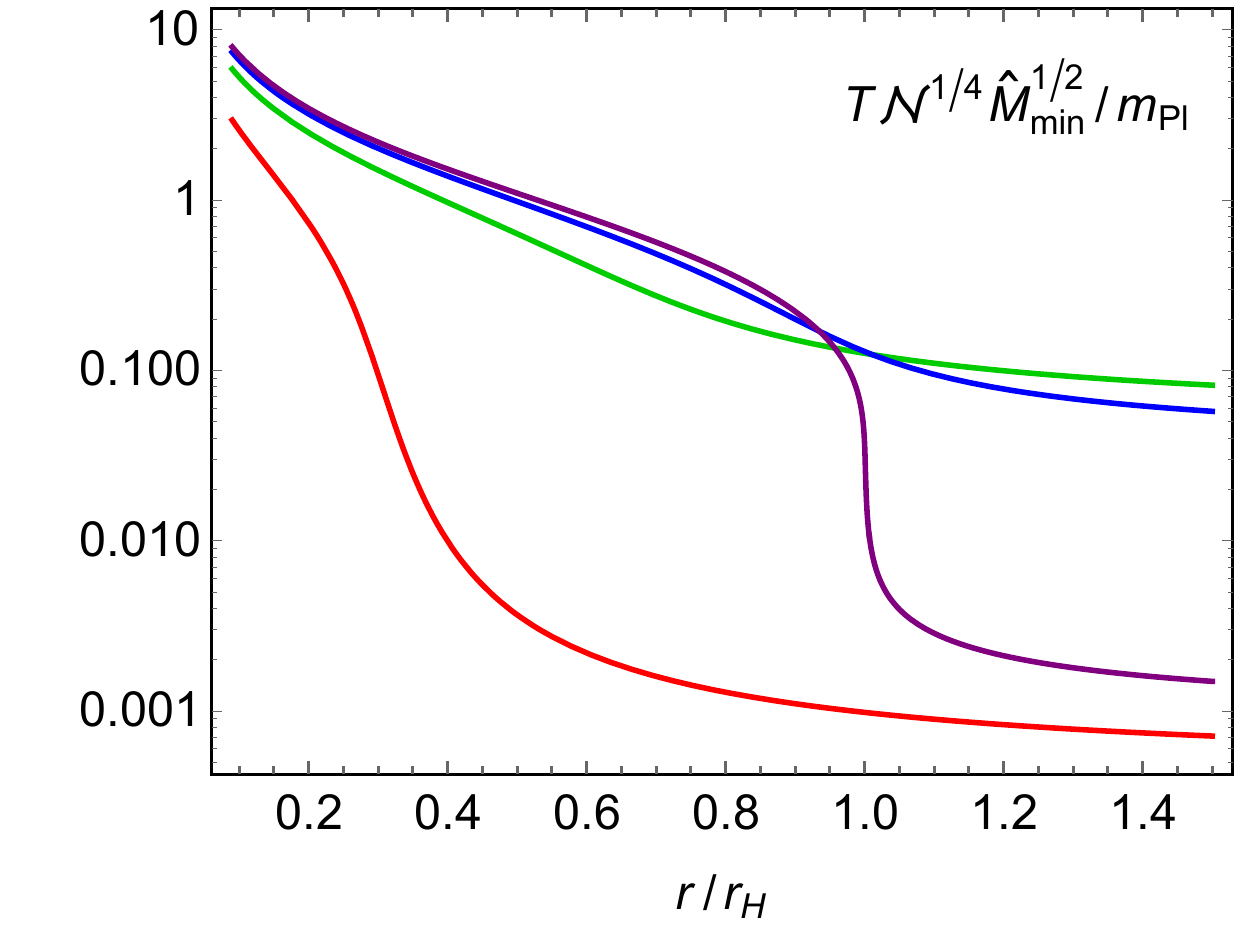}}
\caption{\label{fig:solutions} 
The metric functions $A$, $B$, the Weyl tensor square $C_{\mu\nu\rho\sigma}C^{\mu\nu\rho\sigma}$, and the relativistic thermal gas temperature $T$ as functions of $r/r_H$ for $M/\Mmin\approx 1$ (red), 1.2 (green), 2 (blue), 80 (purple). The black dotted line in the top panels is the Schwarzschild solution. The radius at which the Weyl tensor square vanishes denotes roughly the boundary of the interior.}
\end{figure}

Figure~\ref{fig:solutions} shows how the metric and matter properties vary with the radius $r$ for some typical values of $M/\Mmin$. The plots are arranged to be independent of the values of $\Mmin$ and $\mathcal{N}$, and so the features apply to $\Mmin$ of any size. 
The purple line shows the behavior for a typical large 2-2-hole with $M$ much larger than 
$\Mmin$. Deviations from the Schwarzschild solution occur at a minuscule distance outside $r_H$, and the behavior changes abruptly almost at the same radius. The blue line is for the merger product of 2-2-hole remnants, which is similar to the $M=M_\textrm{peak}$ case (green line) with the maximum $T_\infty$. The sizes of the transition region and the interior region are both at the order of $\mathcal{O}(r_H)$. 
The red line is for a typical 2-2-hole remnant with $\Delta M$ much smaller than $\Mmin$. This corresponds to an interior that is pushed well within the would-be horizon with a size of order $1/\sqrt{a_2}$. 
As for a large 2-2-hole, the interior in this case features an extremely deep gravitational potential as described by $B(r)$. 

In practice, it is numerically demanding to obtain solutions with $M/\Mmin$ being too large or too close to unity. Fortunately, the solutions in these two limits are governed by distinct scaling behaviors that relate the interior solutions at different values of $M/\Mmin$. In the large mass limit, the interior is characterized by $r_H$ and $\lambda_2$, while in the small mass limit the relevant scales are defined by $a_2$ and $b_2$. With the corresponding scaling behaviors, we can then derive the analytical approximations (\ref{eq:LMlimit}) and (\ref{eq:SMlimit}) for the large- and small-mass cases, respectively. 


\section{Constraints on radiation from single remnants}
\label{sec:remrad}

When the 2-2-hole formation mass $\Mini$ is small, the early stage of evaporation ends too early to be constrained. For such cases, constraints on the remnant radiation need to be checked.
As we've discussed in Sec.~\ref{sec:evapo}, the remnant radiation suffers from the complicated low frequency effects, corresponding to a suppressed effective emitted surface. For simplicity we use the would-be horizon area in the Stefan-Boltzmann law (\ref{eq:SBlaw}), which then provides an upper bound on the radiation power. In this section we perform order of magnitude estimates for constraints on the remnant radiation based on (\ref{eq:SBlaw}).  
As we will see, for $\Mmin$ of interest, the remnant contributions are too small to be constrained by the current data. Any further low frequency suppression would make it even smaller.

We start from early universe observations, i.e. BBN and CMB. Constraints on the remnant radiation can still be inferred from that for PBHs by considering the ratio of the emission rate,
\begin{eqnarray}
\frac{\Gamma_{i,22}(t)}{\Gamma_{i,\textrm{BH}}(t)}
=\frac{\left(\textrm{B}_i\,\mathcal{N}_*\right)_{22}}{\left(\textrm{B}_i\,\mathcal{N}_*\right)_\textrm{BH}}\frac{f _{22} \,\rho_\textrm{DM} \,s(t_0)^{-1}\Mmin \,T_\infty^3(t) }{0.4\,\beta _{\text{BH}}\,\gamma^{1/2}g_*^{-1/4}\left(M_{\text{init}}^{\text{BH}}\right)^{-3/2}M_{\text{BH}}^2(t) \,T_{\text{BH}}^3(t)}\,,
\end{eqnarray}
where $n_{22}(t)/s(t)=n_{22}(t_0)/s(t_0)$ and $n_\textrm{BH}(t)/s(t)=n_\textrm{BH}(t_\textrm{init})/s(t_\textrm{init})$ are used. $T_\infty(t)$ is the remnant temperature in (\ref{eq:SMlimitTime}). 
Since $\Gamma_{i,22}(t)$, $T_\infty(t)$  decrease with time and $\Gamma_{i,\textrm{BH}}(t)$, $T_{\text{BH}}(t)$ increase with time, we can find a conservative constraint by comparing a 2-2-hole remnant with a black hole at the earliest time $t_*$ of the relevant physical process with $T_\infty(t_*)=T_{\text{BH}}(t_*)$. This equal-temperature condition determines the remnant mass $\Mmin$ used for comparison for a given  $t_*$ and $M^\textrm{BH}_\textrm{init}$. 
Assuming $\Gamma_{i,22}(t_*)=\Gamma_{i,\textrm{BH}}(t_*)$, the upper bound on the 2-2-hole mass fraction at this $\Mmin$ is 
\begin{eqnarray}\label{eq:f22late}
f_{22}=4.6\times 10^{25} \left(\beta _{\text{BH}}\,\gamma^{1/2}g_*^{-1/4}\right) \left(\frac{M^\textrm{BH}_\textrm{init}}{\textrm{g}}\right)^{1/2}\left(\frac{\Mmin}{\textrm{g}}\right)^{-1}\left(1-\frac{t_*}{\tau_\textrm{BH}}\right)^{2/3},
\end{eqnarray}
with  the upper bound on $\beta _{\text{BH}}\,\gamma^{1/2}g_*^{-1/4}$ for PBHs. This is a conservative estimation since $T_\infty(t), \left(\textrm{B}_i\,\mathcal{N}_*\right)_{22}, \Gamma_{i,22}(t)$ are smaller than the black hole counterpart during the relevant time period $t>t_*$. 

The BBN observations can constrain the remnant radiation when $\Mini\lesssim M_\textrm{BBN}$. For the three processes as we have discussed before, the corresponding range of the remnant mass is $\Mmin\sim \Mp\text{--}2\times 10^{10}\,$g. Using the constraints for PBHs and  (\ref{eq:f22late}), we find that the conservative upper bound on $f_{22}$ in this mass range is larger than unity. In other words, these 2-2-hole remnants can safely evade the BBN constraints on the late-time radiation and account for all of dark matter. 
The CMB observations become relevant for $\Mini \lesssim M_\textrm{rec}$. Similarly, we find no constraint on 2-2-hole remnants with $\Mmin\lesssim 4\times 10^{12}\,$g from PBH studies~\cite{Carr:2009jm,Clark:2016nst}. Heavier remnants with $T_\infty$ much lower than the QCD scale or electron mass have negligible impacts on BBN or CMB.


The present observations can also be used to constrain the remnant radiation. One example is the diffuse $\gamma$-ray background, which has been studied for  PBHs that haven't completed (or just completed) their evaporation by now with mass  around $10^{14}\text{--}10^{15}$\,g~\cite{Page:1976wx,MacGibbon:1991vc,Carr:2009jm}. For 2-2-holes, the photon background receives contribution from both the Milky way at present and the whole universe from early time.

The galactic contribution can be calculated from the equation
\begin{equation}
\label{flux}
\Phi_{\gamma}=\frac{\mathcal{D}}{\Mmin}  \left(\frac{d N_{\gamma}}{dE_{\gamma} dt}\right)\;,
\end{equation}
where the $D$-factor is defined in (\ref{D-factor}) and the photon emission rate is given as
\begin{eqnarray}
\label{emissionrate}
 \left(\frac{d N_{\gamma}}{dE_{\gamma} dt}\right)\approx \frac{\pi^2}{120}2\times 4\pi r_{H}^2 \frac{T_{\infty}^4}{\langle{E}_{\gamma}\rangle^2}\;.
\end{eqnarray}
Here we approximate the spectrum by emission at the average photon energy $\langle{E}_{\gamma}\rangle\approx 5.7 \,T_{\infty}$~\cite{Carr:2009jm}, which ranges from $0.1\,$keV to $10\,$MeV for $\Mmin=\Mp\text{--}10^{22}\,$g. 
For 2-2-holes, the quantity of interest is then  
\begin{eqnarray}
\label{galacticflux}
E_{\gamma}^2\Phi_{\gamma}\approx 1.0\times10^{-7}\;f\;\mathcal{N}^{1/3}\mathcal{N_*}^{-4/3}\left(\frac{\Mmin}{\textrm{g}}\right)^{1/3} \left(\ln\frac{1.8\times10^{56}}{\Mmin/\textrm{g}}\right)^{-7/3}\;\; \mbox{ keV}\;\mbox{sr}^{-1}\mbox{ cm}^{-2}\mbox{ s}^{-1}\;,
\end{eqnarray}
where $T_{\infty}$ is evaluated from (\ref{eq:SMlimitTime}) for $\Delta t \approx t_0$. 
In order to estimate the extragalactic contribution, we can use the corresponding versions of (\ref{flux-EG}) to take into account all the emission from the time of recombination and afterwards. With the Planckian distribution for initial particles, we use $2\pi^2 dN_{\gamma}/(dtdE_{\gamma})=E_{\gamma}(t)^2 \pi r_H^2 /( e^{E_{\gamma}(t)/T_{\infty}(t)}-1)$ for the emission rate, where $E_{\gamma}(t)=(1+z)E_{\gamma}$. We find that the extragalactic contribution is in the same order of magnitude as the galactic one.

Including both contributions, we find that the current existing data for isotropic photon flux~\cite{Strong:2004ry, Sreekumar:1997un} is unable to see a possible 2-2-hole contribution. The largest value for the average photon energy $\braket{E_{\gamma}}\approx 15$ MeV corresponds to $\Mmin\approx \Mp$, resulting in an anticipated flux contribution fifteen orders of magnitude smaller than the observed value. For larger $\Mmin$, even though the relative value of the 2-2-hole flux increases for $\braket{E_{\gamma}}\approx 0.7$ keV ($\Mmin\approx10^{22}\textrm{ g}$), the 2-2-hole contribution is still six orders of magnitude too small. 

In summary, observations from BBN, CMB and diffuse photon flux impose no constraints on the thermal radiation from isolated 2-2-holes remnants with relatively small mass. The larger $\Mmin$ region, on the other hand, has already been strongly constrained by other observations, as we can see in Fig.~\ref{fig:LUcons}.

\raggedright  
\bibliography{References_22_holes}{}

\begin{thebibliography}{85}%
\makeatletter
\providecommand \@ifxundefined [1]{%
 \@ifx{#1\undefined}
}%
\providecommand \@ifnum [1]{%
 \ifnum #1\expandafter \@firstoftwo
 \else \expandafter \@secondoftwo
 \fi
}%
\providecommand \@ifx [1]{%
 \ifx #1\expandafter \@firstoftwo
 \else \expandafter \@secondoftwo
 \fi
}%
\providecommand \natexlab [1]{#1}%
\providecommand \enquote  [1]{``#1''}%
\providecommand \bibnamefont  [1]{#1}%
\providecommand \bibfnamefont [1]{#1}%
\providecommand \citenamefont [1]{#1}%
\providecommand \href@noop [0]{\@secondoftwo}%
\providecommand \href [0]{\begingroup \@sanitize@url \@href}%
\providecommand \@href[1]{\@@startlink{#1}\@@href}%
\providecommand \@@href[1]{\endgroup#1\@@endlink}%
\providecommand \@sanitize@url [0]{\catcode `\\12\catcode `\$12\catcode
  `\&12\catcode `\#12\catcode `\^12\catcode `\_12\catcode `\%12\relax}%
\providecommand \@@startlink[1]{}%
\providecommand \@@endlink[0]{}%
\providecommand \url  [0]{\begingroup\@sanitize@url \@url }%
\providecommand \@url [1]{\endgroup\@href {#1}{\urlprefix }}%
\providecommand \urlprefix  [0]{URL }%
\providecommand \Eprint [0]{\href }%
\providecommand \doibase [0]{http://dx.doi.org/}%
\providecommand \selectlanguage [0]{\@gobble}%
\providecommand \bibinfo  [0]{\@secondoftwo}%
\providecommand \bibfield  [0]{\@secondoftwo}%
\providecommand \translation [1]{[#1]}%
\providecommand \BibitemOpen [0]{}%
\providecommand \bibitemStop [0]{}%
\providecommand \bibitemNoStop [0]{.\EOS\space}%
\providecommand \EOS [0]{\spacefactor3000\relax}%
\providecommand \BibitemShut  [1]{\csname bibitem#1\endcsname}%
\let\auto@bib@innerbib\@empty
\bibitem [{\citenamefont {Abbott}\ \emph {et~al.}(2019)\citenamefont {Abbott}
  \emph {et~al.}}]{LIGOScientific:2019fpa}%
  \BibitemOpen
  \bibfield  {author} {\bibinfo {author} {\bibfnamefont {B.~P.}\ \bibnamefont
  {Abbott}} \emph {et~al.} (\bibinfo {collaboration} {LIGO Scientific,
  Virgo}),\ }\href {\doibase 10.1103/PhysRevD.100.104036} {\bibfield  {journal}
  {\bibinfo  {journal} {Phys. Rev.}\ }\textbf {\bibinfo {volume} {D100}},\
  \bibinfo {pages} {104036} (\bibinfo {year} {2019})},\ \Eprint
  {http://arxiv.org/abs/1903.04467} {arXiv:1903.04467 [gr-qc]} \BibitemShut
  {NoStop}%
\bibitem [{\citenamefont {Cardoso}\ and\ \citenamefont
  {Pani}(2019)}]{Cardoso:2019rvt}%
  \BibitemOpen
  \bibfield  {author} {\bibinfo {author} {\bibfnamefont {V.}~\bibnamefont
  {Cardoso}}\ and\ \bibinfo {author} {\bibfnamefont {P.}~\bibnamefont {Pani}},\
  }\href {\doibase 10.1007/s41114-019-0020-4} {\bibfield  {journal} {\bibinfo
  {journal} {Living Rev. Rel.}\ }\textbf {\bibinfo {volume} {22}},\ \bibinfo
  {pages} {4} (\bibinfo {year} {2019})},\ \Eprint
  {http://arxiv.org/abs/1904.05363} {arXiv:1904.05363 [gr-qc]} \BibitemShut
  {NoStop}%
\bibitem [{\citenamefont {Carr}(1975)}]{Carr:1975qj}%
  \BibitemOpen
  \bibfield  {author} {\bibinfo {author} {\bibfnamefont {B.~J.}\ \bibnamefont
  {Carr}},\ }\href {\doibase 10.1086/153853} {\bibfield  {journal} {\bibinfo
  {journal} {Astrophys. J.}\ }\textbf {\bibinfo {volume} {201}},\ \bibinfo
  {pages} {1} (\bibinfo {year} {1975})}\BibitemShut {NoStop}%
\bibitem [{\citenamefont {Carr}\ \emph {et~al.}(2010)\citenamefont {Carr},
  \citenamefont {Kohri}, \citenamefont {Sendouda},\ and\ \citenamefont
  {Yokoyama}}]{Carr:2009jm}%
  \BibitemOpen
  \bibfield  {author} {\bibinfo {author} {\bibfnamefont {B.~J.}\ \bibnamefont
  {Carr}}, \bibinfo {author} {\bibfnamefont {K.}~\bibnamefont {Kohri}},
  \bibinfo {author} {\bibfnamefont {Y.}~\bibnamefont {Sendouda}}, \ and\
  \bibinfo {author} {\bibfnamefont {J.}~\bibnamefont {Yokoyama}},\ }\href
  {\doibase 10.1103/PhysRevD.81.104019} {\bibfield  {journal} {\bibinfo
  {journal} {Phys. Rev.}\ }\textbf {\bibinfo {volume} {D81}},\ \bibinfo {pages}
  {104019} (\bibinfo {year} {2010})},\ \Eprint {http://arxiv.org/abs/0912.5297}
  {arXiv:0912.5297 [astro-ph.CO]} \BibitemShut {NoStop}%
\bibitem [{\citenamefont {Carr}\ \emph {et~al.}(2016)\citenamefont {Carr},
  \citenamefont {Kuhnel},\ and\ \citenamefont {Sandstad}}]{Carr:2016drx}%
  \BibitemOpen
  \bibfield  {author} {\bibinfo {author} {\bibfnamefont {B.}~\bibnamefont
  {Carr}}, \bibinfo {author} {\bibfnamefont {F.}~\bibnamefont {Kuhnel}}, \ and\
  \bibinfo {author} {\bibfnamefont {M.}~\bibnamefont {Sandstad}},\ }\href
  {\doibase 10.1103/PhysRevD.94.083504} {\bibfield  {journal} {\bibinfo
  {journal} {Phys. Rev.}\ }\textbf {\bibinfo {volume} {D94}},\ \bibinfo {pages}
  {083504} (\bibinfo {year} {2016})},\ \Eprint
  {http://arxiv.org/abs/1607.06077} {arXiv:1607.06077 [astro-ph.CO]}
  \BibitemShut {NoStop}%
\bibitem [{\citenamefont {MacGibbon}(1987)}]{MacGibbon:1987my}%
  \BibitemOpen
  \bibfield  {author} {\bibinfo {author} {\bibfnamefont {J.~H.}\ \bibnamefont
  {MacGibbon}},\ }\href {\doibase 10.1038/329308a0} {\bibfield  {journal}
  {\bibinfo  {journal} {Nature}\ }\textbf {\bibinfo {volume} {329}},\ \bibinfo
  {pages} {308} (\bibinfo {year} {1987})}\BibitemShut {NoStop}%
\bibitem [{\citenamefont {Barrow}\ \emph {et~al.}(1992)\citenamefont {Barrow},
  \citenamefont {Copeland},\ and\ \citenamefont {Liddle}}]{Barrow:1992hq}%
  \BibitemOpen
  \bibfield  {author} {\bibinfo {author} {\bibfnamefont {J.~D.}\ \bibnamefont
  {Barrow}}, \bibinfo {author} {\bibfnamefont {E.~J.}\ \bibnamefont
  {Copeland}}, \ and\ \bibinfo {author} {\bibfnamefont {A.~R.}\ \bibnamefont
  {Liddle}},\ }\href {\doibase 10.1103/PhysRevD.46.645} {\bibfield  {journal}
  {\bibinfo  {journal} {Phys. Rev.}\ }\textbf {\bibinfo {volume} {D46}},\
  \bibinfo {pages} {645} (\bibinfo {year} {1992})}\BibitemShut {NoStop}%
\bibitem [{\citenamefont {Carr}\ \emph {et~al.}(1994)\citenamefont {Carr},
  \citenamefont {Gilbert},\ and\ \citenamefont {Lidsey}}]{Carr:1994ar}%
  \BibitemOpen
  \bibfield  {author} {\bibinfo {author} {\bibfnamefont {B.~J.}\ \bibnamefont
  {Carr}}, \bibinfo {author} {\bibfnamefont {J.~H.}\ \bibnamefont {Gilbert}}, \
  and\ \bibinfo {author} {\bibfnamefont {J.~E.}\ \bibnamefont {Lidsey}},\
  }\href {\doibase 10.1103/PhysRevD.50.4853} {\bibfield  {journal} {\bibinfo
  {journal} {Phys. Rev.}\ }\textbf {\bibinfo {volume} {D50}},\ \bibinfo {pages}
  {4853} (\bibinfo {year} {1994})},\ \Eprint
  {http://arxiv.org/abs/astro-ph/9405027} {arXiv:astro-ph/9405027 [astro-ph]}
  \BibitemShut {NoStop}%
\bibitem [{\citenamefont {Dalianis}\ and\ \citenamefont
  {Tringas}(2019)}]{Dalianis:2019asr}%
  \BibitemOpen
  \bibfield  {author} {\bibinfo {author} {\bibfnamefont {I.}~\bibnamefont
  {Dalianis}}\ and\ \bibinfo {author} {\bibfnamefont {G.}~\bibnamefont
  {Tringas}},\ }\href@noop {} {\  (\bibinfo {year} {2019})},\ \Eprint
  {http://arxiv.org/abs/1905.01741} {arXiv:1905.01741 [astro-ph.CO]}
  \BibitemShut {NoStop}%
\bibitem [{\citenamefont {Chen}\ \emph {et~al.}(2015)\citenamefont {Chen},
  \citenamefont {Ong},\ and\ \citenamefont {Yeom}}]{Chen:2014jwq}%
  \BibitemOpen
  \bibfield  {author} {\bibinfo {author} {\bibfnamefont {P.}~\bibnamefont
  {Chen}}, \bibinfo {author} {\bibfnamefont {Y.~C.}\ \bibnamefont {Ong}}, \
  and\ \bibinfo {author} {\bibfnamefont {D.-h.}\ \bibnamefont {Yeom}},\ }\href
  {\doibase 10.1016/j.physrep.2015.10.007} {\bibfield  {journal} {\bibinfo
  {journal} {Phys. Rept.}\ }\textbf {\bibinfo {volume} {603}},\ \bibinfo
  {pages} {1} (\bibinfo {year} {2015})},\ \Eprint
  {http://arxiv.org/abs/1412.8366} {arXiv:1412.8366 [gr-qc]} \BibitemShut
  {NoStop}%
\bibitem [{\citenamefont {Stelle}(1977)}]{Stelle:1976gc}%
  \BibitemOpen
  \bibfield  {author} {\bibinfo {author} {\bibfnamefont {K.~S.}\ \bibnamefont
  {Stelle}},\ }\href {\doibase 10.1103/PhysRevD.16.953} {\bibfield  {journal}
  {\bibinfo  {journal} {Phys. Rev.}\ }\textbf {\bibinfo {volume} {D16}},\
  \bibinfo {pages} {953} (\bibinfo {year} {1977})}\BibitemShut {NoStop}%
\bibitem [{\citenamefont {Voronov}\ and\ \citenamefont
  {Tyutin}(1984)}]{Voronov:1984kq}%
  \BibitemOpen
  \bibfield  {author} {\bibinfo {author} {\bibfnamefont {B.~L.}\ \bibnamefont
  {Voronov}}\ and\ \bibinfo {author} {\bibfnamefont {I.~V.}\ \bibnamefont
  {Tyutin}},\ }\href@noop {} {\bibfield  {journal} {\bibinfo  {journal} {Yad.
  Fiz.}\ }\textbf {\bibinfo {volume} {39}},\ \bibinfo {pages} {998} (\bibinfo
  {year} {1984})}\BibitemShut {NoStop}%
\bibitem [{\citenamefont {Fradkin}\ and\ \citenamefont
  {Tseytlin}(1982)}]{Fradkin:1981iu}%
  \BibitemOpen
  \bibfield  {author} {\bibinfo {author} {\bibfnamefont {E.~S.}\ \bibnamefont
  {Fradkin}}\ and\ \bibinfo {author} {\bibfnamefont {A.~A.}\ \bibnamefont
  {Tseytlin}},\ }\href {\doibase 10.1016/0550-3213(82)90444-8} {\bibfield
  {journal} {\bibinfo  {journal} {Nucl. Phys.}\ }\textbf {\bibinfo {volume}
  {B201}},\ \bibinfo {pages} {469} (\bibinfo {year} {1982})}\BibitemShut
  {NoStop}%
\bibitem [{\citenamefont {Avramidi}\ and\ \citenamefont
  {Barvinsky}(1985)}]{Avramidi:1985ki}%
  \BibitemOpen
  \bibfield  {author} {\bibinfo {author} {\bibfnamefont {I.~G.}\ \bibnamefont
  {Avramidi}}\ and\ \bibinfo {author} {\bibfnamefont {A.~O.}\ \bibnamefont
  {Barvinsky}},\ }\href {\doibase 10.1016/0370-2693(85)90248-5} {\bibfield
  {journal} {\bibinfo  {journal} {Phys. Lett.}\ }\textbf {\bibinfo {volume}
  {159B}},\ \bibinfo {pages} {269} (\bibinfo {year} {1985})}\BibitemShut
  {NoStop}%
\bibitem [{\citenamefont {Lee}\ and\ \citenamefont {Wick}(1969)}]{Lee:1969fy}%
  \BibitemOpen
  \bibfield  {author} {\bibinfo {author} {\bibfnamefont {T.~D.}\ \bibnamefont
  {Lee}}\ and\ \bibinfo {author} {\bibfnamefont {G.~C.}\ \bibnamefont {Wick}},\
  }\href {\doibase 10.1016/0550-3213(69)90098-4} {\bibfield  {journal}
  {\bibinfo  {journal} {Nucl. Phys.}\ }\textbf {\bibinfo {volume} {B9}},\
  \bibinfo {pages} {209} (\bibinfo {year} {1969})},\ \bibinfo {note}
  {[,83(1969)]}\BibitemShut {NoStop}%
\bibitem [{\citenamefont {Tomboulis}(1977)}]{Tomboulis:1977jk}%
  \BibitemOpen
  \bibfield  {author} {\bibinfo {author} {\bibfnamefont {E.}~\bibnamefont
  {Tomboulis}},\ }\href {\doibase 10.1016/0370-2693(77)90678-5} {\bibfield
  {journal} {\bibinfo  {journal} {Phys. Lett.}\ }\textbf {\bibinfo {volume}
  {70B}},\ \bibinfo {pages} {361} (\bibinfo {year} {1977})}\BibitemShut
  {NoStop}%
\bibitem [{\citenamefont {Grinstein}\ \emph {et~al.}(2009)\citenamefont
  {Grinstein}, \citenamefont {O'Connell},\ and\ \citenamefont
  {Wise}}]{Grinstein:2008bg}%
  \BibitemOpen
  \bibfield  {author} {\bibinfo {author} {\bibfnamefont {B.}~\bibnamefont
  {Grinstein}}, \bibinfo {author} {\bibfnamefont {D.}~\bibnamefont
  {O'Connell}}, \ and\ \bibinfo {author} {\bibfnamefont {M.~B.}\ \bibnamefont
  {Wise}},\ }\href {\doibase 10.1103/PhysRevD.79.105019} {\bibfield  {journal}
  {\bibinfo  {journal} {Phys. Rev.}\ }\textbf {\bibinfo {volume} {D79}},\
  \bibinfo {pages} {105019} (\bibinfo {year} {2009})},\ \Eprint
  {http://arxiv.org/abs/0805.2156} {arXiv:0805.2156 [hep-th]} \BibitemShut
  {NoStop}%
\bibitem [{\citenamefont {Anselmi}\ and\ \citenamefont
  {Piva}(2017)}]{Anselmi:2017yux}%
  \BibitemOpen
  \bibfield  {author} {\bibinfo {author} {\bibfnamefont {D.}~\bibnamefont
  {Anselmi}}\ and\ \bibinfo {author} {\bibfnamefont {M.}~\bibnamefont {Piva}},\
  }\href {\doibase 10.1007/JHEP06(2017)066} {\bibfield  {journal} {\bibinfo
  {journal} {JHEP}\ }\textbf {\bibinfo {volume} {06}},\ \bibinfo {pages} {066}
  (\bibinfo {year} {2017})},\ \Eprint {http://arxiv.org/abs/1703.04584}
  {arXiv:1703.04584 [hep-th]} \BibitemShut {NoStop}%
\bibitem [{\citenamefont {Donoghue}\ and\ \citenamefont
  {Menezes}(2019)}]{Donoghue:2018lmc}%
  \BibitemOpen
  \bibfield  {author} {\bibinfo {author} {\bibfnamefont {J.~F.}\ \bibnamefont
  {Donoghue}}\ and\ \bibinfo {author} {\bibfnamefont {G.}~\bibnamefont
  {Menezes}},\ }\href {\doibase 10.1103/PhysRevD.99.065017} {\bibfield
  {journal} {\bibinfo  {journal} {Phys. Rev.}\ }\textbf {\bibinfo {volume}
  {D99}},\ \bibinfo {pages} {065017} (\bibinfo {year} {2019})},\ \Eprint
  {http://arxiv.org/abs/1812.03603} {arXiv:1812.03603 [hep-th]} \BibitemShut
  {NoStop}%
\bibitem [{\citenamefont {Bender}\ and\ \citenamefont
  {Mannheim}(2008)}]{Bender:2007wu}%
  \BibitemOpen
  \bibfield  {author} {\bibinfo {author} {\bibfnamefont {C.~M.}\ \bibnamefont
  {Bender}}\ and\ \bibinfo {author} {\bibfnamefont {P.~D.}\ \bibnamefont
  {Mannheim}},\ }\href {\doibase 10.1103/PhysRevLett.100.110402} {\bibfield
  {journal} {\bibinfo  {journal} {Phys. Rev. Lett.}\ }\textbf {\bibinfo
  {volume} {100}},\ \bibinfo {pages} {110402} (\bibinfo {year} {2008})},\
  \Eprint {http://arxiv.org/abs/0706.0207} {arXiv:0706.0207 [hep-th]}
  \BibitemShut {NoStop}%
\bibitem [{\citenamefont {Salvio}\ and\ \citenamefont
  {Strumia}(2016)}]{Salvio:2015gsi}%
  \BibitemOpen
  \bibfield  {author} {\bibinfo {author} {\bibfnamefont {A.}~\bibnamefont
  {Salvio}}\ and\ \bibinfo {author} {\bibfnamefont {A.}~\bibnamefont
  {Strumia}},\ }\href {\doibase 10.1140/epjc/s10052-016-4079-8} {\bibfield
  {journal} {\bibinfo  {journal} {Eur. Phys. J.}\ }\textbf {\bibinfo {volume}
  {C76}},\ \bibinfo {pages} {227} (\bibinfo {year} {2016})},\ \Eprint
  {http://arxiv.org/abs/1512.01237} {arXiv:1512.01237 [hep-th]} \BibitemShut
  {NoStop}%
\bibitem [{\citenamefont {Holdom}\ and\ \citenamefont
  {Ren}(2016{\natexlab{a}})}]{Holdom:2015kbf}%
  \BibitemOpen
  \bibfield  {author} {\bibinfo {author} {\bibfnamefont {B.}~\bibnamefont
  {Holdom}}\ and\ \bibinfo {author} {\bibfnamefont {J.}~\bibnamefont {Ren}},\
  }\href {\doibase 10.1103/PhysRevD.93.124030} {\bibfield  {journal} {\bibinfo
  {journal} {Phys. Rev.}\ }\textbf {\bibinfo {volume} {D93}},\ \bibinfo {pages}
  {124030} (\bibinfo {year} {2016}{\natexlab{a}})},\ \Eprint
  {http://arxiv.org/abs/1512.05305} {arXiv:1512.05305 [hep-th]} \BibitemShut
  {NoStop}%
\bibitem [{\citenamefont {Holdom}\ and\ \citenamefont
  {Ren}(2016{\natexlab{b}})}]{Holdom:2016xfn}%
  \BibitemOpen
  \bibfield  {author} {\bibinfo {author} {\bibfnamefont {B.}~\bibnamefont
  {Holdom}}\ and\ \bibinfo {author} {\bibfnamefont {J.}~\bibnamefont {Ren}},\
  }\href {\doibase 10.1142/S0218271816430045} {\bibfield  {journal} {\bibinfo
  {journal} {Int. J. Mod. Phys.}\ }\textbf {\bibinfo {volume} {D25}},\ \bibinfo
  {pages} {1643004} (\bibinfo {year} {2016}{\natexlab{b}})},\ \Eprint
  {http://arxiv.org/abs/1605.05006} {arXiv:1605.05006 [hep-th]} \BibitemShut
  {NoStop}%
\bibitem [{\citenamefont {Salvio}(2018)}]{Salvio:2018crh}%
  \BibitemOpen
  \bibfield  {author} {\bibinfo {author} {\bibfnamefont {A.}~\bibnamefont
  {Salvio}},\ }\href {\doibase 10.3389/fphy.2018.00077} {\bibfield  {journal}
  {\bibinfo  {journal} {Front.in Phys.}\ }\textbf {\bibinfo {volume} {6}},\
  \bibinfo {pages} {77} (\bibinfo {year} {2018})},\ \Eprint
  {http://arxiv.org/abs/1804.09944} {arXiv:1804.09944 [hep-th]} \BibitemShut
  {NoStop}%
\bibitem [{\citenamefont {Holdom}(2002)}]{Holdom:2002xy}%
  \BibitemOpen
  \bibfield  {author} {\bibinfo {author} {\bibfnamefont {B.}~\bibnamefont
  {Holdom}},\ }\href {\doibase 10.1103/PhysRevD.66.084010} {\bibfield
  {journal} {\bibinfo  {journal} {Phys. Rev.}\ }\textbf {\bibinfo {volume}
  {D66}},\ \bibinfo {pages} {084010} (\bibinfo {year} {2002})},\ \Eprint
  {http://arxiv.org/abs/hep-th/0206219} {arXiv:hep-th/0206219 [hep-th]}
  \BibitemShut {NoStop}%
\bibitem [{\citenamefont {Holdom}\ and\ \citenamefont
  {Ren}(2017)}]{Holdom:2016nek}%
  \BibitemOpen
  \bibfield  {author} {\bibinfo {author} {\bibfnamefont {B.}~\bibnamefont
  {Holdom}}\ and\ \bibinfo {author} {\bibfnamefont {J.}~\bibnamefont {Ren}},\
  }\href {\doibase 10.1103/PhysRevD.95.084034} {\bibfield  {journal} {\bibinfo
  {journal} {Phys. Rev.}\ }\textbf {\bibinfo {volume} {D95}},\ \bibinfo {pages}
  {084034} (\bibinfo {year} {2017})},\ \Eprint
  {http://arxiv.org/abs/1612.04889} {arXiv:1612.04889 [gr-qc]} \BibitemShut
  {NoStop}%
\bibitem [{\citenamefont {Holdom}(2019)}]{Holdom:2019ouz}%
  \BibitemOpen
  \bibfield  {author} {\bibinfo {author} {\bibfnamefont {B.}~\bibnamefont
  {Holdom}},\ }in\ \href@noop {} {\emph {\bibinfo {booktitle} {{Scale
  invariance in particle physics and cosmology Geneva (CERN), Switzerland,
  January 28-February 1, 2019}}}}\ (\bibinfo {year} {2019})\ \Eprint
  {http://arxiv.org/abs/1905.08849} {arXiv:1905.08849 [gr-qc]} \BibitemShut
  {NoStop}%
\bibitem [{\citenamefont {Ren}(2019)}]{Ren:2019afg}%
  \BibitemOpen
  \bibfield  {author} {\bibinfo {author} {\bibfnamefont {J.}~\bibnamefont
  {Ren}},\ }\href {\doibase 10.1103/PhysRevD.100.124012} {\bibfield  {journal}
  {\bibinfo  {journal} {Phys. Rev.}\ }\textbf {\bibinfo {volume} {D100}},\
  \bibinfo {pages} {124012} (\bibinfo {year} {2019})},\ \Eprint
  {http://arxiv.org/abs/1905.09973} {arXiv:1905.09973 [gr-qc]} \BibitemShut
  {NoStop}%
\bibitem [{\citenamefont {Bai}\ and\ \citenamefont
  {Orlofsky}(2019)}]{Bai:2019zcd}%
  \BibitemOpen
  \bibfield  {author} {\bibinfo {author} {\bibfnamefont {Y.}~\bibnamefont
  {Bai}}\ and\ \bibinfo {author} {\bibfnamefont {N.}~\bibnamefont {Orlofsky}},\
  }\href@noop {} {\  (\bibinfo {year} {2019})},\ \Eprint
  {http://arxiv.org/abs/1906.04858} {arXiv:1906.04858 [hep-ph]} \BibitemShut
  {NoStop}%
\bibitem [{\citenamefont {Carroll}\ \emph {et~al.}(2009)\citenamefont
  {Carroll}, \citenamefont {Johnson},\ and\ \citenamefont
  {Randall}}]{Carroll:2009maa}%
  \BibitemOpen
  \bibfield  {author} {\bibinfo {author} {\bibfnamefont {S.~M.}\ \bibnamefont
  {Carroll}}, \bibinfo {author} {\bibfnamefont {M.~C.}\ \bibnamefont
  {Johnson}}, \ and\ \bibinfo {author} {\bibfnamefont {L.}~\bibnamefont
  {Randall}},\ }\href {\doibase 10.1088/1126-6708/2009/11/109} {\bibfield
  {journal} {\bibinfo  {journal} {JHEP}\ }\textbf {\bibinfo {volume} {11}},\
  \bibinfo {pages} {109} (\bibinfo {year} {2009})},\ \Eprint
  {http://arxiv.org/abs/0901.0931} {arXiv:0901.0931 [hep-th]} \BibitemShut
  {NoStop}%
\bibitem [{\citenamefont {Conklin}\ \emph {et~al.}(2018)\citenamefont
  {Conklin}, \citenamefont {Holdom},\ and\ \citenamefont
  {Ren}}]{Conklin:2017lwb}%
  \BibitemOpen
  \bibfield  {author} {\bibinfo {author} {\bibfnamefont {R.~S.}\ \bibnamefont
  {Conklin}}, \bibinfo {author} {\bibfnamefont {B.}~\bibnamefont {Holdom}}, \
  and\ \bibinfo {author} {\bibfnamefont {J.}~\bibnamefont {Ren}},\ }\href
  {\doibase 10.1103/PhysRevD.98.044021} {\bibfield  {journal} {\bibinfo
  {journal} {Phys. Rev.}\ }\textbf {\bibinfo {volume} {D98}},\ \bibinfo {pages}
  {044021} (\bibinfo {year} {2018})},\ \Eprint
  {http://arxiv.org/abs/1712.06517} {arXiv:1712.06517 [gr-qc]} \BibitemShut
  {NoStop}%
\bibitem [{\citenamefont {MacGibbon}(1991)}]{MacGibbon:1991tj}%
  \BibitemOpen
  \bibfield  {author} {\bibinfo {author} {\bibfnamefont {J.~H.}\ \bibnamefont
  {MacGibbon}},\ }\href {\doibase 10.1103/PhysRevD.44.376} {\bibfield
  {journal} {\bibinfo  {journal} {Phys. Rev.}\ }\textbf {\bibinfo {volume}
  {D44}},\ \bibinfo {pages} {376} (\bibinfo {year} {1991})}\BibitemShut
  {NoStop}%
\bibitem [{\citenamefont {Barrau}\ \emph {et~al.}(2019)\citenamefont {Barrau},
  \citenamefont {Martineau}, \citenamefont {Moulin},\ and\ \citenamefont
  {Ngono}}]{Barrau:2019cuo}%
  \BibitemOpen
  \bibfield  {author} {\bibinfo {author} {\bibfnamefont {A.}~\bibnamefont
  {Barrau}}, \bibinfo {author} {\bibfnamefont {K.}~\bibnamefont {Martineau}},
  \bibinfo {author} {\bibfnamefont {F.}~\bibnamefont {Moulin}}, \ and\ \bibinfo
  {author} {\bibfnamefont {J.-F.}\ \bibnamefont {Ngono}},\ }\href {\doibase
  10.1103/PhysRevD.100.123505} {\bibfield  {journal} {\bibinfo  {journal}
  {Phys. Rev.}\ }\textbf {\bibinfo {volume} {D100}},\ \bibinfo {pages} {123505}
  (\bibinfo {year} {2019})},\ \Eprint {http://arxiv.org/abs/1906.09930}
  {arXiv:1906.09930 [hep-ph]} \BibitemShut {NoStop}%
\bibitem [{\citenamefont {Barbot}\ and\ \citenamefont
  {Drees}(2003)}]{Barbot:2002gt}%
  \BibitemOpen
  \bibfield  {author} {\bibinfo {author} {\bibfnamefont {C.}~\bibnamefont
  {Barbot}}\ and\ \bibinfo {author} {\bibfnamefont {M.}~\bibnamefont {Drees}},\
  }\href {\doibase 10.1016/S0927-6505(03)00134-8} {\bibfield  {journal}
  {\bibinfo  {journal} {Astropart. Phys.}\ }\textbf {\bibinfo {volume} {20}},\
  \bibinfo {pages} {5} (\bibinfo {year} {2003})},\ \Eprint
  {http://arxiv.org/abs/hep-ph/0211406} {arXiv:hep-ph/0211406 [hep-ph]}
  \BibitemShut {NoStop}%
\bibitem [{\citenamefont {Aloisio}\ \emph {et~al.}(2004)\citenamefont
  {Aloisio}, \citenamefont {Berezinsky},\ and\ \citenamefont
  {Kachelriess}}]{Aloisio:2003xj}%
  \BibitemOpen
  \bibfield  {author} {\bibinfo {author} {\bibfnamefont {R.}~\bibnamefont
  {Aloisio}}, \bibinfo {author} {\bibfnamefont {V.}~\bibnamefont {Berezinsky}},
  \ and\ \bibinfo {author} {\bibfnamefont {M.}~\bibnamefont {Kachelriess}},\
  }\href {\doibase 10.1103/PhysRevD.69.094023} {\bibfield  {journal} {\bibinfo
  {journal} {Phys. Rev.}\ }\textbf {\bibinfo {volume} {D69}},\ \bibinfo {pages}
  {094023} (\bibinfo {year} {2004})},\ \Eprint
  {http://arxiv.org/abs/hep-ph/0307279} {arXiv:hep-ph/0307279 [hep-ph]}
  \BibitemShut {NoStop}%
\bibitem [{\citenamefont {Evans}\ \emph {et~al.}(2016)\citenamefont {Evans},
  \citenamefont {Sanders},\ and\ \citenamefont
  {Geringer-Sameth}}]{Evans:2016xwx}%
  \BibitemOpen
  \bibfield  {author} {\bibinfo {author} {\bibfnamefont {N.~W.}\ \bibnamefont
  {Evans}}, \bibinfo {author} {\bibfnamefont {J.~L.}\ \bibnamefont {Sanders}},
  \ and\ \bibinfo {author} {\bibfnamefont {A.}~\bibnamefont
  {Geringer-Sameth}},\ }\href {\doibase 10.1103/PhysRevD.93.103512} {\bibfield
  {journal} {\bibinfo  {journal} {Phys. Rev.}\ }\textbf {\bibinfo {volume}
  {D93}},\ \bibinfo {pages} {103512} (\bibinfo {year} {2016})},\ \Eprint
  {http://arxiv.org/abs/1604.05599} {arXiv:1604.05599 [astro-ph.GA]}
  \BibitemShut {NoStop}%
\bibitem [{\citenamefont {Pato}\ \emph {et~al.}(2015)\citenamefont {Pato},
  \citenamefont {Iocco},\ and\ \citenamefont {Bertone}}]{Pato:2015dua}%
  \BibitemOpen
  \bibfield  {author} {\bibinfo {author} {\bibfnamefont {M.}~\bibnamefont
  {Pato}}, \bibinfo {author} {\bibfnamefont {F.}~\bibnamefont {Iocco}}, \ and\
  \bibinfo {author} {\bibfnamefont {G.}~\bibnamefont {Bertone}},\ }\href
  {\doibase 10.1088/1475-7516/2015/12/001} {\bibfield  {journal} {\bibinfo
  {journal} {JCAP}\ }\textbf {\bibinfo {volume} {1512}},\ \bibinfo {pages}
  {001} (\bibinfo {year} {2015})},\ \Eprint {http://arxiv.org/abs/1504.06324}
  {arXiv:1504.06324 [astro-ph.GA]} \BibitemShut {NoStop}%
\bibitem [{\citenamefont {Navarro}\ \emph {et~al.}(1997)\citenamefont
  {Navarro}, \citenamefont {Frenk},\ and\ \citenamefont
  {White}}]{Navarro:1996gj}%
  \BibitemOpen
  \bibfield  {author} {\bibinfo {author} {\bibfnamefont {J.~F.}\ \bibnamefont
  {Navarro}}, \bibinfo {author} {\bibfnamefont {C.~S.}\ \bibnamefont {Frenk}},
  \ and\ \bibinfo {author} {\bibfnamefont {S.~D.~M.}\ \bibnamefont {White}},\
  }\href {\doibase 10.1086/304888} {\bibfield  {journal} {\bibinfo  {journal}
  {Astrophys. J.}\ }\textbf {\bibinfo {volume} {490}},\ \bibinfo {pages} {493}
  (\bibinfo {year} {1997})},\ \Eprint {http://arxiv.org/abs/astro-ph/9611107}
  {arXiv:astro-ph/9611107 [astro-ph]} \BibitemShut {NoStop}%
\bibitem [{\citenamefont {Burkert}(1996)}]{Burkert:1995yz}%
  \BibitemOpen
  \bibfield  {author} {\bibinfo {author} {\bibfnamefont {A.}~\bibnamefont
  {Burkert}},\ }\bibfield  {booktitle} {\emph {\bibinfo {booktitle} {{IAU
  Symposium 171: New Light on Galaxy Evolution Heidelberg, Germany, June 26-30,
  1995}}},\ }\href {\doibase 10.1086/309560} {\bibfield  {journal} {\bibinfo
  {journal} {IAU Symp.}\ }\textbf {\bibinfo {volume} {171}},\ \bibinfo {pages}
  {175} (\bibinfo {year} {1996})},\ \bibinfo {note} {[Astrophys.
  J.447,L25(1995)]},\ \Eprint {http://arxiv.org/abs/astro-ph/9504041}
  {arXiv:astro-ph/9504041 [astro-ph]} \BibitemShut {NoStop}%
\bibitem [{\citenamefont {Zaritsky}\ \emph {et~al.}(1989)\citenamefont
  {Zaritsky}, \citenamefont {Olszewski}, \citenamefont {Schommer},
  \citenamefont {Peterson},\ and\ \citenamefont
  {Aaronson}}]{1989ApJ...345..759Z}%
  \BibitemOpen
  \bibfield  {author} {\bibinfo {author} {\bibfnamefont {D.}~\bibnamefont
  {Zaritsky}}, \bibinfo {author} {\bibfnamefont {E.~W.}\ \bibnamefont
  {Olszewski}}, \bibinfo {author} {\bibfnamefont {R.~A.}\ \bibnamefont
  {Schommer}}, \bibinfo {author} {\bibfnamefont {R.~C.}\ \bibnamefont
  {Peterson}}, \ and\ \bibinfo {author} {\bibfnamefont {M.}~\bibnamefont
  {Aaronson}},\ }\href {\doibase 10.1086/167947} {\bibfield  {journal}
  {\bibinfo  {journal} {Astrophys. J.}\ }\textbf {\bibinfo {volume} {345}},\
  \bibinfo {pages} {759} (\bibinfo {year} {1989})}\BibitemShut {NoStop}%
\bibitem [{\citenamefont {Sasaki}\ \emph {et~al.}(2018)\citenamefont {Sasaki},
  \citenamefont {Suyama}, \citenamefont {Tanaka},\ and\ \citenamefont
  {Yokoyama}}]{Sasaki:2018dmp}%
  \BibitemOpen
  \bibfield  {author} {\bibinfo {author} {\bibfnamefont {M.}~\bibnamefont
  {Sasaki}}, \bibinfo {author} {\bibfnamefont {T.}~\bibnamefont {Suyama}},
  \bibinfo {author} {\bibfnamefont {T.}~\bibnamefont {Tanaka}}, \ and\ \bibinfo
  {author} {\bibfnamefont {S.}~\bibnamefont {Yokoyama}},\ }\href {\doibase
  10.1088/1361-6382/aaa7b4} {\bibfield  {journal} {\bibinfo  {journal} {Class.
  Quant. Grav.}\ }\textbf {\bibinfo {volume} {35}},\ \bibinfo {pages} {063001}
  (\bibinfo {year} {2018})},\ \Eprint {http://arxiv.org/abs/1801.05235}
  {arXiv:1801.05235 [astro-ph.CO]} \BibitemShut {NoStop}%
\bibitem [{\citenamefont {Raidal}\ \emph {et~al.}(2019)\citenamefont {Raidal},
  \citenamefont {Spethmann}, \citenamefont {Vaskonen},\ and\ \citenamefont
  {Veermäe}}]{Raidal:2018bbj}%
  \BibitemOpen
  \bibfield  {author} {\bibinfo {author} {\bibfnamefont {M.}~\bibnamefont
  {Raidal}}, \bibinfo {author} {\bibfnamefont {C.}~\bibnamefont {Spethmann}},
  \bibinfo {author} {\bibfnamefont {V.}~\bibnamefont {Vaskonen}}, \ and\
  \bibinfo {author} {\bibfnamefont {H.}~\bibnamefont {Veermäe}},\ }\href
  {\doibase 10.1088/1475-7516/2019/02/018} {\bibfield  {journal} {\bibinfo
  {journal} {JCAP}\ }\textbf {\bibinfo {volume} {1902}},\ \bibinfo {pages}
  {018} (\bibinfo {year} {2019})},\ \Eprint {http://arxiv.org/abs/1812.01930}
  {arXiv:1812.01930 [astro-ph.CO]} \BibitemShut {NoStop}%
\bibitem [{\citenamefont {Vaskonen}\ and\ \citenamefont
  {Veermäe}(2019)}]{Vaskonen:2019jpv}%
  \BibitemOpen
  \bibfield  {author} {\bibinfo {author} {\bibfnamefont {V.}~\bibnamefont
  {Vaskonen}}\ and\ \bibinfo {author} {\bibfnamefont {H.}~\bibnamefont
  {Veermäe}},\ }\href@noop {} {\  (\bibinfo {year} {2019})},\ \Eprint
  {http://arxiv.org/abs/1908.09752} {arXiv:1908.09752 [astro-ph.CO]}
  \BibitemShut {NoStop}%
\bibitem [{\citenamefont {Buitink}\ \emph {et~al.}(2010)\citenamefont
  {Buitink}, \citenamefont {Scholten}, \citenamefont {Bacelar}, \citenamefont
  {Braun}, \citenamefont {de~Bruyn}, \citenamefont {Falcke}, \citenamefont
  {Singh}, \citenamefont {Stappers}, \citenamefont {Strom},\ and\ \citenamefont
  {Yahyaoui}}]{Buitink:2010qn}%
  \BibitemOpen
  \bibfield  {author} {\bibinfo {author} {\bibfnamefont {S.}~\bibnamefont
  {Buitink}}, \bibinfo {author} {\bibfnamefont {O.}~\bibnamefont {Scholten}},
  \bibinfo {author} {\bibfnamefont {J.}~\bibnamefont {Bacelar}}, \bibinfo
  {author} {\bibfnamefont {R.}~\bibnamefont {Braun}}, \bibinfo {author}
  {\bibfnamefont {A.~G.}\ \bibnamefont {de~Bruyn}}, \bibinfo {author}
  {\bibfnamefont {H.}~\bibnamefont {Falcke}}, \bibinfo {author} {\bibfnamefont
  {K.}~\bibnamefont {Singh}}, \bibinfo {author} {\bibfnamefont
  {B.}~\bibnamefont {Stappers}}, \bibinfo {author} {\bibfnamefont {R.~G.}\
  \bibnamefont {Strom}}, \ and\ \bibinfo {author} {\bibfnamefont {R.~a.}\
  \bibnamefont {Yahyaoui}},\ }\href {\doibase 10.1051/0004-6361/201014104}
  {\bibfield  {journal} {\bibinfo  {journal} {Astron. Astrophys.}\ }\textbf
  {\bibinfo {volume} {521}},\ \bibinfo {pages} {A47} (\bibinfo {year}
  {2010})},\ \Eprint {http://arxiv.org/abs/1004.0274} {arXiv:1004.0274
  [astro-ph.HE]} \BibitemShut {NoStop}%
\bibitem [{\citenamefont {Gorham}\ \emph {et~al.}(2019)\citenamefont {Gorham}
  \emph {et~al.}}]{Gorham:2019guw}%
  \BibitemOpen
  \bibfield  {author} {\bibinfo {author} {\bibfnamefont {P.~W.}\ \bibnamefont
  {Gorham}} \emph {et~al.} (\bibinfo {collaboration} {ANITA}),\ }\href
  {\doibase 10.1103/PhysRevD.99.122001} {\bibfield  {journal} {\bibinfo
  {journal} {Phys. Rev.}\ }\textbf {\bibinfo {volume} {D99}},\ \bibinfo {pages}
  {122001} (\bibinfo {year} {2019})},\ \Eprint
  {http://arxiv.org/abs/1902.04005} {arXiv:1902.04005 [astro-ph.HE]}
  \BibitemShut {NoStop}%
\bibitem [{\citenamefont {Gorham}\ \emph {et~al.}(2010)\citenamefont {Gorham}
  \emph {et~al.}}]{Gorham:2010kv}%
  \BibitemOpen
  \bibfield  {author} {\bibinfo {author} {\bibfnamefont {P.~W.}\ \bibnamefont
  {Gorham}} \emph {et~al.} (\bibinfo {collaboration} {ANITA}),\ }\href
  {\doibase 10.1103/PhysRevD.82.022004, 10.1103/PhysRevD.85.049901} {\bibfield
  {journal} {\bibinfo  {journal} {Phys. Rev.}\ }\textbf {\bibinfo {volume}
  {D82}},\ \bibinfo {pages} {022004} (\bibinfo {year} {2010})},\ \bibinfo
  {note} {[Erratum: Phys. Rev.D85,049901(2012)]},\ \Eprint
  {http://arxiv.org/abs/1003.2961} {arXiv:1003.2961 [astro-ph.HE]} \BibitemShut
  {NoStop}%
\bibitem [{\citenamefont {Aartsen}\ \emph {et~al.}(2018)\citenamefont {Aartsen}
  \emph {et~al.}}]{Aartsen:2018vtx}%
  \BibitemOpen
  \bibfield  {author} {\bibinfo {author} {\bibfnamefont {M.~G.}\ \bibnamefont
  {Aartsen}} \emph {et~al.} (\bibinfo {collaboration} {IceCube}),\ }\href
  {\doibase 10.1103/PhysRevD.98.062003} {\bibfield  {journal} {\bibinfo
  {journal} {Phys. Rev.}\ }\textbf {\bibinfo {volume} {D98}},\ \bibinfo {pages}
  {062003} (\bibinfo {year} {2018})},\ \Eprint
  {http://arxiv.org/abs/1807.01820} {arXiv:1807.01820 [astro-ph.HE]}
  \BibitemShut {NoStop}%
\bibitem [{\citenamefont {Aartsen}\ \emph
  {et~al.}(2015{\natexlab{a}})\citenamefont {Aartsen} \emph
  {et~al.}}]{Aartsen:2015xup}%
  \BibitemOpen
  \bibfield  {author} {\bibinfo {author} {\bibfnamefont {M.~G.}\ \bibnamefont
  {Aartsen}} \emph {et~al.} (\bibinfo {collaboration} {IceCube}),\ }\href
  {\doibase 10.1103/PhysRevD.91.122004} {\bibfield  {journal} {\bibinfo
  {journal} {Phys. Rev.}\ }\textbf {\bibinfo {volume} {D91}},\ \bibinfo {pages}
  {122004} (\bibinfo {year} {2015}{\natexlab{a}})},\ \Eprint
  {http://arxiv.org/abs/1504.03753} {arXiv:1504.03753 [astro-ph.HE]}
  \BibitemShut {NoStop}%
\bibitem [{\citenamefont {Aartsen}\ \emph
  {et~al.}(2015{\natexlab{b}})\citenamefont {Aartsen} \emph
  {et~al.}}]{Aartsen:2015knd}%
  \BibitemOpen
  \bibfield  {author} {\bibinfo {author} {\bibfnamefont {M.~G.}\ \bibnamefont
  {Aartsen}} \emph {et~al.} (\bibinfo {collaboration} {IceCube}),\ }\href
  {\doibase 10.1088/0004-637X/809/1/98} {\bibfield  {journal} {\bibinfo
  {journal} {Astrophys. J.}\ }\textbf {\bibinfo {volume} {809}},\ \bibinfo
  {pages} {98} (\bibinfo {year} {2015}{\natexlab{b}})},\ \Eprint
  {http://arxiv.org/abs/1507.03991} {arXiv:1507.03991 [astro-ph.HE]}
  \BibitemShut {NoStop}%
\bibitem [{\citenamefont {Aartsen}\ \emph
  {et~al.}(2015{\natexlab{c}})\citenamefont {Aartsen} \emph
  {et~al.}}]{Aartsen:2014qna}%
  \BibitemOpen
  \bibfield  {author} {\bibinfo {author} {\bibfnamefont {M.~G.}\ \bibnamefont
  {Aartsen}} \emph {et~al.} (\bibinfo {collaboration} {IceCube}),\ }\href
  {\doibase 10.1140/epjc/s10052-015-3330-z} {\bibfield  {journal} {\bibinfo
  {journal} {Eur. Phys. J.}\ }\textbf {\bibinfo {volume} {C75}},\ \bibinfo
  {pages} {116} (\bibinfo {year} {2015}{\natexlab{c}})},\ \Eprint
  {http://arxiv.org/abs/1409.4535} {arXiv:1409.4535 [astro-ph.HE]} \BibitemShut
  {NoStop}%
\bibitem [{\citenamefont {Gondolo}\ \emph {et~al.}(1993)\citenamefont
  {Gondolo}, \citenamefont {Gelmini},\ and\ \citenamefont
  {Sarkar}}]{Gondolo:1991rn}%
  \BibitemOpen
  \bibfield  {author} {\bibinfo {author} {\bibfnamefont {P.}~\bibnamefont
  {Gondolo}}, \bibinfo {author} {\bibfnamefont {G.}~\bibnamefont {Gelmini}}, \
  and\ \bibinfo {author} {\bibfnamefont {S.}~\bibnamefont {Sarkar}},\ }\href
  {\doibase 10.1016/0550-3213(93)90199-Y} {\bibfield  {journal} {\bibinfo
  {journal} {Nucl. Phys.}\ }\textbf {\bibinfo {volume} {B392}},\ \bibinfo
  {pages} {111} (\bibinfo {year} {1993})},\ \Eprint
  {http://arxiv.org/abs/hep-ph/9209236} {arXiv:hep-ph/9209236 [hep-ph]}
  \BibitemShut {NoStop}%
\bibitem [{\citenamefont {Aab}\ \emph {et~al.}(2017)\citenamefont {Aab} \emph
  {et~al.}}]{Aab:2016agp}%
  \BibitemOpen
  \bibfield  {author} {\bibinfo {author} {\bibfnamefont {A.}~\bibnamefont
  {Aab}} \emph {et~al.} (\bibinfo {collaboration} {Pierre Auger}),\ }\href
  {\doibase 10.1088/1475-7516/2017/04/009} {\bibfield  {journal} {\bibinfo
  {journal} {JCAP}\ }\textbf {\bibinfo {volume} {1704}},\ \bibinfo {pages}
  {009} (\bibinfo {year} {2017})},\ \Eprint {http://arxiv.org/abs/1612.01517}
  {arXiv:1612.01517 [astro-ph.HE]} \BibitemShut {NoStop}%
\bibitem [{\citenamefont {Abbasi}\ \emph {et~al.}(2019)\citenamefont {Abbasi}
  \emph {et~al.}}]{Abbasi:2018ywn}%
  \BibitemOpen
  \bibfield  {author} {\bibinfo {author} {\bibfnamefont {R.~U.}\ \bibnamefont
  {Abbasi}} \emph {et~al.} (\bibinfo {collaboration} {Telescope Array}),\
  }\href {\doibase 10.1016/j.astropartphys.2019.03.003} {\bibfield  {journal}
  {\bibinfo  {journal} {Astropart. Phys.}\ }\textbf {\bibinfo {volume} {110}},\
  \bibinfo {pages} {8} (\bibinfo {year} {2019})},\ \Eprint
  {http://arxiv.org/abs/1811.03920} {arXiv:1811.03920 [astro-ph.HE]}
  \BibitemShut {NoStop}%
\bibitem [{\citenamefont {Aab}\ \emph {et~al.}(2013)\citenamefont {Aab} \emph
  {et~al.}}]{ThePierreAuger:2013eja}%
  \BibitemOpen
  \bibfield  {author} {\bibinfo {author} {\bibfnamefont {A.}~\bibnamefont
  {Aab}} \emph {et~al.} (\bibinfo {collaboration} {Pierre Auger}),\ }in\ \href
  {http://lss.fnal.gov/archive/2013/conf/fermilab-conf-13-285-ad-ae-cd-td.pdf}
  {\emph {\bibinfo {booktitle} {{Proceedings, 33rd International Cosmic Ray
  Conference (ICRC2013): Rio de Janeiro, Brazil, July 2-9, 2013}}}}\ (\bibinfo
  {year} {2013})\ \Eprint {http://arxiv.org/abs/1307.5059} {arXiv:1307.5059
  [astro-ph.HE]} \BibitemShut {NoStop}%
\bibitem [{\citenamefont {Abu-Zayyad}\ \emph {et~al.}(2015)\citenamefont
  {Abu-Zayyad} \emph {et~al.}}]{Abu-Zayyad:2013qwa}%
  \BibitemOpen
  \bibfield  {author} {\bibinfo {author} {\bibfnamefont {T.}~\bibnamefont
  {Abu-Zayyad}} \emph {et~al.} (\bibinfo {collaboration} {Telescope Array}),\
  }\href {\doibase 10.1016/j.astropartphys.2014.05.002} {\bibfield  {journal}
  {\bibinfo  {journal} {Astropart. Phys.}\ }\textbf {\bibinfo {volume} {61}},\
  \bibinfo {pages} {93} (\bibinfo {year} {2015})},\ \Eprint
  {http://arxiv.org/abs/1305.7273} {arXiv:1305.7273 [astro-ph.HE]} \BibitemShut
  {NoStop}%
\bibitem [{\citenamefont {{Vainer}}\ and\ \citenamefont
  {{Naselskii}}(1978)}]{1978SvA....22..138V}%
  \BibitemOpen
  \bibfield  {author} {\bibinfo {author} {\bibfnamefont {B.~V.}\ \bibnamefont
  {{Vainer}}}\ and\ \bibinfo {author} {\bibfnamefont {P.~D.}\ \bibnamefont
  {{Naselskii}}},\ }\href@noop {} {\bibfield  {journal} {\bibinfo  {journal}
  {Sov.Astron.}\ }\textbf {\bibinfo {volume} {22}},\ \bibinfo {pages} {138}
  (\bibinfo {year} {1978})}\BibitemShut {NoStop}%
\bibitem [{\citenamefont {Miyama}\ and\ \citenamefont
  {Sato}(1978)}]{10.1143/PTP.59.1012}%
  \BibitemOpen
  \bibfield  {author} {\bibinfo {author} {\bibfnamefont {S.}~\bibnamefont
  {Miyama}}\ and\ \bibinfo {author} {\bibfnamefont {K.}~\bibnamefont {Sato}},\
  }\href {\doibase 10.1143/PTP.59.1012} {\bibfield  {journal} {\bibinfo
  {journal} {Progress of Theoretical Physics}\ }\textbf {\bibinfo {volume}
  {59}},\ \bibinfo {pages} {1012} (\bibinfo {year} {1978})},\ \Eprint
  {http://arxiv.org/abs/http://oup.prod.sis.lan/ptp/article-pdf/59/3/1012/5242831/59-3-1012.pdf}
  {http://oup.prod.sis.lan/ptp/article-pdf/59/3/1012/5242831/59-3-1012.pdf}
  \BibitemShut {NoStop}%
\bibitem [{\citenamefont {{Zeldovich}}\ \emph {et~al.}(1977)\citenamefont
  {{Zeldovich}}, \citenamefont {{Starobinskii}}, \citenamefont {{Khlopov}},\
  and\ \citenamefont {{Chechetkin}}}]{1977SvAL....3..110Z}%
  \BibitemOpen
  \bibfield  {author} {\bibinfo {author} {\bibfnamefont {I.~B.}\ \bibnamefont
  {{Zeldovich}}}, \bibinfo {author} {\bibfnamefont {A.~A.}\ \bibnamefont
  {{Starobinskii}}}, \bibinfo {author} {\bibfnamefont {M.~I.}\ \bibnamefont
  {{Khlopov}}}, \ and\ \bibinfo {author} {\bibfnamefont {V.~M.}\ \bibnamefont
  {{Chechetkin}}},\ }\href@noop {} {\bibfield  {journal} {\bibinfo  {journal}
  {Soviet Astronomy Letters}\ }\textbf {\bibinfo {volume} {3}},\ \bibinfo
  {pages} {110} (\bibinfo {year} {1977})}\BibitemShut {NoStop}%
\bibitem [{\citenamefont {{Vainer}}\ \emph {et~al.}(1978)\citenamefont
  {{Vainer}}, \citenamefont {{Dryzhakova}},\ and\ \citenamefont
  {{Naselskii}}}]{1978SvAL....4..185V}%
  \BibitemOpen
  \bibfield  {author} {\bibinfo {author} {\bibfnamefont {B.~V.}\ \bibnamefont
  {{Vainer}}}, \bibinfo {author} {\bibfnamefont {O.~V.}\ \bibnamefont
  {{Dryzhakova}}}, \ and\ \bibinfo {author} {\bibfnamefont {P.~D.}\
  \bibnamefont {{Naselskii}}},\ }\href@noop {} {\bibfield  {journal} {\bibinfo
  {journal} {Soviet Astronomy Letters}\ }\textbf {\bibinfo {volume} {4}},\
  \bibinfo {pages} {185} (\bibinfo {year} {1978})}\BibitemShut {NoStop}%
\bibitem [{\citenamefont {Kohri}\ and\ \citenamefont
  {Yokoyama}(2000)}]{Kohri:1999ex}%
  \BibitemOpen
  \bibfield  {author} {\bibinfo {author} {\bibfnamefont {K.}~\bibnamefont
  {Kohri}}\ and\ \bibinfo {author} {\bibfnamefont {J.}~\bibnamefont
  {Yokoyama}},\ }\href {\doibase 10.1103/PhysRevD.61.023501} {\bibfield
  {journal} {\bibinfo  {journal} {Phys. Rev.}\ }\textbf {\bibinfo {volume}
  {D61}},\ \bibinfo {pages} {023501} (\bibinfo {year} {2000})},\ \Eprint
  {http://arxiv.org/abs/astro-ph/9908160} {arXiv:astro-ph/9908160 [astro-ph]}
  \BibitemShut {NoStop}%
\bibitem [{\citenamefont {{Zeldovich}}\ and\ \citenamefont
  {{Starobinskii}}(1976)}]{1976Zel}%
  \BibitemOpen
  \bibfield  {author} {\bibinfo {author} {\bibfnamefont {I.~B.}\ \bibnamefont
  {{Zeldovich}}}\ and\ \bibinfo {author} {\bibfnamefont {A.~A.}\ \bibnamefont
  {{Starobinskii}}},\ }\href@noop {} {\bibfield  {journal} {\bibinfo  {journal}
  {ZhETF Pisma Redaktsiiu}\ }\textbf {\bibinfo {volume} {24}},\ \bibinfo
  {pages} {616} (\bibinfo {year} {1976})}\BibitemShut {NoStop}%
\bibitem [{\citenamefont {Carr}(1976)}]{Carr:1976zz}%
  \BibitemOpen
  \bibfield  {author} {\bibinfo {author} {\bibfnamefont {B.~J.}\ \bibnamefont
  {Carr}},\ }\href {\doibase 10.1086/154351} {\bibfield  {journal} {\bibinfo
  {journal} {Astrophys. J.}\ }\textbf {\bibinfo {volume} {206}},\ \bibinfo
  {pages} {8} (\bibinfo {year} {1976})}\BibitemShut {NoStop}%
\bibitem [{\citenamefont {Zeldovich}(1976)}]{Zeldovich:1976vw}%
  \BibitemOpen
  \bibfield  {author} {\bibinfo {author} {\bibfnamefont {{\relax Ya}.~B.}\
  \bibnamefont {Zeldovich}},\ }\href@noop {} {\bibfield  {journal} {\bibinfo
  {journal} {Pisma Zh. Eksp. Teor. Fiz.}\ }\textbf {\bibinfo {volume} {24}},\
  \bibinfo {pages} {29} (\bibinfo {year} {1976})}\BibitemShut {NoStop}%
\bibitem [{\citenamefont {Toussaint}\ \emph {et~al.}(1979)\citenamefont
  {Toussaint}, \citenamefont {Treiman}, \citenamefont {Wilczek},\ and\
  \citenamefont {Zee}}]{Toussaint:1978br}%
  \BibitemOpen
  \bibfield  {author} {\bibinfo {author} {\bibfnamefont {D.}~\bibnamefont
  {Toussaint}}, \bibinfo {author} {\bibfnamefont {S.~B.}\ \bibnamefont
  {Treiman}}, \bibinfo {author} {\bibfnamefont {F.}~\bibnamefont {Wilczek}}, \
  and\ \bibinfo {author} {\bibfnamefont {A.}~\bibnamefont {Zee}},\ }\href
  {\doibase 10.1103/PhysRevD.19.1036} {\bibfield  {journal} {\bibinfo
  {journal} {Phys. Rev.}\ }\textbf {\bibinfo {volume} {D19}},\ \bibinfo {pages}
  {1036} (\bibinfo {year} {1979})}\BibitemShut {NoStop}%
\bibitem [{\citenamefont {Turner}(1979)}]{Turner:1979bt}%
  \BibitemOpen
  \bibfield  {author} {\bibinfo {author} {\bibfnamefont {M.~S.}\ \bibnamefont
  {Turner}},\ }\href {\doibase 10.1016/0370-2693(79)90095-9} {\bibfield
  {journal} {\bibinfo  {journal} {Phys. Lett.}\ }\textbf {\bibinfo {volume}
  {89B}},\ \bibinfo {pages} {155} (\bibinfo {year} {1979})}\BibitemShut
  {NoStop}%
\bibitem [{\citenamefont {Grillo}(1980)}]{Grillo:1980rt}%
  \BibitemOpen
  \bibfield  {author} {\bibinfo {author} {\bibfnamefont {A.~F.}\ \bibnamefont
  {Grillo}},\ }\href {\doibase 10.1016/0370-2693(80)90897-7} {\bibfield
  {journal} {\bibinfo  {journal} {Phys. Lett.}\ }\textbf {\bibinfo {volume}
  {94B}},\ \bibinfo {pages} {364} (\bibinfo {year} {1980})}\BibitemShut
  {NoStop}%
\bibitem [{\citenamefont {Zhang}\ \emph {et~al.}(2007)\citenamefont {Zhang},
  \citenamefont {Chen}, \citenamefont {Kamionkowski}, \citenamefont {Si},\ and\
  \citenamefont {Zheng}}]{Zhang:2007zzh}%
  \BibitemOpen
  \bibfield  {author} {\bibinfo {author} {\bibfnamefont {L.}~\bibnamefont
  {Zhang}}, \bibinfo {author} {\bibfnamefont {X.}~\bibnamefont {Chen}},
  \bibinfo {author} {\bibfnamefont {M.}~\bibnamefont {Kamionkowski}}, \bibinfo
  {author} {\bibfnamefont {Z.-g.}\ \bibnamefont {Si}}, \ and\ \bibinfo {author}
  {\bibfnamefont {Z.}~\bibnamefont {Zheng}},\ }\href {\doibase
  10.1103/PhysRevD.76.061301} {\bibfield  {journal} {\bibinfo  {journal} {Phys.
  Rev.}\ }\textbf {\bibinfo {volume} {D76}},\ \bibinfo {pages} {061301}
  (\bibinfo {year} {2007})},\ \Eprint {http://arxiv.org/abs/0704.2444}
  {arXiv:0704.2444 [astro-ph]} \BibitemShut {NoStop}%
\bibitem [{\citenamefont {Barnacka}\ \emph {et~al.}(2012)\citenamefont
  {Barnacka}, \citenamefont {Glicenstein},\ and\ \citenamefont
  {Moderski}}]{Barnacka:2012bm}%
  \BibitemOpen
  \bibfield  {author} {\bibinfo {author} {\bibfnamefont {A.}~\bibnamefont
  {Barnacka}}, \bibinfo {author} {\bibfnamefont {J.~F.}\ \bibnamefont
  {Glicenstein}}, \ and\ \bibinfo {author} {\bibfnamefont {R.}~\bibnamefont
  {Moderski}},\ }\href {\doibase 10.1103/PhysRevD.86.043001} {\bibfield
  {journal} {\bibinfo  {journal} {Phys. Rev.}\ }\textbf {\bibinfo {volume}
  {D86}},\ \bibinfo {pages} {043001} (\bibinfo {year} {2012})},\ \Eprint
  {http://arxiv.org/abs/1204.2056} {arXiv:1204.2056 [astro-ph.CO]} \BibitemShut
  {NoStop}%
\bibitem [{\citenamefont {Graham}\ \emph {et~al.}(2015)\citenamefont {Graham},
  \citenamefont {Rajendran},\ and\ \citenamefont {Varela}}]{Graham:2015apa}%
  \BibitemOpen
  \bibfield  {author} {\bibinfo {author} {\bibfnamefont {P.~W.}\ \bibnamefont
  {Graham}}, \bibinfo {author} {\bibfnamefont {S.}~\bibnamefont {Rajendran}}, \
  and\ \bibinfo {author} {\bibfnamefont {J.}~\bibnamefont {Varela}},\ }\href
  {\doibase 10.1103/PhysRevD.92.063007} {\bibfield  {journal} {\bibinfo
  {journal} {Phys. Rev.}\ }\textbf {\bibinfo {volume} {D92}},\ \bibinfo {pages}
  {063007} (\bibinfo {year} {2015})},\ \Eprint
  {http://arxiv.org/abs/1505.04444} {arXiv:1505.04444 [hep-ph]} \BibitemShut
  {NoStop}%
\bibitem [{\citenamefont {Capela}\ \emph {et~al.}(2013)\citenamefont {Capela},
  \citenamefont {Pshirkov},\ and\ \citenamefont {Tinyakov}}]{Capela:2013yf}%
  \BibitemOpen
  \bibfield  {author} {\bibinfo {author} {\bibfnamefont {F.}~\bibnamefont
  {Capela}}, \bibinfo {author} {\bibfnamefont {M.}~\bibnamefont {Pshirkov}}, \
  and\ \bibinfo {author} {\bibfnamefont {P.}~\bibnamefont {Tinyakov}},\ }\href
  {\doibase 10.1103/PhysRevD.87.123524} {\bibfield  {journal} {\bibinfo
  {journal} {Phys. Rev.}\ }\textbf {\bibinfo {volume} {D87}},\ \bibinfo {pages}
  {123524} (\bibinfo {year} {2013})},\ \Eprint {http://arxiv.org/abs/1301.4984}
  {arXiv:1301.4984 [astro-ph.CO]} \BibitemShut {NoStop}%
\bibitem [{\citenamefont {Niikura}\ \emph {et~al.}(2019)\citenamefont {Niikura}
  \emph {et~al.}}]{Niikura:2017zjd}%
  \BibitemOpen
  \bibfield  {author} {\bibinfo {author} {\bibfnamefont {H.}~\bibnamefont
  {Niikura}} \emph {et~al.},\ }\href {\doibase 10.1038/s41550-019-0723-1}
  {\bibfield  {journal} {\bibinfo  {journal} {Nat. Astron.}\ }\textbf {\bibinfo
  {volume} {3}},\ \bibinfo {pages} {524} (\bibinfo {year} {2019})},\ \Eprint
  {http://arxiv.org/abs/1701.02151} {arXiv:1701.02151 [astro-ph.CO]}
  \BibitemShut {NoStop}%
\bibitem [{\citenamefont {Smyth}\ \emph {et~al.}(2020)\citenamefont {Smyth},
  \citenamefont {Profumo}, \citenamefont {English}, \citenamefont {Jeltema},
  \citenamefont {McKinnon},\ and\ \citenamefont
  {Guhathakurta}}]{Smyth:2019whb}%
  \BibitemOpen
  \bibfield  {author} {\bibinfo {author} {\bibfnamefont {N.}~\bibnamefont
  {Smyth}}, \bibinfo {author} {\bibfnamefont {S.}~\bibnamefont {Profumo}},
  \bibinfo {author} {\bibfnamefont {S.}~\bibnamefont {English}}, \bibinfo
  {author} {\bibfnamefont {T.}~\bibnamefont {Jeltema}}, \bibinfo {author}
  {\bibfnamefont {K.}~\bibnamefont {McKinnon}}, \ and\ \bibinfo {author}
  {\bibfnamefont {P.}~\bibnamefont {Guhathakurta}},\ }\href {\doibase
  10.1103/PhysRevD.101.063005} {\bibfield  {journal} {\bibinfo  {journal}
  {Phys. Rev.}\ }\textbf {\bibinfo {volume} {D101}},\ \bibinfo {pages} {063005}
  (\bibinfo {year} {2020})},\ \Eprint {http://arxiv.org/abs/1910.01285}
  {arXiv:1910.01285 [astro-ph.CO]} \BibitemShut {NoStop}%
\bibitem [{\citenamefont {Katz}\ \emph {et~al.}(2018)\citenamefont {Katz},
  \citenamefont {Kopp}, \citenamefont {Sibiryakov},\ and\ \citenamefont
  {Xue}}]{Katz:2018zrn}%
  \BibitemOpen
  \bibfield  {author} {\bibinfo {author} {\bibfnamefont {A.}~\bibnamefont
  {Katz}}, \bibinfo {author} {\bibfnamefont {J.}~\bibnamefont {Kopp}}, \bibinfo
  {author} {\bibfnamefont {S.}~\bibnamefont {Sibiryakov}}, \ and\ \bibinfo
  {author} {\bibfnamefont {W.}~\bibnamefont {Xue}},\ }\href {\doibase
  10.1088/1475-7516/2018/12/005} {\bibfield  {journal} {\bibinfo  {journal}
  {JCAP}\ }\textbf {\bibinfo {volume} {1812}},\ \bibinfo {pages} {005}
  (\bibinfo {year} {2018})},\ \Eprint {http://arxiv.org/abs/1807.11495}
  {arXiv:1807.11495 [astro-ph.CO]} \BibitemShut {NoStop}%
\bibitem [{\citenamefont {Conroy}\ \emph {et~al.}(2011)\citenamefont {Conroy},
  \citenamefont {Loeb},\ and\ \citenamefont {Spergel}}]{Conroy:2010bs}%
  \BibitemOpen
  \bibfield  {author} {\bibinfo {author} {\bibfnamefont {C.}~\bibnamefont
  {Conroy}}, \bibinfo {author} {\bibfnamefont {A.}~\bibnamefont {Loeb}}, \ and\
  \bibinfo {author} {\bibfnamefont {D.}~\bibnamefont {Spergel}},\ }\href
  {\doibase 10.1088/0004-637X/741/2/72} {\bibfield  {journal} {\bibinfo
  {journal} {Astrophys. J.}\ }\textbf {\bibinfo {volume} {741}},\ \bibinfo
  {pages} {72} (\bibinfo {year} {2011})},\ \Eprint
  {http://arxiv.org/abs/1010.5783} {arXiv:1010.5783 [astro-ph.GA]} \BibitemShut
  {NoStop}%
\bibitem [{\citenamefont {Ibata}\ \emph {et~al.}(2013)\citenamefont {Ibata},
  \citenamefont {Nipoti}, \citenamefont {Sollima}, \citenamefont {Bellazzini},
  \citenamefont {Chapman},\ and\ \citenamefont {Dalessandro}}]{Ibata:2012eq}%
  \BibitemOpen
  \bibfield  {author} {\bibinfo {author} {\bibfnamefont {R.}~\bibnamefont
  {Ibata}}, \bibinfo {author} {\bibfnamefont {C.}~\bibnamefont {Nipoti}},
  \bibinfo {author} {\bibfnamefont {A.}~\bibnamefont {Sollima}}, \bibinfo
  {author} {\bibfnamefont {M.}~\bibnamefont {Bellazzini}}, \bibinfo {author}
  {\bibfnamefont {S.}~\bibnamefont {Chapman}}, \ and\ \bibinfo {author}
  {\bibfnamefont {E.}~\bibnamefont {Dalessandro}},\ }\href {\doibase
  10.1093/mnras/sts302} {\bibfield  {journal} {\bibinfo  {journal} {Mon. Not.
  Roy. Astron. Soc.}\ }\textbf {\bibinfo {volume} {428}},\ \bibinfo {pages}
  {3648} (\bibinfo {year} {2013})},\ \Eprint {http://arxiv.org/abs/1210.7787}
  {arXiv:1210.7787 [astro-ph.CO]} \BibitemShut {NoStop}%
\bibitem [{\citenamefont {Salvio}\ and\ \citenamefont
  {Strumia}(2014)}]{Salvio:2014soa}%
  \BibitemOpen
  \bibfield  {author} {\bibinfo {author} {\bibfnamefont {A.}~\bibnamefont
  {Salvio}}\ and\ \bibinfo {author} {\bibfnamefont {A.}~\bibnamefont
  {Strumia}},\ }\href {\doibase 10.1007/JHEP06(2014)080} {\bibfield  {journal}
  {\bibinfo  {journal} {JHEP}\ }\textbf {\bibinfo {volume} {06}},\ \bibinfo
  {pages} {080} (\bibinfo {year} {2014})},\ \Eprint
  {http://arxiv.org/abs/1403.4226} {arXiv:1403.4226 [hep-ph]} \BibitemShut
  {NoStop}%
\bibitem [{\citenamefont {Carney}\ \emph {et~al.}(2019)\citenamefont {Carney},
  \citenamefont {Ghosh}, \citenamefont {Krnjaic},\ and\ \citenamefont
  {Taylor}}]{Carney:2019pza}%
  \BibitemOpen
  \bibfield  {author} {\bibinfo {author} {\bibfnamefont {D.}~\bibnamefont
  {Carney}}, \bibinfo {author} {\bibfnamefont {S.}~\bibnamefont {Ghosh}},
  \bibinfo {author} {\bibfnamefont {G.}~\bibnamefont {Krnjaic}}, \ and\
  \bibinfo {author} {\bibfnamefont {J.~M.}\ \bibnamefont {Taylor}},\
  }\href@noop {} {\  (\bibinfo {year} {2019})},\ \Eprint
  {http://arxiv.org/abs/1903.00492} {arXiv:1903.00492 [hep-ph]} \BibitemShut
  {NoStop}%
\bibitem [{\citenamefont {Cai}\ and\ \citenamefont {Wang}(2020)}]{Cai:2019igo}%
  \BibitemOpen
  \bibfield  {author} {\bibinfo {author} {\bibfnamefont {R.-G.}\ \bibnamefont
  {Cai}}\ and\ \bibinfo {author} {\bibfnamefont {S.-J.}\ \bibnamefont {Wang}},\
  }\href {\doibase 10.1103/PhysRevD.101.043508} {\bibfield  {journal} {\bibinfo
   {journal} {Phys. Rev.}\ }\textbf {\bibinfo {volume} {D101}},\ \bibinfo
  {pages} {043508} (\bibinfo {year} {2020})},\ \Eprint
  {http://arxiv.org/abs/1910.07981} {arXiv:1910.07981 [astro-ph.CO]}
  \BibitemShut {NoStop}%
\bibitem [{\citenamefont {Wang}\ \emph {et~al.}(2019)\citenamefont {Wang},
  \citenamefont {Terada},\ and\ \citenamefont {Kohri}}]{Wang:2019kaf}%
  \BibitemOpen
  \bibfield  {author} {\bibinfo {author} {\bibfnamefont {S.}~\bibnamefont
  {Wang}}, \bibinfo {author} {\bibfnamefont {T.}~\bibnamefont {Terada}}, \ and\
  \bibinfo {author} {\bibfnamefont {K.}~\bibnamefont {Kohri}},\ }\href
  {\doibase 10.1103/PhysRevD.99.103531} {\bibfield  {journal} {\bibinfo
  {journal} {Phys. Rev.}\ }\textbf {\bibinfo {volume} {D99}},\ \bibinfo {pages}
  {103531} (\bibinfo {year} {2019})},\ \Eprint
  {http://arxiv.org/abs/1903.05924} {arXiv:1903.05924 [astro-ph.CO]}
  \BibitemShut {NoStop}%
\bibitem [{\citenamefont {Salvio}\ and\ \citenamefont
  {Veermäe}(2019)}]{Salvio:2019llz}%
  \BibitemOpen
  \bibfield  {author} {\bibinfo {author} {\bibfnamefont {A.}~\bibnamefont
  {Salvio}}\ and\ \bibinfo {author} {\bibfnamefont {H.}~\bibnamefont
  {Veermäe}},\ }\href@noop {} {\  (\bibinfo {year} {2019})},\ \Eprint
  {http://arxiv.org/abs/1912.13333} {arXiv:1912.13333 [gr-qc]} \BibitemShut
  {NoStop}%
\bibitem [{\citenamefont {Clark}\ \emph {et~al.}(2017)\citenamefont {Clark},
  \citenamefont {Dutta}, \citenamefont {Gao}, \citenamefont {Strigari},\ and\
  \citenamefont {Watson}}]{Clark:2016nst}%
  \BibitemOpen
  \bibfield  {author} {\bibinfo {author} {\bibfnamefont {S.}~\bibnamefont
  {Clark}}, \bibinfo {author} {\bibfnamefont {B.}~\bibnamefont {Dutta}},
  \bibinfo {author} {\bibfnamefont {Y.}~\bibnamefont {Gao}}, \bibinfo {author}
  {\bibfnamefont {L.~E.}\ \bibnamefont {Strigari}}, \ and\ \bibinfo {author}
  {\bibfnamefont {S.}~\bibnamefont {Watson}},\ }\href {\doibase
  10.1103/PhysRevD.95.083006} {\bibfield  {journal} {\bibinfo  {journal} {Phys.
  Rev.}\ }\textbf {\bibinfo {volume} {D95}},\ \bibinfo {pages} {083006}
  (\bibinfo {year} {2017})},\ \Eprint {http://arxiv.org/abs/1612.07738}
  {arXiv:1612.07738 [astro-ph.CO]} \BibitemShut {NoStop}%
\bibitem [{\citenamefont {Page}\ and\ \citenamefont
  {Hawking}(1976)}]{Page:1976wx}%
  \BibitemOpen
  \bibfield  {author} {\bibinfo {author} {\bibfnamefont {D.~N.}\ \bibnamefont
  {Page}}\ and\ \bibinfo {author} {\bibfnamefont {S.~W.}\ \bibnamefont
  {Hawking}},\ }\href {\doibase 10.1086/154350} {\bibfield  {journal} {\bibinfo
   {journal} {Astrophys. J.}\ }\textbf {\bibinfo {volume} {206}},\ \bibinfo
  {pages} {1} (\bibinfo {year} {1976})}\BibitemShut {NoStop}%
\bibitem [{\citenamefont {MacGibbon}\ and\ \citenamefont
  {Carr}(1991)}]{MacGibbon:1991vc}%
  \BibitemOpen
  \bibfield  {author} {\bibinfo {author} {\bibfnamefont {J.~H.}\ \bibnamefont
  {MacGibbon}}\ and\ \bibinfo {author} {\bibfnamefont {B.~J.}\ \bibnamefont
  {Carr}},\ }\href {\doibase 10.1086/169909} {\bibfield  {journal} {\bibinfo
  {journal} {Astrophys. J.}\ }\textbf {\bibinfo {volume} {371}},\ \bibinfo
  {pages} {447} (\bibinfo {year} {1991})}\BibitemShut {NoStop}%
\bibitem [{\citenamefont {Strong}\ \emph {et~al.}(2004)\citenamefont {Strong},
  \citenamefont {Moskalenko},\ and\ \citenamefont {Reimer}}]{Strong:2004ry}%
  \BibitemOpen
  \bibfield  {author} {\bibinfo {author} {\bibfnamefont {A.~W.}\ \bibnamefont
  {Strong}}, \bibinfo {author} {\bibfnamefont {I.~V.}\ \bibnamefont
  {Moskalenko}}, \ and\ \bibinfo {author} {\bibfnamefont {O.}~\bibnamefont
  {Reimer}},\ }\href {\doibase 10.1086/423196} {\bibfield  {journal} {\bibinfo
  {journal} {Astrophys. J.}\ }\textbf {\bibinfo {volume} {613}},\ \bibinfo
  {pages} {956} (\bibinfo {year} {2004})},\ \Eprint
  {http://arxiv.org/abs/astro-ph/0405441} {arXiv:astro-ph/0405441 [astro-ph]}
  \BibitemShut {NoStop}%
\bibitem [{\citenamefont {Sreekumar}\ \emph {et~al.}(1998)\citenamefont
  {Sreekumar} \emph {et~al.}}]{Sreekumar:1997un}%
  \BibitemOpen
  \bibfield  {author} {\bibinfo {author} {\bibfnamefont {P.}~\bibnamefont
  {Sreekumar}} \emph {et~al.} (\bibinfo {collaboration} {EGRET}),\ }\href
  {\doibase 10.1086/305222} {\bibfield  {journal} {\bibinfo  {journal}
  {Astrophys. J.}\ }\textbf {\bibinfo {volume} {494}},\ \bibinfo {pages} {523}
  (\bibinfo {year} {1998})},\ \Eprint {http://arxiv.org/abs/astro-ph/9709257}
  {arXiv:astro-ph/9709257 [astro-ph]} \BibitemShut {NoStop}%
\end{thebibliography}%
\bibliographystyle{apsrev4-1}

\end{document}